\documentclass[prc,twocolumn,showpacs,nofootinbib]{revtex4-1}
\usepackage[utf8]{inputenc}
\usepackage{color}
\usepackage{amsmath}
\usepackage{url}
\usepackage{graphicx}
\usepackage{amssymb}
\usepackage[dvipsnames]{xcolor}
\usepackage{appendix}
\usepackage[section]{placeins}
\usepackage{multirow}
\usepackage{bm}
\usepackage{soul}
\usepackage{array}
\usepackage{lipsum}
\usepackage{relsize}
\usepackage{amsfonts}
\usepackage[normalem]{ulem}
\newcommand{\rhozero}{\rho_{\raisebox{-2.0pt}{\tiny\!0}}}

\newcommand\numberthis{\addtocounter{equation}{1}\tag{\theequation}}
\begin{document}

\title{From noise to information: The transfer function formalism for uncertainty quantification on nuclear density reconstruction}
\author{P. G. Giuliani$^{1,2}$}\email{pgg15@my.fsu.edu}\email{giulianp@frib.msu.edu}
\author{J. Piekarewicz$^1$}\email{jpiekarewicz@fsu.edu}

\affiliation{$^1$Department of Physics, Florida State University, 
               Tallahassee, FL 32306, USA}
\affiliation{$^2$Department of Statistics and Probability, and Facility for Rare Isotope Beams, Michigan State University, East Lansing, Michigan 48824, USA}

\date{\today}

\begin{abstract}

\textbf{Background:} The neutron distribution of neutron-rich nuclei provides critical information on the structure of finite nuclei and neutron stars. Parity violating experiments --- such as PREX and CREX --- provide a clean and largely model-independent determination of neutron densities. Such experiments, however, are challenging and expensive which is why sound statistical arguments are required to maximize the information gained.

\textbf{Purpose:} To introduce a new framework, ``the transfer function formalism", aimed at uncertainty quantification, model selection, and experimental design in the context of neutron densities.

\textbf{Methods:} The transfer functions (TFs) are built analytically by expressing the linear response of the objective function (e.g., $\chi^2$) to small perturbations of the data. Using the TF formalism, we are able to analyze the expected overall uncertainty --- quantified in terms of bias and variance --- of the mean square radius and interior density of $^{48}$Ca and $^{208}$Pb.

\textbf{Results:} Using relativistic mean field models as a proxy for the weak-charge density --- and assuming that a total of five measurements could be performed on the weak form factor of $^{48}$Ca and $^{208}$Pb --- we identify the optimal models and experimental locations that minimize the uncertainty in the extraction of the radius and interior density. We also explore the use of the TF formalism to understand the influence of prior distributions for the model parameters, as well as the optimization of model hyperparameters not constrained by the data.

\textbf{Conclusions:} We establish how the choice of experimental locations and the model that is used can have a significant impact on the final uncertainties of the extracted quantities of interest. For challenging experiments such as CREX and PREX, a proper quantification of such uncertainties is critical. We have demonstrated how the TF formalism provides several advantages for this type of analysis.
\end{abstract}

\maketitle

\section{Introduction}

Nuclear saturation, the existence of an equilibrium density, is a hallmark of the nuclear dynamics. Shortly after Chadwick's discovery of the neutron, the semi-empirical  mass formula of Bethe and Weizs\"acker\,\cite{Weizsacker:1935,Bethe:1936} was conceived to predict the binding energy of atomic nuclei. Using only a handful of parameters, the semi-empirical mass formula provides a remarkably good description of the masses of stable nuclei by regarding the atomic nucleus as an incompressible quantum drop consisting of $Z$ protons and $N$ neutrons ($A\!=\!Z\!+\!N$). Among the earliest predictions of the semi-empirical mass formula was the $A^{1/3}$ scaling of the nuclear size. Indeed, assuming an incompressible drop at an equilibrium (or ``saturation") density of $\rhozero\!\approx\!0.15\,{\rm fm}^{-3}$,  yields a root-mean-square radius of:
\begin{equation}
 R(A)=r_{{}_{\!0}}A^{1/3}\,, \;{\rm where}\;
 r_{{}_{\!0}}\!=\!\sqrt[3]{\frac{3}{4\pi\rhozero}}\!\approx 1.17\,{\rm fm}\,.
 \label{Radius}
\end{equation}
While the description of atomic nuclei as an incompressible quantum drop has stood the test of time, we now know that at a finer scale the distribution of nucleons is much more interesting and complex. Shell corrections, deformation, and pairing correlations are all important dynamical effects that impact the spatial distribution in atomic nuclei.
To date, the most precise knowledge of the nuclear density comes from mapping the charge distribution of atomic nuclei \cite{Fricke:1995,Angeli:2013}. Starting with the pioneering work of Hofstadter in the late 1950's \cite{Hofstadter:1956qs} and continuing to this day, elastic electron scattering has painted the most accurate picture of the atomic nucleus. Our knowledge of the nuclear size, surface thickness, and saturation density all originate from such studies that have provided some of the most stringent constrains on nuclear properties. For example, the root-mean-square charge radius of 
$^{208}$Pb is known with exquisite precision: $R_{\rm ch}\!=\!5.5012(13)\,{\rm fm}$\,\cite{Angeli:2013}. 

Electron scattering is an ideal tool to map the charge distribution because the electromagnetic interaction is well known and
the coupling (``fine structure") constant is small. So, in a plane wave impulse approximation, the differential cross section for 
the elastic scattering of an electron from a spinless target may be written as follows\,\cite{Walecka:2001}:
\begin{equation}
    \left(\frac{d\sigma}{d\Omega}\right) =  
    \left(\frac{d\sigma}{d\Omega}\right)_{\rm\!\!Mott}\!\!\!\!
    Z^{2}F_{\rm ch}^{2}(Q^2),
\label{dedOEM}
\end{equation}
where $Z$ is the electric charge of the nucleus and $Q^{2}\!\ge\!0$ is the square of the space-like 
four-momentum transfer. The Mott cross section represents the scattering of a relativistic (massless) 
electron from a spinless and structureless target, and is given exclusively in terms of kinematical 
variables and the fine structure constant. Deviations from the structureless limit are encoded in the 
charge form factor, which has been normalized to one at zero momentum transfer 
$F_{\rm ch}(Q^2\!=\!0)\!=\!1$. The distribution of electric charge in a nucleus---which is carried mostly 
by the protons---is obtained from the Fourier transform of the charge form factor.

This favorable situation stands in stark contrast to our knowledge of the distribution of weak charge, 
which is dominated by the neutrons because the weak charge of the proton is 
small\,\cite{Androic:2013rhu,Androic:2018kni}. Probing neutron densities has traditionally relied on 
hadronic experiments involving strongly interacting probes, such as pions, protons, and alpha
particles, that are hindered by uncontrolled approximations related to the reaction mechanism, 
medium-modifications to the underlying two-body interaction, and hadronic distortions. For a recent 
review on this topic see Ref.\,\cite{Thiel:2019tkm} and references contained therein. For symmetric 
($N\!=\!Z$) nuclei, the expectation is that both proton and neutron densities will have the same shape, 
with the proton distribution extending slightly farther out because of the Coulomb repulsion. However, 
for heavy neutron-rich nuclei---which best illustrate the notion of nuclear saturation---the excess 
neutrons are pushed out against surface tension, creating a neutron-rich skin. Indeed, the interior 
baryon density of ${}^{208}$Pb is expected to be fairly constant and close to $\rhozero$.  As such, 
the interior baryon density of $^{208}$Pb may provide the physical observable that is most closely 
related to $\rhozero$\,\cite{Horowitz:2020evx}.

It is also possible to measure weak charge densities with much smaller systematic uncertainties by 
relying on electroweak probes that offer a clean and model-independent alternative to strongly interacting 
probes. However, this requires a more challenging and sophisticated class of experiment, such as 
coherent elastic neutrino-nucleus scattering (CEvNS) or parity violating electron nucleus scattering. 
The enormous advantage of these electroweak experiments is that the weak $Z^0$ boson couples 
preferentially to neutrons because of the small weak charge of the 
proton\,\cite{Androic:2013rhu,Androic:2018kni}. For example, in the case of CEvNS the cross section 
is directly proportional to the square of the weak charge form factor. 
That is\,\cite{Scholberg:2005qs,Yang:2019pbx}, 
\begin{equation}
   \left(\frac{d\sigma}{dT}\right) = \frac{G_{\!F}^{2}}{8\pi} M
   \left[2 - 2 \frac{T}{E} - \frac{MT}{E^{2}} \right]
   Q_{\rm wk}^{2}F_{\rm wk}^{2}(Q^{2}),
\label{CEvens}
\end{equation}
where $G_{\!F}$ is Fermi's constant, $Q_{\rm wk}\!=\!-\!N\!+\!(1\!-\!4\sin^{2}\!\theta_{\rm W})Z$ is the 
weak charge of the nucleus written in terms of the weak-mixing angle, and the weak form factor has 
been normalized to $F_{\rm wk}^{2}(Q^{2}\!=\!0)\!=\!1$. The remaining quantities are all of kinematical 
origin: $E$ is the incident neutrino energy, $T$ the nuclear recoil energy, and $Q^{2}\!=\!2MT$. In 
particular, at forward angles the differential cross section is proportional to the square of the weak 
charge of the nucleus $Q_{\rm wk}^{2}\!\approx\!N^{2}$. The approximate $N^{2}$ scaling is the 
hallmark of the coherent reaction and the main reason for the identification by Freedman of CEvNS 
as having favorable cross sections\,\cite{Freedman:1973yd}, even if it took more than four decades 
for its experimental confirmation\,\cite{Akimov:2017ade,Akimov:2019rhz}.

Although CEvNS holds enormous promise in the determination of neutron densities, the parity-violating 
electron program has become a precision tool in the determination of both hadronic/nuclear structure 
and electroweak physics. Following the 30-year old idea by Donnelly, Dubach, and Sick who proposed 
the use of parity violating electron scattering (PVES) as a clean probe of neutron 
densities\,\cite{Donnelly:1989qs}, the pioneering Lead Radius EXperiment (PREX) at the Jefferson 
Laboratory (JLab) extracted the weak radius of ${}^{208}$Pb, providing for the first time model-independence 
evidence in favor of a neutron-rich skin\,\cite{Abrahamyan:2012gp,Horowitz:2012tj}. To reach the original 
goal of a $\pm0.06\,{\rm fm}$ determination of the weak radius of ${}^{208}$Pb, the follow-up PREX-II campaign has now been completed and has delivered on the promise to determine
the neutron radius of ${}^{208}$Pb with a precision that is about 3 times better than the original
PREX measurement. By combining both experiments the following value for the neutron skin thickness
of ${}^{208}$Pb was  reported\,\cite{Adhikari:2021phr}:
$R_{\rm skin}=R_{n}-R_{p}=(0.283\pm0.071)\,{\rm fm}$. This result challenges several
experimental measurements and theoretical predictions that systematically underestimate
the newly reported value of $R_{\rm skin}$\,\cite{Thiel:2019tkm}. At the same time, the ongoing 
CREX campaign will provide the first electroweak determination of the weak radius of 
$^{48}$Ca\,\cite{CREX:2013,Horowitz:2013wha}. Beyond JLab, the Mainz Energy recovery 
Superconducting Accelerator (MESA), envisioned to start operations by 2023\,\cite{Becker:2018ggl},
may be able to determine the weak radius of both $^{48}$Ca and ${}^{208}$Pb with increased 
precision\,\cite{Thiel:2019tkm}. Besides its intrinsic value as a fundamental nuclear-structure observable, 
the neutron skin thickness of ${}^{208}$Pb, defined as the difference between the neutron 
and proton root-mean-square radii $R_{\rm skin}\!\equiv\!R_{n}\!-\!R_{p}$, is strongly correlated to the 
slope of the symmetry energy at saturation
density\,\cite{Brown:2000,Furnstahl:2001un,Centelles:2008vu,RocaMaza:2011pm}. The symmetry energy at saturation density is a fundamental 
parameter of the equation of state of neutron-rich matter that impacts the structure, composition, and cooling 
mechanism of neutron stars\,\cite{Horowitz:2000xj,Horowitz:2001ya,Carriere:2002bx,
Steiner:2004fi,Erler:2012qd,Chen:2014sca,Chen:2014mza}.

A parity violating asymmetry emerges from the difference in the scattering between right- and left-handed 
polarized electrons. In a plane wave impulse approximation, the parity violating asymmetry from a spinless 
target may be written as follows\,\cite{Donnelly:1989qs}:
\begin{equation}
  A_{PV}(Q^{2})  = - \frac{G_{\!F}Q^{2}}{4\pi\alpha\sqrt{2}}
                              \frac{Q_{\rm wk}F_{\rm wk}(Q^{2})}{ZF_{\rm ch}(Q^{2})},
\label{APV}
\end{equation}
where $\alpha$ is the fine structure constant and the nuclear contribution enters as the ratio of the weak to
the charge form factor. Given that $F_{\rm ch}$ is known from (parity conserving) electron scattering 
measurements, the parity violating asymmetry determines the weak form factor which, in turn, is dominated 
by the neutron distribution. 

To date, PREX, PREX-II, and CREX have focused on extracting the weak radius $R_{\rm wk}$ from a single 
measurement at a relatively low momentum transfer. Yet additional features of the weak charge density can 
be revealed by measuring the parity violating asymmetry at higher momentum transfers. In particular, if 
$A_{PV}$ could be measured at several momentum transfers, then the entire weak charge form factor and 
its associated density could be determined. Such experimental program may required measurements of 
$A_{PV}$ at about six values of $Q^{2}$, a task that may be feasible for ${}^{48}$Ca\,\cite{Lin:2015ata}. 
For ${}^{208}$Pb, such a task is significantly more challenging given that at high momentum transfer the 
elastic cross section is small because of the strong suppression from the nuclear form factor. Nevertheless, 
with two experimental points it may be sufficient to gain valuable insights into the weak charge form factor of
 ${}^{208}$Pb over a significant range of momentum transfers\,\cite{Piekarewicz:2016vbn,Horowitz:2020evx}. 
 Regardless, with asymmetries of the order of one part per 
million\,\cite{Abrahamyan:2012gp,Horowitz:2012tj}, PVES experiments are both highly expensive and 
enormously challenging, so robust statistical arguments---above and beyond a compelling physics case---should 
be made in the quest for an optimal experimental design. Such is the central goal of the present manuscript.

In this paper, we present a novel statistical analysis--the ``transfer function formalism"--inspired from the treatment 
of noise in signal processing theory\,\cite{aastrom2010feedback}. In such a framework, the transfer function is 
a general function that models a device output for each possible input. In our particular case, we define the 
transfer function in terms of coefficients that encode the linear part of the response of the fitted model parameters 
to small changes in the data inputs. We have already implemented an early version of these ideas to estimate the 
bias and variance of models within the proton puzzle context\,\cite{higinbotham2018bias} and in 
Ref.\,\cite{gueye2020dispersive} to estimate the effect of dispersive corrections on the $^{12}$C elastic cross section.

Within the transfer function formalism, the noise is propagated in the measured observable to the uncertainty in the 
quantity of interest. Given that each single measurement in the data has an associated transfer function, an important
feature of the formalism is that we can identify those \textit{critical points}, if any, that are responsible for driving most 
of the uncertainty. For example, in this manuscript we are interested in quantifying the statistical error in the extracted 
weak charge radii of both ${}^{48}$Ca and ${}^{208}$Pb from the experimental error in their corresponding weak charge 
form factor.  Values of the form factor with higher transfer functions will propagate their errors more efficiently to the total 
variance of the calculated weak charge radii. Using the transfer function ({$\mathcal{T\!F}$}) formalism, we aim to 
quantify the ability of seven different models to accurately determine both the interior (saturation) density and mean square radius of the weak charge distribution. Given that the electric charge distribution of both nuclei is accurately known, 
we are able to validate our formalism against known data before making predictions for the unknown weak charge 
distribution. 

The performance of the seven models is evaluated in terms of bias and variance\,\cite{hastie2009elements},
similar to the approach implemented in\,\cite{Yan:2018bez,Higinbotham:2018jfh} to extract the charge radius of 
the proton from electron scattering data. The ``bias-variance trade-off" is an important concept in statistics and 
machine learning that addresses the complexity of a model. If the model is too simple, it will result in a poor 
description of the data (underfit=high bias). If the model is too complex, it will be extremely sensitive to the 
random dispersion in the data (overfit=high variance). The bias-variance trade-off is the inevitable conflict that
ensues when trying to simultaneously minimize these two critical sources of error. 

The rest of this paper is organized as follows. Sec.~\ref{Sec: Theoretical Brackground} includes a brief review of the 
main concepts involved in the discussion of nuclear form factors and density distributions. We also discuss 
statistical concepts related to our proposed formalism, such as Bayesian inference and bias-variance trade-off. 
Sec.~\ref{Sec: TF formalism} presents a detailed account of the transfer function formalism and how it is 
implemented in the context of the bias-variance trade-off. 
Sec.~\ref{Sec: Results} contains a compilation of our main results. We start this section by testing and 
validating our method using the experimentally known charge densities of both $^{48}$Ca and $^{208}$Pb 
as a proxy for the unknown weak charge densities. 
Finally, Sec.~\,\ref{Sec: Conclusions} presents our final remarks and vision for the future.  In addition, we 
provide several appendices that contain useful information in the form of supporting tables and figures, as 
well as mathematical proofs of the central concepts that have been developed.

The core idea of the transfer function formalism is that for small perturbations in the input of a system, the response of the system is perturbed a \textit{proportional} amount. This idea is clearly not new and it has been implemented in many scientific and engineering problems for centuries (consider for example the concept of Green's functions). On the statistics front, we have found several related concepts such as the adjoint method (page 203 \cite{sullivan2015introduction}), the influence functions (page 45 \cite{huber2004robust}), and the sensitivity of the system response (Sec. III F in \cite{cacuci2005sensitivity}), for example. However, despite our best efforts, we were not able to find a direct application to model selection, the analysis of the influence of priors, and the description of both bias and variance, such as the one we developed in this work.

\section{Theoretical Background}\label{Sec: Theoretical Brackground}

\subsection{Nuclear Density and Form Factor}

The electric charge density $\rho_{\rm ch}({\bf r})$ and the weak charge density $\rho_{\rm wk}({\bf r})$ 
describe the spatial distribution of electric charge and weak charge in the atomic nucleus, respectively. 
In the case of $\rho_{\rm ch}({\bf r})$, elastic electron scattering experiments determine the ground state
charge density by measuring the differential cross section, which for a spinless nucleus is given by 
Eq.\,(\ref{dedOEM}). In the case of the weak charge density, the aim is to extract the weak charge 
form factor from measuring the parity violating asymmetry given by Eq.\eqref{APV}.

Having extracted the corresponding form factors $F_\text{ch}$ and $F_\text{wk}$ from experiments, 
the nuclear charge density and weak charge density are obtained trough a Fourier transform. To 
simplify the notation, no subscripts (either ${\rm ``ch"}$ or ${\rm ``wk"}$) will be included henceforth, 
except when this omission may create confusion. The density and form factor are related as follows:
\begin{equation}
  \rho({\bf r}) = \int \frac{d^{3}q}{(2\pi)^{3}}
   e^{i{\bf q} \cdot {\bf r}} F({\bf q}), 
    \label{Rhoch0} 
\end{equation}
where ${|\bf q|}\!=\!q=\!\!\sqrt{Q^{2}}$ in the limit in which the nuclear recoil can be ignored. For a 
spinless nucleus the density distribution is spherically symmetric so it becomes
\begin{equation}
  \rho(r) = \frac{1}{2\pi^{2}r}\int_{0}^{\infty} 
   \!\!F(q) \sin(qr)qdq.
  \label{Rhoch1}
\end{equation}
Alternatively, the inverse Fourier transform can be written as:
\begin{align}
  F(q) = \frac{4\pi}{q}\int_{0}^{\infty} 
   \!\!\rho(r) \sin(qr)rdr.
  \label{FFch}
\end{align}
Note that we have adopted the following normalization condition for both electric and weak distributions: 
\begin{equation}
    F(q\!=\!0) = \int\!\rho(r) d^3r=1.
 \label{Eq:Charge Normalization}   
\end{equation}
Finally, the mean-squared radius of the spatial distribution is given by:
\begin{equation}
  R^2\equiv \langle r^2 \rangle = 
  \int\!\rho(r) r^{2}d^3r = 4 \pi\!\int_{0}^{\infty} \rho(r) r^4 dr.
  \label{MSR}  
\end{equation}

\subsection{Models, parameters, and errors}
\label{Sec: Models, parameters and errors}

Several parametrizations (or models) exist in the literature to describe nuclear densities and 
their associated form factors\,\cite{de1987nuclear}. In this paper, we study the performance 
of seven models in total: Fourier Bessel\,\cite{dreher1974determination}, Helm\,\cite{helm1956inelastic}, 
Symmetrized Fermi Function (SF)\,\cite{sprung1997symmetrized} of two, three and four parameters, 
and two hybrid models obtained from combining the SF  with a Fourier Bessel expansion (SF+B) 
and the SF with a sum of Gaussians (SF+G). Note that we did not consider the original Sum of 
Gaussians model\,\cite{sick1974model} since certain conditions were difficult to implement within
the transfer function formalism. Moreover, we found that for the small (5) number of data points here 
considered, the Sum of Gaussians did not provide a good fit to the data. Appendix\,\ref{App: Models Details} 
describes in detail the seven models employed in this work.

We assume that we have collected $J$ experimental data points that we write as 
$\boldsymbol{Y}\!=\!\{(q_j,y_j,\sigma _{j})\}$, where $q_j$ is the $j$th value of the momentum transfer, 
$y_j$ is the value of the form factor at $q_j$, and $\sigma _{j}$ is the associated experimental error.
In turn, we refer to the set of $K$ calibration parameters of any particular model as 
$\boldsymbol{\omega}\!=\!\{\omega_k\}$. Finally, we denote as $m\!=\!m(\boldsymbol{\omega})$ 
the quantity of interest that we want to estimate from the given experimental data. Such quantity, 
for example, the mean square radius of the weak-charge distribution, depends  on the selection 
of experimental points through the fitted parameters $\boldsymbol{\omega}$.

\subsubsection{Standard Fitting Protocol}
\label{Sec: Standard Fit}

A traditional approach used to estimate the optimal set of parameters $\boldsymbol{\omega}$ that 
best describes the observed data, is to minimize the sum of the squares of the residuals between 
the experiment and the model predictions. The residuals are contained in an objective (or cost) 
function $\chi^2$ defined as follows:
\begin{equation}\label{Eq: chi^2 Definition}
    \chi^2= \sum_{j=1}^J \frac{(F(q_j,\boldsymbol{ \omega}) -y_j)^2}{\sigma _{\!j}^2},
\end{equation}
where $F(q_j,\boldsymbol{\omega})$ represents the model predictions of the form factor. 
The optimal set of fitted parameters is obtained by minimizing the objective function and is denoted 
by $\boldsymbol{\omega_0}\equiv \text{argmin}(\chi^2)$. Fundamental to the quantification of the 
model uncertainties is the behavior of the objective function in the vicinity of the optimal point 
$\boldsymbol{\omega_0}$. Such a behavior is imprinted in the Hessian matrix of $\chi^2$ which 
is computed from its second derivatives evaluated at the optimal value. That is, matrix elements 
of the $K\!\times\!K$ Hessian matrix are given by:
\begin{align*}
   & \mathcal{H}_{i,k}\equiv \frac{1}{2}\Big(\frac{\partial^2 \chi^2}{\partial \omega_i \partial \omega_k}\Big)_0 =\numberthis \label{Eq: chi2 Hessian} \\ &\sum_{j=1}^J \frac{1}{\sigma_j^2}\Bigg[\Big(\frac{\partial F(q_j,\boldsymbol{\omega})}{\partial \omega_i}\Big)\Big(\frac{\partial F(q_j,\boldsymbol{\omega})}{\partial \omega_k}\Big) \\
   &+ (F(q_j,\boldsymbol{\omega})-y_j)\frac{\partial^2 F(q_j,\boldsymbol{\omega})}{\partial \omega_i \partial \omega_k}\Bigg]_0.
\end{align*}
The inverse of the Hessian matrix $\mathcal{H}^{-1}$, often called the error or covariance matrix, is used 
to estimate uncertainty and correlations associated with the fitted parameters as well as with other 
quantities\,\cite{Bevington2003}. For example, the square of the standard error (or standard deviation) of
$m(\boldsymbol{\omega})$ is given by
\begin{equation}\label{Eq: Standard Fit}
\Delta m^2=\nabla m \mathcal{H}^{-1} \nabla m \Big|_{\boldsymbol{\omega_0}},
\end{equation}
where $\nabla m$ is the gradient of $m$ with respect to the parameters $\omega_k$, and all quantities 
are evaluated at $\boldsymbol{\omega}=\boldsymbol{\omega_0}$. 

\subsubsection{Bayesian Approach}
\label{Sec: Bayesian Approach}

An alternative framework to estimate model parameters and to quantify their statistical properties which 
has been gaining popularity in the physics community is the Bayesian approach\,\cite{Gregory:2005,Stone:2013}. 
Within this framework, the posterior distribution of model parameters $\boldsymbol{\omega}$ given the 
experimental data $\boldsymbol{Y}$ is given by Bayes' theorem:
\begin{equation}
    P(\boldsymbol{\omega}|\boldsymbol{Y})=
    \frac{P(\boldsymbol{Y}|\boldsymbol{\omega})
    P(\boldsymbol{\omega})}{P(\boldsymbol{Y})},
 \label{BayesPosterior}   
\end{equation}
where $P(\boldsymbol{Y}|\boldsymbol{\omega})$ is the likelihood that a given set of model parameters describes 
the experimental data, $P(\boldsymbol{\omega})$ is the prior distribution of model parameters, and 
$P(\boldsymbol{Y})$ is the evidence, which can be treated as a normalization constant to enforce 
$\int P(\boldsymbol{\omega}|\boldsymbol{Y})d\boldsymbol{\omega}=1$. The prior distribution encapsulates our 
prior knowledge (or beliefs) of the distribution of model parameters. Such prior beliefs will be refined as a 
result of the additional experimental information contained in the likelihood, which ultimately yields an updated 
distribution of model parameters $P(\boldsymbol{\omega}|\boldsymbol{Y})$.

Once the posterior distribution $P(\boldsymbol{\omega}|\boldsymbol{Y})$ is obtained, the average value of any quantity $m$ and its associated error may be estimated from integrating over the probability distribution. That is,
\begin{subequations}
\begin{align} \label{Eq: Bayesian Fit}
    &\langle m\rangle = \int m(\boldsymbol{\omega}) P(\boldsymbol{\omega}|\boldsymbol{Y}) d\boldsymbol{\omega}, \\
   & \Delta m^2 = \int \Big(m(\boldsymbol{\omega})-\langle m\rangle\Big)^2 P(\boldsymbol{\omega}|\boldsymbol{Y}) 
   d\boldsymbol{\omega},\label{Eq: Bayesian Var}
\end{align}
\end{subequations}
where $\langle m \rangle$ denotes the average---or central---value of $m$. In the case of the likelihood, it is often 
assumed that it is related to the $\chi^2$ function introduced in Eq.\eqref{Eq: chi^2 Definition} as follows: 
\begin{equation}
 P(\boldsymbol{Y}|\boldsymbol{\omega}) = e^{-\chi^{2}(\boldsymbol{Y}\!,\boldsymbol{\omega})/2}.
 \label{Likelihood} 
\end{equation}
Hence, reference to the maximum likelihood is equivalent to the minimum value of $\chi^2$. For the prior 
distribution it is common to assume an uncorrelated Gaussian distribution of model parameters, namely,
\begin{subequations}
\begin{align}\label{Eq: Prior Form}
    P(\boldsymbol{\omega}) = e^{-\phi^2(\boldsymbol{\omega})/2},   \ \text{where} \\
    \phi^2(\boldsymbol{\omega})= \sum_{k=1}^{K} 
    \left(\frac{\omega_k-\omega_k^0}{\sigma_k}\right)^{2},\label{Eq: prior phi}
\end{align}

\end{subequations}
where $\omega_k^0$ is our prior estimate for the central value of $\omega_k$ and $\sigma_k$ is the 
estimated uncertainty. Small values of $\sigma_k$ will make the distribution sharply peaked around
$\omega_k^0$ and  the fitting procedure more ``prior driven". Conversely, large values of $\sigma_k$ 
reflect a large uncertainty in the model parameters so the fitting procedure becomes more ``data driven". 
Under the prior and likelihood definitions, the posterior distribution takes the following form:
\begin{align}\label{Eq: chi2 tilde}
 P(\boldsymbol{\omega}|\boldsymbol{Y}) = 
 e^{-\widetilde\chi^2(\boldsymbol{Y},\boldsymbol{\omega})/2} =
 e^{-\left(\chi^2(\boldsymbol{Y},\boldsymbol{\omega})+\phi^2(\boldsymbol{\omega})\right)/2},
\end{align}
where $\widetilde\chi^2$ now encodes contributions from both the likelihood and the prior. For an optimal 
point $\boldsymbol{\omega_0}\!=\!\text{argmin}(\widetilde\chi^2)$, the behavior of $\widetilde\chi^2$ 
around the minimum is encoded in the augmented Hessian matrix $\widetilde{\mathcal{H}}$ defined as:
\begin{equation}\label{Eq: chi2 tilde Hessian}
    \widetilde{\mathcal{H}}_{i,k}\equiv \frac{1}{2}\Big(\frac{\partial^2 \widetilde\chi^2}{\partial \omega_i \partial \omega_k}\Big)_0 = \mathcal{H}_{i,k}+\delta_{ik} \frac{1}{\sigma_k^2},
\end{equation}
where $\mathcal{H}$ is the Hessian of $\chi^2$ defined in Eq.\eqref{Eq: chi2 Hessian} and $\delta_{ik}$ is the 
Kronecker delta. If the adopted prior includes correlations between the different parameters, then Eq.\eqref{Eq: prior phi} 
will be written as a quadratic form $\phi^2(\boldsymbol{\omega})\!=\!\boldsymbol{ \omega} \Sigma^{-1} \boldsymbol{\omega}$, 
where the matrix $\Sigma$ contains the (prior) covariances between parameters. In such a case 
Eq.\eqref{Eq: chi2 tilde Hessian} would have to be modified accordingly.

\subsection{Bias, Variance, and MSE}\label{Sec: ScoreMetric}

Our objective is to identify which of the seven models defined in Sec.~\ref{Sec: Models, parameters and errors}
will best perform---using a criterion to be precisely defined shortly---in extracting the radius and interior density
when faced with real experimental data on the weak charge form factor. Given that the experimental results 
have yet to be published, we rely on synthetic data generated by a set of five covariant energy density functionals
that we refer to as \textit{generators}: $^n F_\text{true}(q)$ ($n\!=\!1,\ldots,5$). The particular set of accurately
calibrated  functionals are: RMF012, RMF016 (commonly referred to as ``FSUGarnet"), RMF022, RMF028 
and RMF032\,\cite{Chen:2014mza}. The main difference among these generators is the assumed value for the 
yet to be accurately determined neutron skin thickness of $^{208}$Pb; for example, RMF022 predicts a neutron 
skin thickness of $\sim\!0.22$\,fm. For each data point generated for the weak charge form factor there is an 
associated error $\sigma_j$ which resembles realistic experimental uncertainties. Once a generator is 
selected, any observable of interest $m$ can be calculated directly from the synthetic data. 

As in Refs\,\cite{yan2018robust,higinbotham2018bias}, we evaluate the performance of each of the seven 
models using a bias-variance trade-off criterion. Bias is understood as the discrepancy between the true value 
of $m$ (coming from one of the generators $^n F_\text{true}$) and the extracted value. In contrast, the 
variance is the spread in the extracted value of $m$ as given by the square of the standard deviation (SD);
see Eqs.~\eqref{Eq: Standard Fit} and~\eqref{Eq: Bayesian Var}. Thus, we quantify the performance of a model
by combining the bias and variance into the Mean Squared Error (MSE) defined as
\begin{equation}\label{Eq:MSE}
    \text{MSE}^2(m,\boldsymbol{q},\boldsymbol{\sigma},n) \equiv \text{Bias}^2 +\text{Variance}.
\end{equation}
Note that we have highlighted the dependence of the MSE on the quantity $m$, the locations of the 
momentum-transfer points $\boldsymbol{q}$, the associated experimental errors $\boldsymbol{\sigma}$,
and the generator index $n$. The MSE is a good indicator of the score, as it captures the bias vs variance
trade-off often present in predictive models across the fields of statistics and machine 
learning \cite{hastie2009elements}. Finally, we define the squared average of the MSE by combining 
the predictions from the $n$ different generators:
\begin{equation}\label{Eq: Average MSE}
    \langle\text{MSE}\rangle^2(m,\boldsymbol{q},\boldsymbol{\sigma})= \frac{1}{N}\sum_{n=1}^N \text{MSE}^2(m,\boldsymbol{q},\boldsymbol{\sigma},n).
\end{equation}
The same formula may be used to obtain the squared average of the bias and variance from the
different ``truths" (generators).

\begin{figure}[t]
	\centering
		\includegraphics[width=0.4\textwidth]{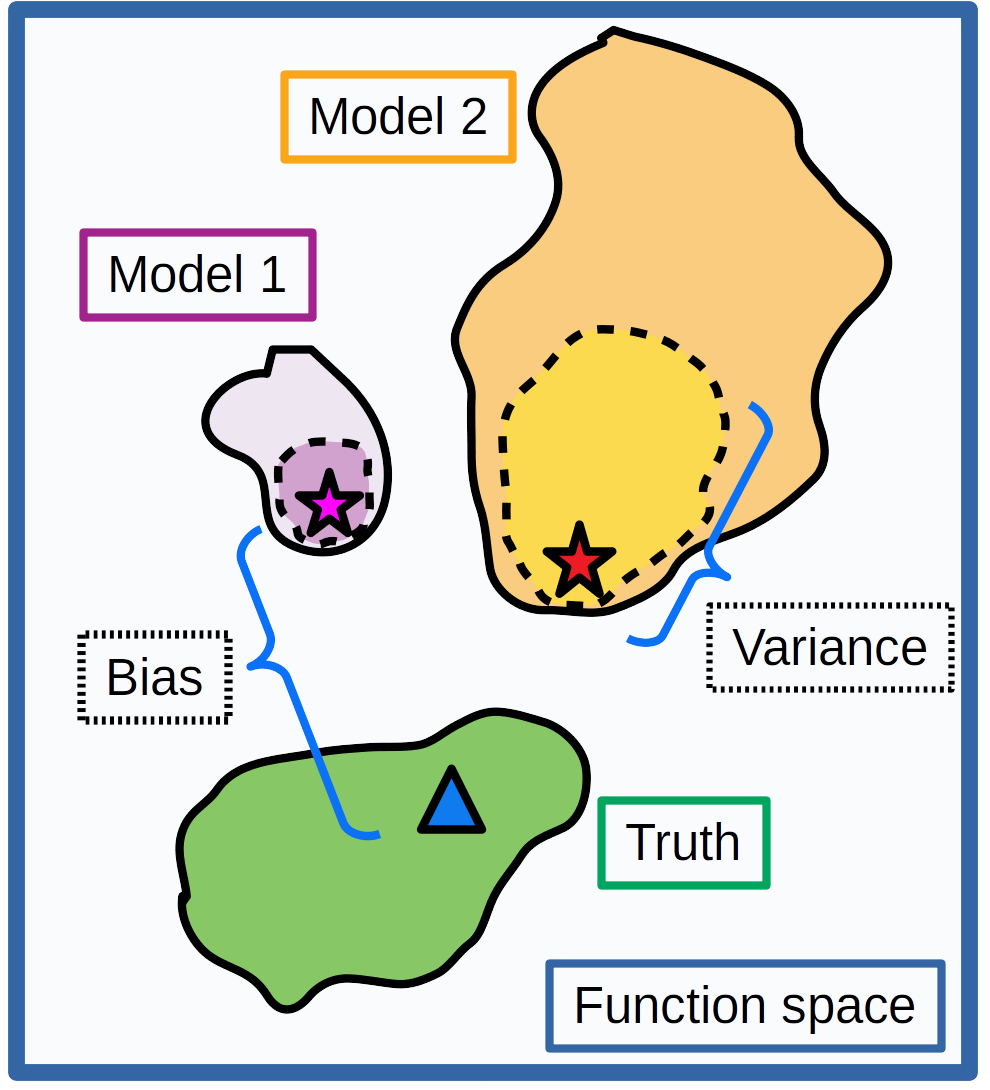}
	\caption{ Abstract representation of the impact of bias and variance on recovering information. The entire function space is represented by the blue enclosing square. The green blob represents the collections of all the truths (generators) such as RMF012 while the blue triangle is one of its members. The purple and yellow blobs represents all the possible members (for different parameters sets) from model 1 and model 2 respectively. The purple and red stars are two particular members of those groups. The bias is shown as the distance between the recovered members (the stars) of each model and the blue triangle. The variance is shown as the dashed contours surrounding each star.}\label{fig:FunctionSpace}
\end{figure}

An abstract representation of these concepts is illustrated in Fig.~\ref{fig:FunctionSpace}. On the entire function space 
depicted with the blue surrounding box, the truth region (in green) is assumed to be spanned by the set of all generators, 
with the blue triangle within this region representing a single member of such family (for example RMF022). 
The set of possible functions adopted to reproduce the data are also displayed. For example, Model 1 (in purple) 
could be the Symmetrized Fermi function whereas Model 2 (in orange) could be the Bessel expansion. In turn, the 
purple and red stars are the members of these respective families that are obtained after fitting the data generated 
by the blue triangle. The corresponding stars are associated with specific values of their 
parameters $\boldsymbol{\omega}$. Under some metric which depends on our choice for $m$, the ``distance" from
the stars to the triangle will represent the bias. In the example, the bias is larger for Model 1. Due to the unavoidable
errors in the experimental data, there will be uncertainty in the exact location of both stars. This uncertainty is 
represented by the dashed contour which size illustrates the variance for each model; in this example the variance
is larger for Model 2. Once we allow the blue triangle to explore the ``truth space", the combination of the accumulated 
bias and variance makes the score, as indicated in Eq.(\ref{Eq: Average MSE}). The task is to identify the model with 
the best score, which emerges from a compromise between the bias and variance.

A possible approach to calculate the bias and variance for each model would be to create many noisy realizations 
of the data to accumulate enough statistics and then apply the standard fitting protocol described in 
Sec.~\ref{Sec: Standard Fit} \cite{yan2018robust}. An alternative approach would be to directly compute the Bayesian 
integrals highlighted in Sec.~\ref{Sec: Bayesian Approach}\,\cite{piekarewicz2016power}. In the following section we 
present a third option: a new formalism that---under certain assumptions---can speed up these calculations, aid in 
the identification of ``critical" points in the data, provide a highly intuitive picture of the propagation of the uncertainty, 
and be extended from model selection to model building.

\section{Transfer Function Formalism}\label{Sec: TF formalism}

We want to understand how the uncertainty---both in terms of bias and variance---gets propagated from the 
experimental data $\boldsymbol{Y}$ to the observable of interest $m$. To do so, we invoke the ``transfer 
functions", a central concept in signal processing and control theory\,\cite{aastrom2010feedback}: if we can make 
a linear map connecting an arbitrary change in the input to the associated change in the output, then 
analyzing the dynamic response of the system becomes straightforward. Note that for nonlinear systems such 
map is not possible. Nevertheless, if the changes in the input are ``small", then linearizing the system around 
its equilibrium point might suffice for most practical purposes\,\cite{aastrom2010feedback}.

In our case, the ``system" is the $\chi^2$ fit in which the inputs are the experimental data $\boldsymbol{Y}$ and 
the output could be either the model parameters $\boldsymbol{\omega}$ or any quantity $m$. Under the transfer 
function formalism we assume that, once the minimum $\boldsymbol{\omega_0}$ of $\chi^2(\boldsymbol{\omega})$ 
is found, then small changes in the value of the data $y_j$ will also produce small changes in both the parameters
and any observable $m$. That is, we assume that the response of the system to the perturbation is linear. To this
end, our main objective is to write:
\begin{align}\label{Eq: delta m from delta ys}
    \delta m = \sum_{j=1} ^J \mathcal{T\!F}^{m}_{\!j} \ \delta y_j,
\end{align}
where $\delta m$ is the small change in the observable $m$ in response to small changes $\delta y_j$ 
in the experimental data $y_j$. The Transfer Functions (TF), denoted by $\mathcal{T\!F}^{m}_j$, encode the changes in $m$ as a result of a change $\delta y_j$ in a given individual input $y_j$. That is,
there is a total of $J$ transfer functions for each observable $m$. The adopted notation uses a subscript for the
$j$th observation $y_j$ and a superscript for the responding quantity $m$. We can 
now expand $\mathcal{T\!F}^{m}_j$ in terms of the model's parameters as follows:
\begin{equation}
   \mathcal{T\!F}^{m}_j\equiv \frac{\partial m}{\partial y_j} =  
   \sum_{k=1}^K \frac{\partial m}{\partial \omega_k} \frac{\partial \omega_k}{\partial y_j} = 
   \nabla m \cdot \mathcal{T\!F}^{\boldsymbol{\omega}}_j, \numberthis\label{Eq: TF m Definition}
\end{equation}
where $\mathcal{T\!F}^{\boldsymbol{\omega}}_j$ is a $K$-dimensional vector with its components
being the transfer functions connecting a small change in each observation $y_j$ to the response 
of the $k$th model parameter $\omega_k$. That is, in analogy to Eq.(\ref{Eq: delta m from delta ys}) 
we obtains:
\begin{equation}\label{Eq: TF for parameters}
    \delta \omega_k = \sum_{j=1}^J\mathcal{T\!F}^{\omega_k}_j \delta y_j.
\end{equation}
As we show in Appendix~\ref{Sec: App Math Proofs}, the general expression for 
$\mathcal{T\!F}^{\boldsymbol{\omega}}_j$ is given by:
\begin{equation}\label{Eq: TF omega Definition}
    \mathcal{T\!F}^{\boldsymbol{\omega}}_j=  \mathcal{H}^{-1} \nabla\!F(q_j,\boldsymbol{\omega})\sigma_j^{-2},
\end{equation}
where the gradient $\nabla\!F(q_j,\boldsymbol{\omega})$ is taken with respect to the model parameters 
$\boldsymbol{\omega}$.  In the following subsections, we use the transfer functions to calculate both the 
variance and bias of any quantity of interest $m$.

\subsection{Variance calculation}\label{Sec: Variance}
Eq.~\eqref{Eq: delta m from delta ys} allows us to write the linear response of $\emph{any}$ quantity 
$ \delta m$ to a given set of small changes in the observations $\boldsymbol{\delta y}$. We interpret 
the errors in the experimental data as independent, Gaussian distributed random variables with mean zero and standard deviation $\sigma _{\!j}$. Hence, in this scenario, if we identify the perturbations $\delta y_j$ as these Gaussian independent experimental errors, the variance in $\delta m$ may be obtained by adding each 
term in Eq.\eqref{Eq: delta m from delta ys} in quadrature:
\begin{equation}\label{Eq: TF Variance}
     \Delta m^2 =  \sum_{j=1} ^J \Big(\mathcal{T\!F}^{m}_j\Big)^2 \sigma_{\!j}^2,
\end{equation}
where the transfer functions $\mathcal{T\!F}_j^{m}$ are evaluated in the model's parameters that
are obtained from the original central values of the experimental points $\boldsymbol{Y}$ ($\boldsymbol{\delta y}=\boldsymbol{0}$). This is analogous to a Taylor series expansion in which the derivatives of the expanded function are evaluated at the unperturbed variable. Note that
Eq.\eqref{Eq: delta m from delta ys} may still be used to calculate the variance even in the more 
general case when there are correlations or the distribution is not Gaussian. However, in this case 
we would have to perform the appropriate integrals on $\delta m$, as a function of \boldsymbol{$\delta y$}, times 
the joint probability distribution $P(\boldsymbol{\delta y})$.

One of the main advantages of Eq.\eqref{Eq: TF Variance} is that it separates, up to some degree, the contribution from 
each observation $y_j$ to the entire variance $\Delta m^2$. As we show in 
Sec.~\ref{Sec: Ca and Pb Charge Example}, this separation allows us to identify those data points 
having undue influence on the variance. This information could be valuable in experimental design 
through the optimal allocation of resources, such as beam time in scattering experiments (see 
Sec. IV in Ref.\,\cite{lin2015full}). Note that the Hessian $\mathcal{H}^{-1}$ in 
$\mathcal{T\!F}_j^{m}$ effectively mixes all observations, so it is not possible to cleanly isolate the 
contribution from each data point. Nevertheless, Eq.\eqref{Eq: TF Variance} provides a more 
efficient and natural way of addressing the influence of each data point as compared to other 
well-known approaches, such as those represented by Eqs.~\eqref{Eq: Standard Fit} 
and~\eqref{Eq: Bayesian Fit}. We also note that in comparing the variance calculated in 
Eq.\eqref{Eq: TF Variance} to that obtained from the standard approach in Eq.\eqref{Eq: Standard Fit},
the results are identical in the limit in which the nonlinear part of the Hessian matrix 
[the terms proportional to second derivatives of $F$ in Eq.\eqref{Eq: chi2 Hessian}] may be ignored. We give a formal proof of this statement in Appendix~\ref{Sec: App Math Proofs}.
In cases in which the model parametrizations depend nonlinearly on the model parameters, then the
variances will differ. 

So, which (if any) of the two approaches is correct in the event that the calculated variances 
differ from each other? Although the answer is not obvious, the transfer function formalism 
seems to be in agreement with those analyses in which many realizations of the data are 
generated via Monte Carlo sampling\,\cite{yan2018robust}. The traditional approach in Eq.~\eqref{Eq: Standard Fit} deviates from the observed Monte Carlo results, an issue generally discussed in statistics under the name of ``model misspecification" (for more information on this topic see theorem 5.23 and example 5.25 in Ref.\,\cite{van2000asymptotic}). However, 
we note that the accuracy of both approaches deteriorates as the errors in the data become 
large enough for the nonlinearities to become important. In such a case, the Gaussian 
approximation, namely, the notion that the entire $\chi^2$ landscape may be described 
by the second derivatives at the minimum, is no longer valid.

As a final remark, we note that the variance computed as in Eq.\eqref{Eq: TF Variance} 
changes with the location of the momentum transfers $q_j$. This change happens not
only because the experimental errors $\sigma_j$ may change with $q_j$, but also 
because the transfer functions themselves depend on the location of $q_j$.
Indeed, by exploring the available $q$-range, we could find the optimal locations 
that minimize the variance of the quantity of interest. In this way, we can answer a fundamental
question in experimental design: \textit{given the available resources, 
how do we select the optimal locations of $q_j$ to minimize the statistical 
uncertainty?}\cite{piekarewicz2016power}. When exploring the $q$-range we
must be aware that the fitted parameters $\boldsymbol{\omega}$ will also
change, which in turn will impact the value of each of the transfer functions
$\mathcal{T\!F}_{\!j}^m$ introduced in Eq.\eqref{Eq: TF m Definition}. This
suggests the need to refit the optimal parameters every time a new set of $q_j$
is considered. As we shall see below, one of the important results of the present 
formalism is that, under certain assumptions, re-fitting may be skipped
altogether.

\subsection{Bias calculation and the Central Function}\label{Sec:Central Function and Bias}

In this section we study the bias as explained in Sec.~\ref{Sec: ScoreMetric}. That is, the 
discrepancy between the true value of the observable of interest $m$ and the one extracted 
by the model. A traditional way of calculating the bias would be to fit the model parameters 
to the data $\boldsymbol{Y}\!=\!\{(q_j,y_j,\sigma _{j})\}$ by minimizing Eq.\eqref{Eq: chi^2 Definition},
and then calculate $m(\boldsymbol{\omega_0})$. Alternatively, we may compute $\langle m\rangle$ 
from Eq.\eqref{Eq: Bayesian Fit}. In both cases the bias is obtained by subtracting the true value 
$m_\text{true}$. Regardless of the approach, we must either refit the model parameters or 
perform the integrals over the posterior distribution for every combination of points $q_j$ that
we want to test. The main reason to explore the behavior of both the bias and the variance 
as we change the $q_j$ locations is that we may be interested in finding the optimal locations
that minimize the mean squared error defined in Eq.\eqref{Eq:MSE}. As we will show shortly,
once we cast the bias calculation under the TF framework, it is possible to avoid refitting as
we explore different sets of $q_j$ locations. 

As indicated in Eq.\eqref{Eq: TF Variance}, the main sources that contribute to the variance 
are the individual data errors $\sigma_{\!j}$, which get propagated to the quantity of interest
through the transfer functions. To write the corresponding expression for the bias in the context
of the TF formalism, we must identify the sources that replace $\sigma_{\!j}$ in the variance 
equation. To do so, we first study how the fitted parameters obtained from minimizing $\chi^2$ 
evolve in the parameter space, as the observations $q_j$ move in their available momentum transfer range. We refer henceforth to the obtained parameters for a given set of locations $q_j$ as the 
``empirical" parameters $\boldsymbol{\omega_e}$. Note that we employ the specific notation $\boldsymbol{\omega_e}$,
rather than the more general $\boldsymbol{\omega_0}$ defined after Eq.\eqref{Eq: chi^2 Definition}. 
The set $\boldsymbol{\omega_e}$ refers exclusively to parameters obtained directly from data (or pseudo data) without any perturbation, while the set $\boldsymbol{\omega_0}$ represents the minimum of $\chi^2$ in any situation, even when we perturb the data 
by small amounts \boldsymbol{$\delta y$}.

\begin{figure}[t!]
        \centering
                \includegraphics[width=0.45\textwidth]{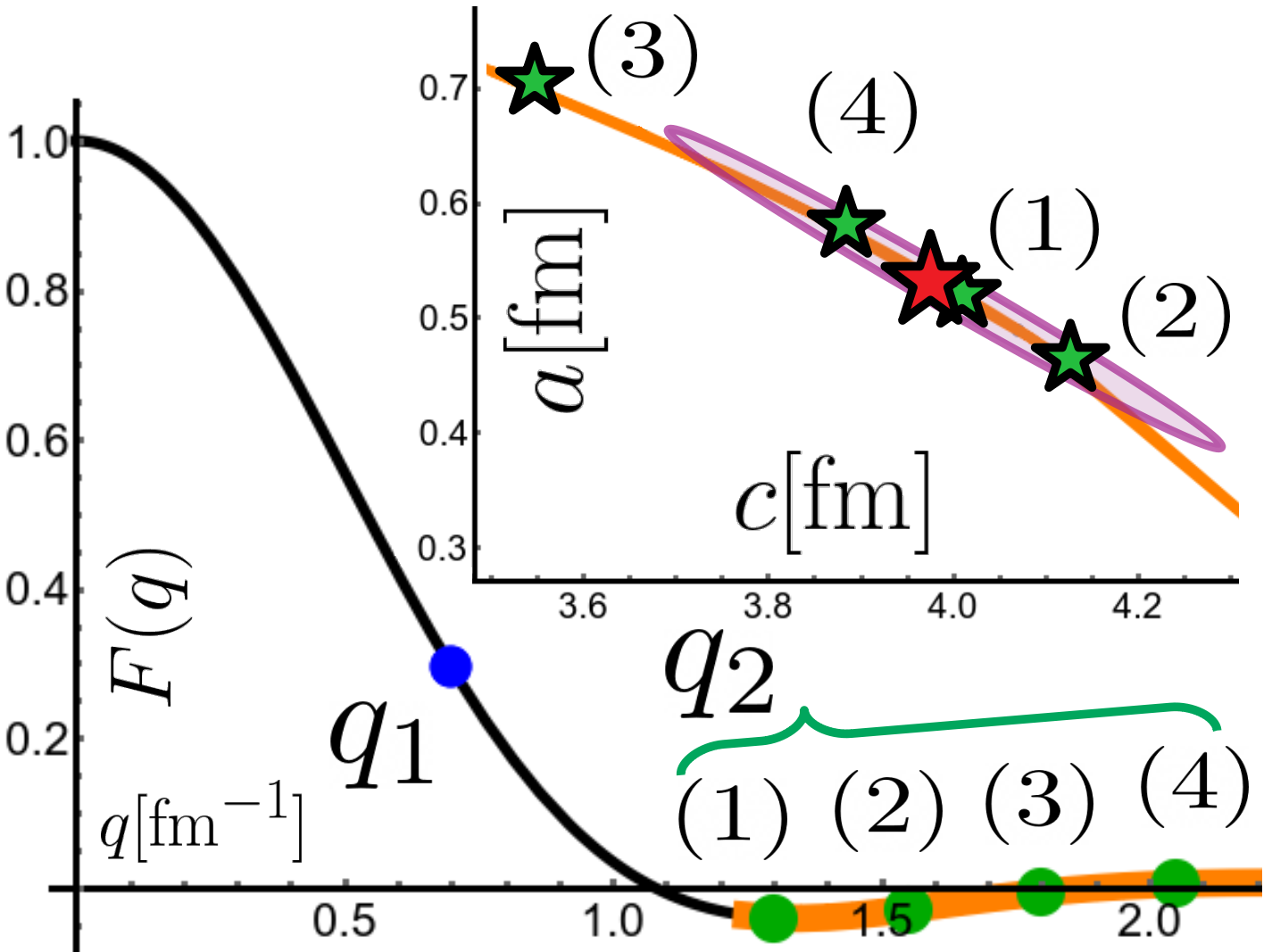}
        \caption{Evolution of $\boldsymbol{\omega_e}$ as the second location $q_2$ is moved in the orange region. The model being fit is the Symmetrized Fermi function of two parameters: $\boldsymbol{\omega}=[c,a]$. The inset plot shows how the orange curve gets mapped into the parameter space, highlighting four locations in green. Although not easy to observe in the plot the mapping ``folds into itself" in the parameter space. The red star represents the central parameters (see the text before Eq.~\eqref{Eq: Etas Omegas}), while the purple ellipse represents the $95\%$ confidence interval for a fit using $\boldsymbol{q}=\{0.7,1.8\}$ fm$^{-1}$ but with $y$ values dictated by the central function (see the text after Eq.~\eqref{Eq: Etas Omegas} and in Fig~\ref{Fig: Etas Example SF}). }\label{Fig: Stars Moving}
\end{figure}

As an example, we show in Fig.~\ref{Fig: Stars Moving} how $\boldsymbol{\omega_e}$ 
evolves as the location of a single measurement changes. In this case, the model being fitted is the 
two-parameter ($\boldsymbol{\omega}=[c,a]$) symmetrized Fermi function. We assume that measurements 
can be made at two different values of the momentum transfer, one fixed at $q_{1}\!=\!0.7$ fm$^{-1}$ and
the other one $q_2$ that is allowed to move along the orange curve in the $[1.25,\ 2.2]$ fm$^{-1}$range . 
The value of the weak form factor of $^{48}$Ca at each of these two points is predicted using the
generator RMF012\,\cite{Chen:2014mza}. The four possible locations of the second point $q_2$, labeled 
respectively as $1,2,3$, and $4$, are displayed as green circles on the orange curve. Each of these locations, 
in combination with $q_1$ determines a single optimized value $\boldsymbol{\omega_e}$. The associated
values of $\boldsymbol{\omega_e}$, one for each choice of $q_2$, are displayed in the inset as the green 
stars in the parameter space. We expect that as the value of the second point changes, so will the value of 
any derived quantity $m$, which will ultimately result in a change to the bias.

\begin{figure}[t!]
        \centering
                \includegraphics[width=0.45\textwidth]{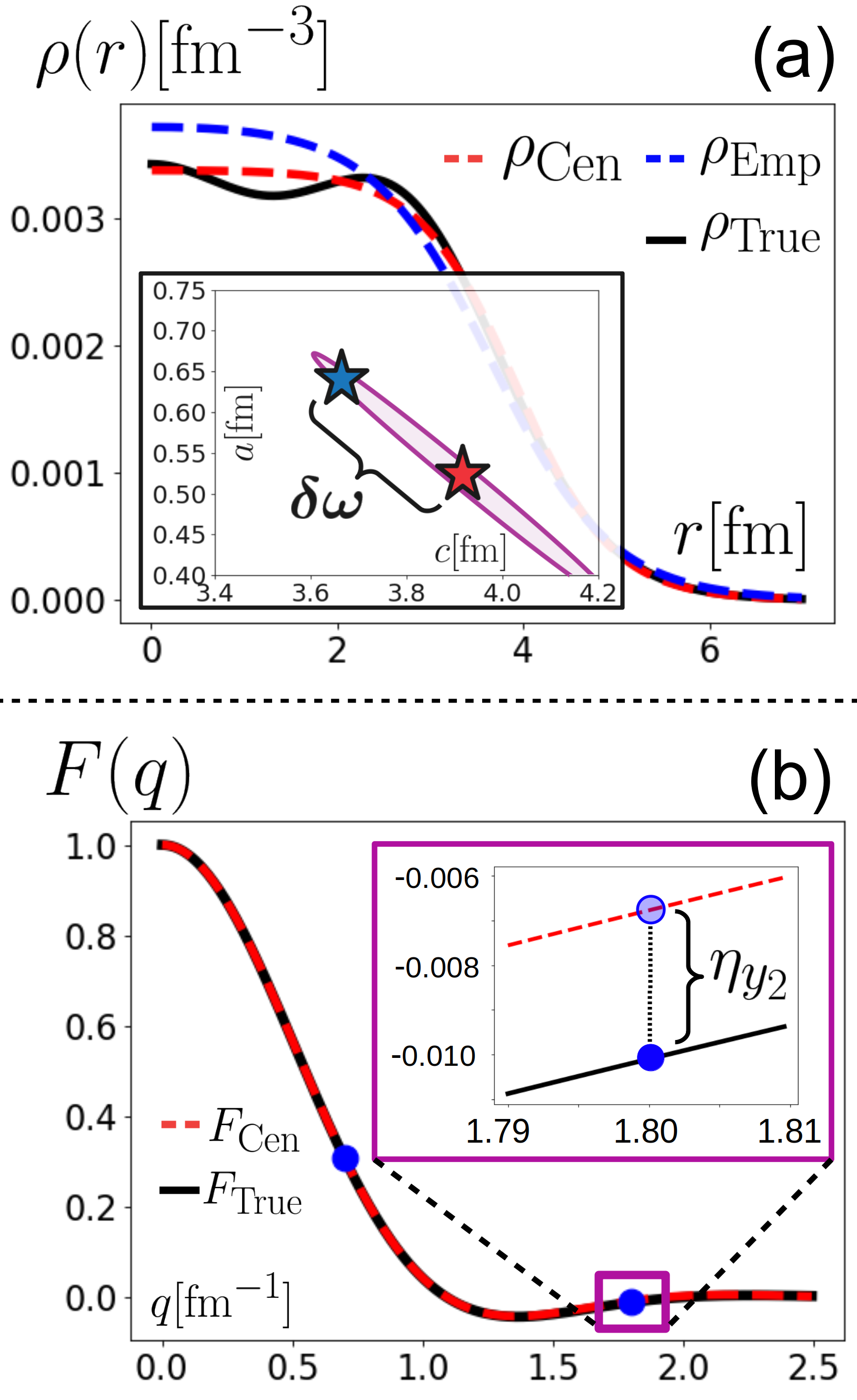}
        \caption{Example of the change in the central parameters driven by deviations between the central and the true form factors. (a) Reconstructed density from $\boldsymbol{\omega_c}$, reconstructed density from  $\boldsymbol{\omega_e}$ using Eq\eqref{Eq: Etas Omegas}, and the true density (red, blue, and black lines, respectively). The inset plot shows the locations in the parameter space of $\boldsymbol{\omega_c}$ and the estimated $\boldsymbol{\omega_e}$ (red and blue stars, respectively). Also shown in purple is the $95\%$ confidence ellipse when fitting the parameters to the central data $\{[q_1, F(q_1,\boldsymbol{\omega_c})],[q_2, F(q_2,\boldsymbol{\omega_c})]\}$ using the errors described in Sec~\ref{Sec: Results}. (b) Central form factor $F(q,\boldsymbol{\omega_c})$ and true form factor (red dashed and black lines, respectively). The inset plot shows a close-up of the difference between these two curves ($\eta_2$) in the neighborhood of $q_2=1.8$ fm$^{-1}$. }\label{Fig: Etas Example SF}
\end{figure}

As can be seen in Fig.~\ref{Fig: Stars Moving} the four empirical parameters $\boldsymbol{\omega_e}$ displayed with the green stars, do not move 
too far away from some \emph{central} value $\boldsymbol{\omega_c}$ (shown as a red star). From this central location and using the transfer 
functions, we can describe the entire \textit{trajectory} of the empirical parameters---particularly how they deviate from the central value 
$\boldsymbol{\omega_c}$ to first order ($ \boldsymbol{\omega_e}\!\approx\!\boldsymbol{\omega_c} +\boldsymbol{\delta \omega}$) for a given set 
of  $J$ observations. That is,
\begin{align}\label{Eq: Etas Omegas}
   & \boldsymbol{\delta \omega} = \sum_{j=1}^J \mathcal{T\!F}^{\boldsymbol{\omega}}_j \ \eta_{j}, \\
  &  \eta_{j} \equiv y_j - F(q_j, \boldsymbol{\omega_c}).
\end{align}

We refer to $\eta_{j}$ as the quantity that is now driving the change in the parameters to distinguish it from an arbitrary perturbation $\delta y_j$. 
The transfer functions in Eq.\eqref{Eq: Etas Omegas} are evaluated at the central parameters $\boldsymbol{\omega_c}$ and in the ``data" 
created by $F(q,\boldsymbol{\omega_c})$. The main idea that we are exploiting is that the minimum of $\chi^2_c$ defined as:
\begin{equation}\label{Eq: chi^2 central}
    \chi^2_c(\boldsymbol{\omega})= \sum_{j=1}^J \frac{\Big(F(q_j,\boldsymbol{ \omega}) -F(q_j,\boldsymbol{ \omega_c})\Big)^2}{\sigma _{\!j}^2},
\end{equation}
where is $\boldsymbol{ \omega_c}\!=\!\text{argmin}(\chi^2_c)$ since $\chi^2_c(\boldsymbol{\omega_c})\!=\!0$. This expression for $\chi^2_c$ is identical 
to the one defined in Eq.~\eqref{Eq: chi^2 Definition}, but with the real observations $y_j$ replaced by $F(q_j,\boldsymbol{ \omega_c})$. From 
Eq.\eqref{Eq: Etas Omegas} we can say that, if the values of $F(q_j,\boldsymbol{ \omega_c})$ are perturbed such that 
$F(q_j,\boldsymbol{ \omega_c})\!\rightarrow\!F(q_j,\boldsymbol{ \omega_c})\!+\!\eta_j\!=\!y_j$, then the central parameters $\boldsymbol{\omega_c}$ 
will respond by moving by $\boldsymbol{\omega_c}\rightarrow\boldsymbol{\omega_c}+\boldsymbol{\delta \omega}\approx \boldsymbol{\omega_e}$.

If this last approximation, $ \boldsymbol{\omega_e}\!\approx\!\boldsymbol{\omega_c} +\boldsymbol{\delta \omega}$, is accurately enough for our 
purposes, then we can say that $m_e\!=\!m(\boldsymbol{\omega_e})$, may be approximated by the central value 
$m_c\!=\!m(\boldsymbol{\omega_c})$ plus a small correction $\delta m$:
\begin{equation}
    m_e = m_c +  \delta m = m_c + \sum_{j=1}^J \mathcal{T\!F}^{m}_j \ \eta_{j}. 
\end{equation}
With these tools at hand, we can write the bias for the quantity of interest $m$ as follows:
\begin{align*}
   \text{Bias}\ (m) &\equiv m_{e} - m_{t} 
   =\Big[m_c  +   \sum_{j=1}^J \mathcal{T\!F}^{m}_j \ \eta_{j} \Big] - m_{t},\numberthis\label{Eq: TF Bias}
\end{align*}
where $m_t$ is the true value of $m$ and the $\mathcal{T\!F}_j^{m}$ are evaluated at the central parameters $\boldsymbol{\omega_c}$. We note that, if $m_c-m_t$ is negligible, the bias is completely driven by the $\eta_j$, analogous to how the variance in Eq.~\eqref{Eq: TF Variance} was driven by the errors $\sigma_j$.

Using the same model and generator as in Fig.~\ref{Fig: Stars Moving}, we display in Fig.~\ref{Fig: Etas Example SF} an estimate of the bias using 
the interior density of ${}^{48}$Ca as the observable of interest. As in Fig.~\ref{Fig: Stars Moving}, we keep the value of the first point fixed at 
$q_1=0.7$ fm$^{-1}$ and select the second point at $q_2=1.8$ fm$^{-1}$, which corresponds to the third point in Fig.~\ref{Fig: Stars Moving}. 
Eq.\eqref{Eq: Etas Omegas} is then used to approximate the empirical parameters, which in turn provide an estimate for the empirical density 
$\rho(r)_\text{Emp}$, which is depicted as the blue dashed line in Fig.~\ref{Fig: Etas Example SF}(a). The deviation of the empirical density from 
the central density $\rho(r)_\text{Cen}$ can be understood in terms of the $\eta_j$'s: the difference between the central form factor 
$F(q,\boldsymbol{\omega_c})$ and the true form factor $F(q)_\text{True}$ evaluated at $q_1$ and $q_2$. To appreciate these minor differences, we 
enlarge a window around $q_2$ and show $\eta_2\!=\!y_2\!-\!F(q_2, \boldsymbol{\omega_c})$ on the inset of Fig.~\ref{Fig: Etas Example SF}(b). The 
hollow blue circle corresponds to $F(q_2,\boldsymbol{\omega_c})$ while the filled one to $F(q_2)_\text{True}$. 

Under the linear approximation assumed in Eq.\eqref{Eq: Etas Omegas}, these $\eta_j$ will move the central $\boldsymbol{\omega_c}$ [red star 
in the inset of Figure (a)] towards the approximated empirical parameters $\boldsymbol{\omega_e}$ (blue star). Since this is a nonlinear model, 
Eq.\eqref{Eq: Etas Omegas} is indeed just an approximation and the change in the parameters in this case was under-predicted. This can be seen 
when comparing the position of the blue star in Fig.~\ref{Fig: Etas Example SF} with the green star (3) in Fig.~\ref{Fig: Stars Moving}. Nevertheless, 
the empirical density is not too different from the density shown in Fig.~\ref{Fig: Etas Example SF} (a).

We close this section by discussing the selection of $\boldsymbol{\omega_c}$. In principle, the precise location of $\boldsymbol{\omega_c}$ 
should not have a significant impact on our calculations provided that the actual change $\boldsymbol{\delta \omega}$ is linear in $\eta_{j}$. 
In the interest of clarity, and given that the experimental observable is the form factor but we are interested in extracting the spatial density, 
we distinguish between two main choices for $\boldsymbol{\omega_c}$:
\begin{enumerate}
\item Central Function Fit: We define $\boldsymbol{\omega_c}$ as the value that minimizes the $L^2$ norm between the
model $F(q,\boldsymbol{\omega})$ and the true function $F(q)_\text{true}$ in the momentum transfer space $q$, as  in 
Figures~\ref{Fig: Stars Moving} and~\ref{Fig: Etas Example SF}. The expectation is that $\boldsymbol{\omega_c}$ 
should be relatively close to most of the possible obtainable parameters for different locations of the data. We refer to these 
parameters as $\boldsymbol{\omega_\text{Cen}}$.

\item Optimal Fit: We define $\boldsymbol{\omega_c}$ as the parameters that make the central estimation $m_c$ as close as 
possible to the true value $m_ t$. For example, if we are interested in modeling the interior density, $\boldsymbol{\omega_c}$ 
should be chosen by fitting the models directly to the spatial density, effectively minimizing the $L^2$ norm between the model density $\rho(r,\boldsymbol{\omega})$ and the true density $\rho(r)_\text{true}$ . Note that this procedure is not feasible in the 
case of real data given that scattering experiments can only access the form factor directly and not the density. However, the 
advantage of this option is that, if $m_c\!-\!m_{t}$ is negligible, then the total bias is dominated by the $\eta_{j}$, making easier the 
search for the optimal locations. We refer to these parameters as $\boldsymbol{\omega_\text{Opt}}$, and will use them 
extensively in Sec.~\ref{Sec: Rec Bias And Opt Func}.
\end{enumerate}

\subsection{Mean Squared Error}

Having constructed the bias and variance within the TF formalism, we write the Mean Squared Error (MSE) as:
\begin{align*}
     &\text{MSE}^2= \numberthis\label{Eq: TF MSE}\\
     &\Big( (m_c- m_{t})+ \sum_{j=1}^J \Big[\mathcal{T\!F}_j^{(m)}\Big]\eta_{j}\Big)^2+ \sum_{j=1}^J \Big[\mathcal{T\!F}_j^{(m)}\Big]^2\sigma_{j}^2.   
\end{align*}
Recall that the MSE is the quantity that we aim to optimize in an effort to find a compromise between the bias 
and the variance. For the specific quantity of interest $m$, the MSE will depend on the selected data points 
$q_{j}$ (e.g., the momentum-transfer points), the 
associated errors $\sigma_j$, and the input values $y_{j}$ (e.g., the weak form factor),  with the last quantity drawn from experimental data or pseudo data 
generated by mean field models. The equations developed in the TF framework enables us to address the 
expected MSE for a given set of experimental data and then report which model has the lowest error, as 
implemented in Ref\,\cite{yan2018robust}. However, if the experiment is still in its design phase, then the 
TF formalism may be used to optimize the MSE not only with respect to the model, but also relative to the 
location of the data and the distribution of errors. 

Naturally, a unique set of central parameters $\boldsymbol{\omega_c}$ will be associated to a given 
model (e.g., Fourier-Bessel) and generator (e.g. RMF012). Thus, unless we suspect that variations
under different choices of model and generator are negligible, each MSE should be calculated with 
its own parameters $\boldsymbol{\omega_c}$. Indeed, these parameters $\boldsymbol{\omega_c}$ 
are necessary for the numerical calculation of each $\mathcal{T\!F}^{m}$. Moreover, it is important to 
note that the $\mathcal{T\!F}_j^m$ from the bias term in Eq.\eqref{Eq: TF MSE} are evaluated at the
central parameters $\boldsymbol{\omega_c}$, while the $\mathcal{T\!F}_j^m$ associated with the 
variance are not. From our construction in Sec.~\ref{Sec: Variance}, these $\mathcal{T\!F}_j^m$ should 
be evaluated at the parameters associated with the observed data (the empirical parameters 
$\boldsymbol{\omega_e}$ defined in Sec.~\ref{Sec:Central Function and Bias}). There are two 
options on how to obtain $\boldsymbol{\omega_e}$. One may select $\boldsymbol{\omega_e}$ 
directly from Eq.\eqref{Eq: Etas Omegas} in the event that the linear relationship encoded in the
equation provides a good approximation. However, if we suspect that the linear approximation 
is not accurate, for example when dealing with a strongly non-linear model, then we should resort
to a numerical algorithm informed by the data $\boldsymbol{Y}$ every time the data locations change. 
This will allow to calculate the bias directly from the empirical parameters with no need for 
Eq.\eqref{Eq: TF Bias}. Indeed, to guarantee numerical accuracy, we use this last option for all 
the nonlinear models that we explore in this paper, while we resort to Eq.\eqref{Eq: TF Bias} for 
linear models. For example, in the case shown in Fig.~\ref{Fig: Etas Example SF}, the calculated 
change in $m=\rho(0)$ from the central value using the TF, underestimates the true change by 
around $30\%$. In Sec.~\ref{Sec: Rec Bias And Opt Func} we describe an important implementation
of Eq.\eqref{Eq: TF Bias} that would not be possible with a numerical optimizer and which can be 
useful even when dealing with nonlinear models.

\subsection{Priors under the TF formalism}\label{Sec: Priors under TF}

If we have Gaussian priors of the form presented in Eq.~\eqref{Eq: Prior Form}, then we can treat each prior term as a pseudo observation. These priors act in the same way as true observations in $\chi^2$: $(F(\boldsymbol{\omega},q_j)-y_j)/(\sigma_j^2)$, by pulling the value of $\boldsymbol{\omega}$ in a particular direction in the parameter space. The new $\widetilde{\mathcal{H}}$ defined in Eq.~\eqref{Eq: chi2 tilde Hessian} should be used when calculating the observation's transfer functions defined in Eq.~\eqref{Eq: TF omega Definition}.

The effect of the priors will not only be the conversion of $\mathcal{H}$ to the new $\widetilde{\mathcal{H}}$, but each prior estimate value $\omega_k^0$ will have its own transfer function as if it were an observation:

\begin{equation}
    \mathcal{T\!F}_k^{\boldsymbol{\omega}}\equiv \frac{\partial\boldsymbol{\omega}}{\partial \omega_k^0}  = \widetilde{\mathcal{H}}^{-1}I_k\sigma_k^{-2},
\end{equation}

where $I_k$ is the k-th column of the identity matrix of size $K\times K$ (a vector with 0 in every entry except with a 1 on entry $k$). $I_k\sigma_k^{-2}$ is the analogous of $\nabla F(\omega,q_j)\sigma_j^{-2}$ when calculating $\mathcal{T\!F}_j^{m}$. We use the sub index $k$ to denote that what we are perturbing is not $y_j$, but rather the prior estimate value $\omega_k^0$. In the case where the prior contains correlations then Eq.~\eqref{Eq: prior phi} is written as quadratic form $\phi^2(\boldsymbol{\omega}) = \boldsymbol{ \omega} \Sigma^{-1} \boldsymbol{\omega}$. In this case $I_k\sigma_k^{-2}$ will be replaced by the $k$-th column of the matrix $\Sigma^{-1}$.

These transfer functions of the priors ``observations" will appear at the same level as regular observations in the variance and bias equations~\eqref{Eq: TF Variance} and~\eqref{Eq: TF Bias}. For the bias part, the associated $\eta_j$ -which we will call $\tilde \eta_k$- is defined as the difference between the value of $[\boldsymbol{\omega_{c}}]_k$ (the $k$ entry of the central parameters) and the prior ``observation" $\omega^0_k$.

\subsection{Reconstruction Bias and the Optimal Function}\label{Sec: Rec Bias And Opt Func}

In this section, we describe the estimation of a non-intuitive bias which we call the reconstruction bias, that strongly depends on the $q_j$ locations.

This reconstruction bias is closely related to what we observed in the example in Fig.~\ref{Fig: Etas Example SF}. When dealing with incomplete data (a few $q_j$ points on the entire form factor curve, for example), the empirical parameters $\boldsymbol{\omega_e}$ we recover might deviate considerably from the best parameters that reproduce the entire true function ($\boldsymbol{\omega}_\text{Cen}$ in the case of $F(q)_\text{true}$ or $\boldsymbol{\omega}_\text{Opt}$ in the case of $\rho(r)_\text{true}$). As a consequence, the second term inside the brackets in Eq.~\eqref{Eq: TF Bias} could grow substantially.  This will result in a significant bias even in flexible models which in principle could reproduce the true function almost perfectly. 

\begin{figure*}[!htbp]
	\centering
		\includegraphics[width=1\textwidth]{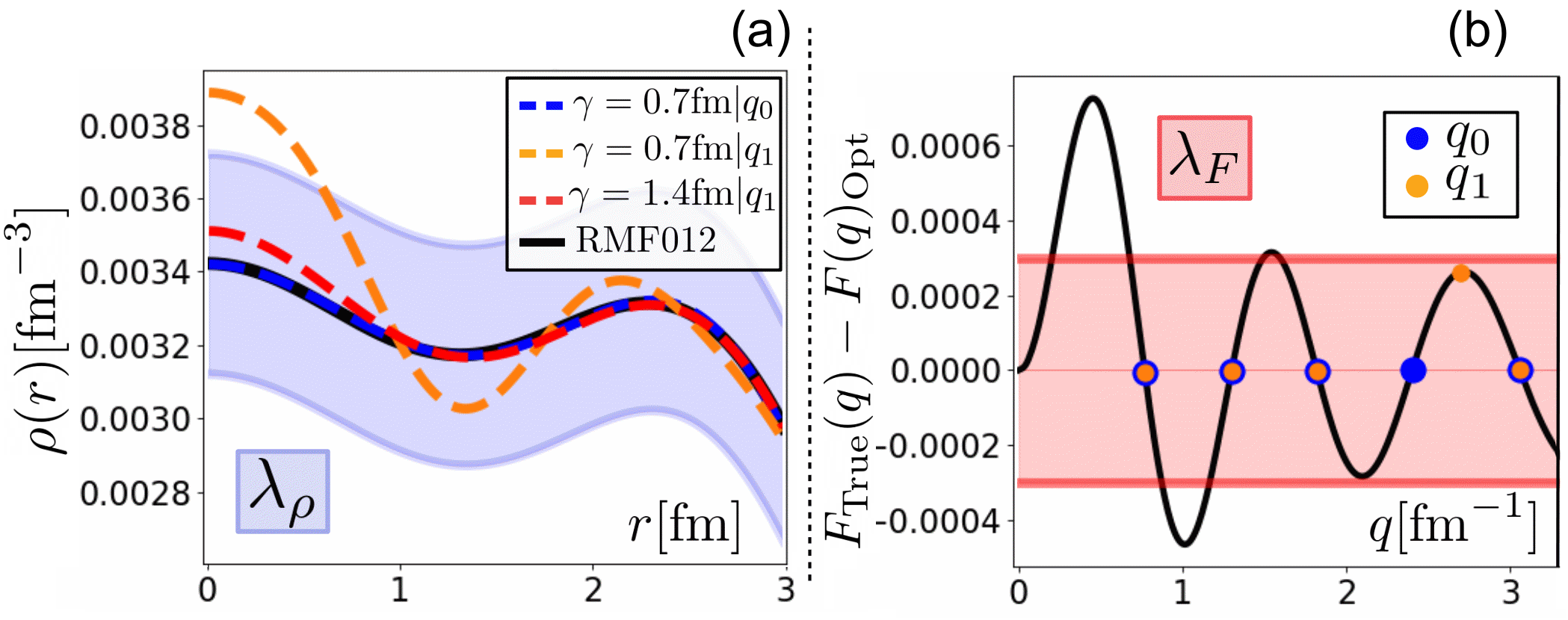}
	\caption{(a) Recovered densities using the SF+G model for two different data locations. The generated (``true") $^{48}$Ca weak density is shown in black, while the orange and blue dashed lines are the SF+G obtained densities for the two data sets $\boldsymbol{q_0}$ and $\boldsymbol{q_1}$. The red dashed line is obtained by the same SF+G model on $\boldsymbol{q_1}$, but with the hyperparameter $\gamma$ set to $1.4$ fm instead of $0.7$ fm. The blue band is associated with the scale $\lambda_\rho$. (b) The locations of the two data sets in orange and blue. These data sets only differ on the location of the fourth point $q_4$. The black line shows the difference between the true function (RMF012) and the optimal function ($F(q_j,\boldsymbol{\omega_\text{Opt}})$). The red band represents the scale $\lambda_F$. }\label{Fig: Ca48 Reconstruction Bias 1}
\end{figure*}

For illustration purposes, in this section we use the generator RMF012. Fig.~\ref{Fig: Ca48 Reconstruction Bias 1} (a) shows the recovered $^{48}$Ca weak density using the SF+G model with two sets, $\boldsymbol{q_0}$ and $\boldsymbol{q_1}$, of five data points each (blue and orange dashed lines). The first data set is $\boldsymbol{q_0}=[0.77,1.30,1.82,2.41,3.06]$ fm$^{-1}$, while the second one is identical to the first except for the fourth location: $\boldsymbol{q_1}=[0.77,1.30,1.82,2.70,3.06]$ fm$^{-1}$, as seen in Fig.~\ref{Fig: Ca48 Reconstruction Bias 1} (b).

The blue and orange SF+G model in Fig.~\ref{Fig: Ca48 Reconstruction Bias 1} (a) has its hyperparameter controlling the size of the Gaussians set to $\gamma=0.7$ fm, close to the nucleon size (Appendix~\ref{App: Models Details} shows a detailed description of the SF+G model and its hyperparameters). The orange curve has a clear bias in the interior density. This is the reconstruction bias. It is not the same type of bias showed, for example, by the SF model which by definition has a flat interior and can not reproduce the interior structure of $^{48}$Ca.

To better analyze this phenomenon, we use the optimal function, i.e., the parameter set $\boldsymbol{\omega_\text{Opt}}$ from the SF+G model that creates the density in the $r$ space that is closest to the true density. By definition, any deviation from $\boldsymbol{\omega}_\text{Opt}$ will result in a stronger bias. We want to understand this increase in bias in terms of the difference $\boldsymbol{\delta \omega} \equiv \boldsymbol{\omega}-\boldsymbol{\omega_\text{Opt}}$. To simplify our analysis, we just focus on $\rho(0)$.

Fig.~\ref{Fig: Ca48 Reconstruction Bias 1} (b) shows the difference in momentum space between the optimal function $F(q,\boldsymbol{\omega_{\text{Opt}}})$ and the true (RMF2012) $F(q)_\text{True}$. The blue points are situated exactly at the locations where both functions have the same value, while the fourth orange point is at a place where these functions differ. 

Similar to what we developed in Sec.~\ref{Sec:Central Function and Bias}, we can imagine that our data are currently centered at the optimal function $F(q,\boldsymbol{\omega_{\text{Opt}}})$ (the minimum of $\chi^2$ is currently at the optimal parameters). The $y_j$ values are slightly perturbed from their starting values by small quantities $\eta_j$ now defined as:

\begin{align}
\eta_{j} \equiv F(q_j)_\text{True}- F(q_j, \boldsymbol{\omega_\text{Opt}}).
\end{align}

Using our TF formalism, we can write how much $\rho(0)$ changes to first order due to these displacements, when compared to the predicted $\rho(0)$ by the optimal function. Since just $\eta_{y_4}$ is nonzero, we have:

\begin{equation}\label{Eq: Delta Rho Eta4}
    \delta \rho(0) =  \Big[\mathcal{T\!F}_4^{(\rho(0))}\Big]\eta_{y_4}.
\end{equation}

Therefore, if we move $q_4$ around in the q-space while leaving the other four $q_j$ in place, those locations with a high product value $\mathcal{T\!F}_4^{(\rho(0))}\eta_{y_4}$ will create a strong bias. Such is the case showed in Fig.~\ref{Fig: Ca48 Reconstruction Bias 1} (a) by the orange curve. In this particular example, if we want to maintain a bias of less than five percent ($\delta\rho(0)<5\%$), we can only locate $q_4$ in around $15\%$ of the possible momentum transfer range $[0-3.5]$ fm$^{-1}$ (see Appendix~\ref{App: Details about the SF+G plots} for more details). 

Let us call $\lambda_F$ the expected scale for the size of $\eta_j$ for our range of $q_j$ values. Let us call $\lambda_\rho$ the desired threshold we want for our accuracy in the estimation of $\rho(0)$. Fig.~\ref{Fig: Ca48 Reconstruction Bias 1} shows these two scales as the red and blue bands, respectively. Replacing the transfer function by its explicit expression in Eq.~\eqref{Eq: Delta Rho Eta4}, we will maintain that threshold in $\rho(0)$ as long as:

\begin{equation}\label{Eq: Rule of Thumb}
     |\nabla \rho (0) \mathcal{H}^{-1}\nabla F(q_4)/\sigma_4^2|\leq \frac{\lambda_\rho}{\lambda_F}.
\end{equation}

For our particular problem, we can set $\lambda_F\approx 3\times10^{-4}$ and $\lambda_\rho\approx 1.5\times 10^{-4}$ fm$^{-3}$ (roughly $5\%$ of $\rho(0)$), which makes the ratio $\frac{\lambda_\rho}{\lambda_F}=\frac{1}{2}$ fm$^{-3}$. For the SF+G model at $q_4=2.7$ fm$^{-1}$, the product $\mathcal{T\!F}_4^{(\rho(0))}\eta_4\approx 1.3$ fm$^{-3}$, which implies that the reconstruction bias falls outside of our tolerable range $\lambda_\rho$. 

In an actual experiment, we would not know ahead of time the optimal $q_j$ locations where $\mathcal{T\!F}_j^{(\rho(0))}\eta_{j}$ is small. Therefore, we could not use a model like SF+G with such a limited $q$ range and strong reconstruction bias.

For the SF+G, the situation seems to be mainly driven by the first Gaussian with $R_1=0$, which scales as $\gamma^{-3}$ in the $\rho$ space. Based on this, we decided to double the size of $\gamma$, from $\gamma=0.7$ fm, to $\gamma=1.4$ fm (which is the value we use in Sec.~\ref{Sec: Weak Charge Results}).

Using this new value of $\gamma$, the new transfer function product value at $q_4=2.7$ fm$^{-1}$ is $\mathcal{T\!F}_4^{(\rho(0))}\eta_4\approx 0.3$ fm$^{-3}$ and the reconstruction bias is reduced considerably\footnote{To be rigorous, we should now move the other $q_j$ values to the locations where the new optimal model is equal to the true function. Since they are almost in the same location, we decided to keep them in the same place to simplify the discussion.}. This is shown by the red dashed line in Fig.~\ref{Fig: Ca48 Reconstruction Bias 1} (a). Moreover, with this new value of $\gamma$, $q_4$ can be allocated in around $30\%$ of the possible momentum transfer range $[0-3.5]$ fm$^{-1}$ while maintaining a bias of less than five percent (see Appendix~\ref{App: Details about the SF+G plots} for more details).

We close this section with two important remarks regarding this type of analyses. First, they could ultimately serve not only to model selection, but to model building. In many cases, a hyperparameter (such as $\gamma$) might be fixed to a sub-optimal value that hinders rather than helps the extraction of information from experimental data.

Second, these analyses can give an estimate of the impact of the reconstruction bias which is impossible to get by just focusing on the statistical errors in experimental data. Consider the purple $95\%$ confidence ellipse in Figures~\ref{Fig: Stars Moving} and~\ref{Fig: Etas Example SF} centered at $\boldsymbol{\omega_c}$, the red star. This ellipse does not contain the actual estimated parameters from the data, i.e., the green star (3) in Fig.~\ref{Fig: Stars Moving} (let us recall that the blue star in Fig.~\ref{Fig: Etas Example SF} is just the linear approximation). The reverse is also true: the ellipse centered at the true empirical parameters (not shown) will not contain the red star, which reproduces the true weak charge density in Fig.~\ref{Fig: Etas Example SF} better than the approximated empirical blue density.  

The errors $\sigma_j$ and the deviations $\eta_j$ are two unrelated scales. Confidence ellipses are usually related to the errors $\sigma_j$ but the reconstruction bias is related to the $\eta_j$. There is no reason for the ellipse obtained from the true data to contain $\boldsymbol{\omega}_\text{Opt}$, i.e., the set of parameters in our model that best describe the real curve that generated that data. However, this is often the assumed scenario when extracting information from experiments.

\section{Results: Analyzing charge and weak charge densities}\label{Sec: Results}

In this section, we discuss in detail the process used to select the optimal models and the impact that varying the locations 
of the selected momentum transfers $q_j$ will have on the extracted densities of both $^{48}$Ca and $^{208}$Pb. In particular,
we are interested in describing the root mean square radius and interior density of the charge and weak charge distributions. 

The calculation of the MSE for the charge radius is straightforward as it involves a single, well-defined quantity. For the interior
density, we allocate $30$ grid points between $r\!=\!0$\,fm and $r\!=\!3$\,fm for $^{48}$Ca, and between $r\!=\!0$\,fm and 
$r\!=\!5$\,fm for $^{208}$Pb. The MSE for the interior density is then constructed by averaging in quadrature the single MSE 
for each individual point. That is,
\begin{equation}
    \text{MSE}[\text{Interior}]^2 \equiv \frac{1}{30}\sum_{i=1}^{30} \text{MSE}[\rho(r_i)]^2.
    \label{Eq: Grid Points}
 \end{equation}
We then combine both the radius and interior MSE into a single quantity known as the Figure of Merit (FOM):
\begin{equation}
    \text{FOM}^2\equiv \left(\frac{\text{MSE}[\text{Radius}]}{\Delta R}\right)^2 + 
    \left(\frac{\text{MSE}[\text{Interior}]}{\Delta \rho}\right)^2,
    \label{Eq: FOM}
\end{equation}
where $\Delta \rho$ and $\Delta R$ are natural scales associated to each quantity; roughly $5\%-10\%$ and $1\%$ for the
interior density and radius, respectively; see Table\,\ref{Tab: Natural Scales}. By adjusting these scales, the FOM could be 
made more sensitive to the radius or the interior density. Note that for both $^{48}$Ca and $^{208}$Pb the densities have 
been normalized to $1$ rather than to the number of nucleons.

\vspace{10pt}
{\renewcommand{\arraystretch}{1.2}
\begin{table}[h]
\centering
\begin{tabular}{l|c|c|}
\cline{2-3}
                              & $\Delta \rho$ [fm$^{-3}$] & $\Delta R$ [fm]\\ \hline
\multicolumn{1}{|l|}{$^{48}$Ca} & 0.00015  & 0.04   \\ \hline
\multicolumn{1}{|l|}{$^{208}$Pb}    & 0.00008  & 0.06   \\ \hline
\end{tabular}\caption{Natural scales for the uncertainties in the interior density and radius for $^{48}$Ca and $^{208}$Pb. 
}\label{Tab: Natural Scales}
\end{table}}

To simplify the analysis, we assume that the errors in the experimental data for both the charge and weak-charge form factor 
depend only on their assumed value at the selected momentum transfers. For example, following\,\cite{piekarewicz2016power}, 
we assume a constant value of $\sigma(q)\!=\!0.005$ for $^{208}$Pb. For the case of $^{48}$Ca we adopt the prescription given
in Ref\,\cite{lin2015full} for the errors at the selected five momentum transfers $[q_1,q_2,q_3,q_4,q_5]$. When required, a simple
function of the form $\sigma(q)\!=\!\text{Max}(0.00057,0.0081\!-\!0.003q)$ is used to interpolate between the selected $q$ values. 
Once the data set and the selected model are specified, the FOM depends solely on the location of the momentum transfers. In the following subsections we illustrate how the value of the FOM can be minimized by optimizing such locations and how the TF 
formalism is the ideal tool for interpreting the results and further reducing the uncertainties, for example, by identifying critical measurements for error reduction.


\subsection{Electric Charge Densities}\label{Sec: Ca and Pb Charge Example}

To test the new formalism we start by analyzing the well known experimentally determined electric charge density 
of $^{48}$Ca and $^{208}$Pb\,\cite{DeJager:1987qc,Fricke:1995}. We illustrate the power and flexibility of the 
transfer functions formalism by describing the $^{48}$Ca data using a Fourier--Bessel expansion and a SF model
in the case of $^{208}$Pb. Following the prescription of Ref.\,\cite{lin2015full}, we assign the starting values of the
momentum transfer for $^{48}$Ca at $\boldsymbol{q_0}\!=\![0.9,1.35,1.8,2.24,2.69]\,{\rm fm}^{-1}$. Similarly, for the 
case of $^{208}$Pb, we fixed the two starting locations as in Ref.\,\cite{piekarewicz2016power} at
$\boldsymbol{q_0}\!=\![0.5,0.8]\,{\rm fm}^{-1}$.

\begin{figure*}[!htbp]
	\centering
		\includegraphics[width=0.9\textwidth]{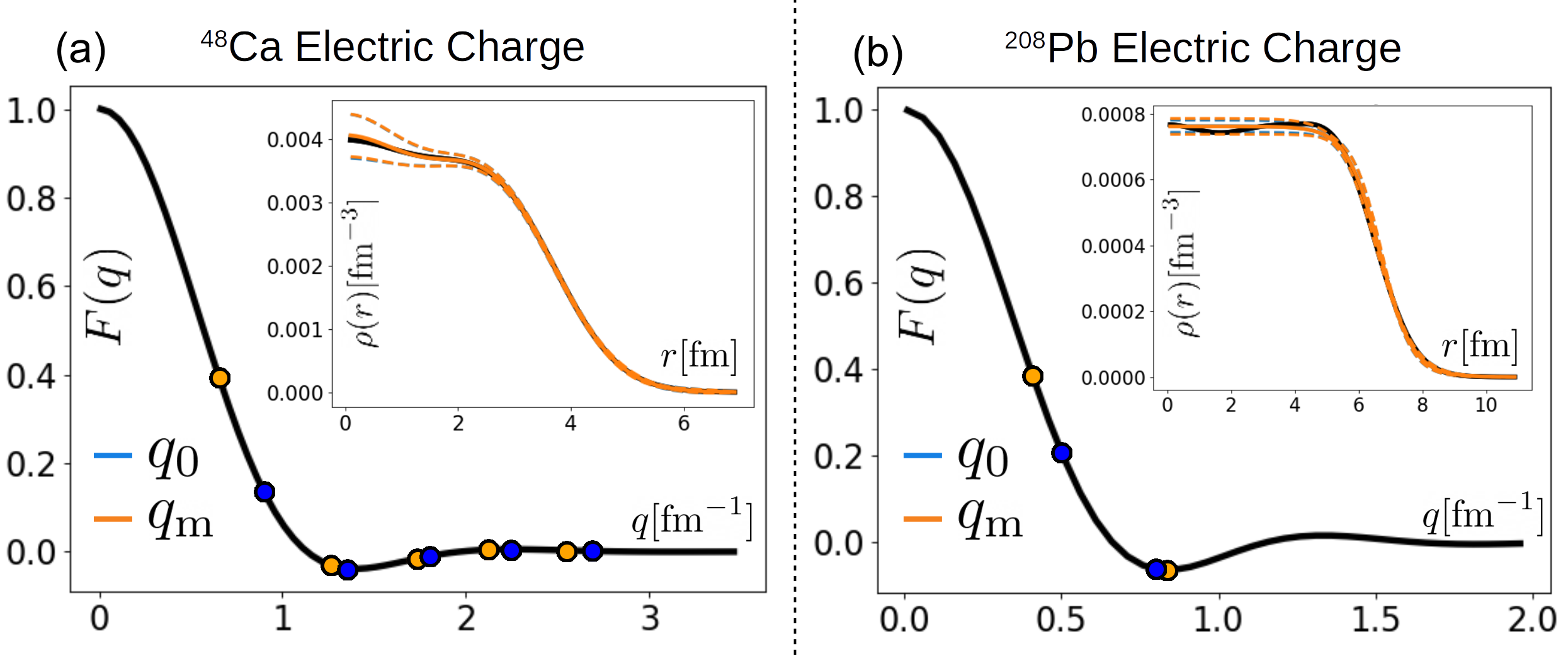}
	\caption{(Color online). Optimization of the FOM for $^{48}$Ca using a Fourier--Bessel expansion (a)
        and for $^{208}$Pb using the Symmetrized-Fermi function (b). The original locations of the momentum 
        transfer  $q_0$ (blue points) are displaced to $q_m$ (orange points) to minimize the FOM. The inset 
        plots show the reconstructed charge densities with their respective error bands.}\label{Fig: Example Ca Pb Ch}
\end{figure*}

\begin{figure*}[!htbp]
	\centering
		\includegraphics[width=0.9\textwidth]{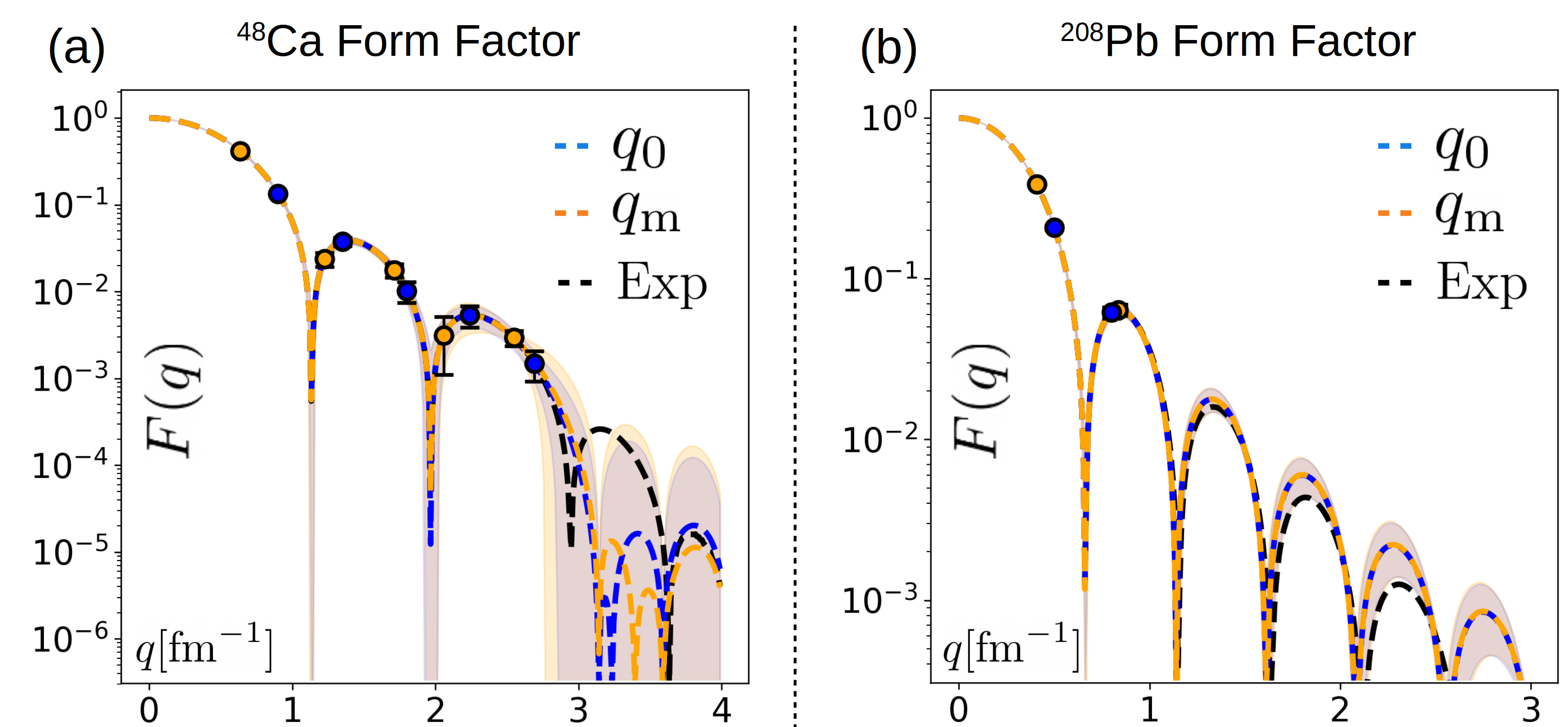}
	\caption{(Color online). Same as in Fig.~\ref{Fig: Example Ca Pb Ch}, but now for the absolute value
	of the form factor using a logarithmic scale. The black curve represents the respective experimental 
	charge form factor\,\cite{DeJager:1987qc,Fricke:1995} and for clarity, the central curves are displayed 
	as dashed lines, whereas the error bands are shown as colored bands. }
	\label{Fig: Log Form Factor}
\end{figure*}

We show in Fig.~\ref{Fig: Example Ca Pb Ch} results obtained before and after optimizing the location of the 
momentum-transfer points. We display the original locations in blue and their shift to their optimal locations in 
orange,  where the FOM is minimized subject to the following constraints: $q_j\leq3.5\,{\rm fm}^{-1}$ for 
$^{48}$Ca and $q_j\leq2\,{\rm fm}^{-1}$ for $^{208}$Pb. Beyond these limits, we assume that the experimental 
challenge to measure such small cross sections can not be met. This may be better appreciated by displaying
the form factor in a logarithmic plot, as in Fig.~\ref{Fig: Log Form Factor}; note that the cross section is
proportional to the \emph{square} of the form factor. Note that the minimization of the FOM was done 
by running the python numpy optimization library with the `TNC' method for 10 different seeds including the 
original $\boldsymbol{q_0}$ choice. In Table\,\ref{Tab: MSE Ca and Pb results}, we show results for the MSE 
for both nuclei in terms of their natural scales. The MSE in the interior was not substantially reduced for 
$^{48}$Ca and it even increased by $\sim\!20\%$ for $^{208}$Pb, as can be seen by the slightly larger
error bands in Fig.~\ref{Fig: Example Ca Pb Ch}. On the other hand, the MSE for the radius was improved by $\sim30\%$ for $^{48}$Ca and by $\sim\!20\%$ for $^{208}$Pb. These results are driven by our selection
of scales which favored an improvement in the radius rather than in the interior density. Also, the radius is an 
easier quantity to constrain than the interior density. Note, however, that even minimizing the FOM with only 
the interior term does not significantly improve the interior density.

\vspace{10pt}
{\renewcommand{\arraystretch}{1.2}
\begin{table}[]
\centering
\begin{tabular}{l|l|l|l|l|}
\cline{2-5}
                         & \multicolumn{2}{l|}{$^{48}$Ca} & \multicolumn{2}{l|}{$^{208}$Pb} \\
                         & Int       & Rad      & Int     & Rad     \\ \hline
\multicolumn{1}{|l|}{MSE ($q_0$)} & 1.27           & 1.37        & 0.27         & 0.77       \\ \hline
\multicolumn{1}{|l|}{MSE ($q_m$)} & 1.26           & 0.94        & 0.32         & 0.63       \\ \hline
\end{tabular}\caption{MSE results for the interior (Int) and radius (Rad) for $^{48}$Ca and $^{208}$Pb for their respective original locations $q_0$ and optimal locations $q_m$. Each quantity has been divided by its respective natural scale as defined in Table~\ref{Tab: Natural Scales}.}\label{Tab: MSE Ca and Pb results}
\end{table}}

Finally, listed in Tables\,\ref{Tab: Ca TFs Example} and~\ref{Tab: Pb TFs Example} in Appendix\,\ref{App: Tables Ca and Pb example} 
are the numerical values of the $\mathcal{T\!F}_j$ times the respective error $\sigma_j$ for the density at $r\!=\!0$ fm and the radius 
for two sets of locations of the momentum transfer, namely, original $\boldsymbol{q_0}$ and optimal $\boldsymbol{q_m}$. These individual 
values illustrate how much each measurement is currently impacting the variance in the radius and in the density at $r\!=\!0$.
Note that in Eq.\eqref{Eq: TF Variance} each term $\mathcal{T\!F}_j\sigma_j$ is added in quadrature. Therefore, the final variance 
is not linear on each component. Indeed, the quadrature equation will enhance the effect of bigger numbers with respect to their 
smaller counterparts. For example, in the case $^{48}$Ca with the optimized set, the variance in $\rho(0)$ is dominated by the
observations at $q_2$ and $q_3$, whereas for the radius the variance is largely driven by the form factor at $q_1$. A similar
analysis for $^{208}$Pb reveals that the variance in $\rho(0)$ is driven by $q_2$, whereas the measurement at $q_1$ dominates
the variance in the radius. Given that the radius is obtained from the slope of the form factor at zero momentum transfer and the
interior density is controlled by the large-q behavior of the form factor, the previous results are fully consistent with our expectations.
Note that as the errors in the observations change, these statements might no longer hold true. Our main conclusion is that to 
reduce the final variance on each quantity $m$ within this hypothetical experimental design---and to first approximation---these 
are the critical data locations that should be targeted for error reduction. 


\subsection{Weak Charge Densities}\label{Sec: Weak Charge Results}

We now proceed to compare the performance of each of the seven models mentioned in 
Sec.~\ref{Sec: Models, parameters and errors} in reproducing the interior density and 
radius of the weak charge distribution of $^{48}$Ca and $^{208}$Pb. 
Appendix~\ref{App: Tables Ca and Pb example} presents the corresponding analysis for 
the charge densities.

Given that there is no experimental information on the weak charge form factors of $^{48}$Ca 
and $^{208}$Pb, we use Eq.\eqref{Eq: Average MSE} to calculate the squared average MSE 
from the five different generators obtained from Ref.\,\cite{Chen:2014mza}, namely, RMF012, 
RMF016, RMF022, RMF028 and RMF032. As in the previous section, we start with five fixed 
locations $\boldsymbol{q_0}$ and then optimize these values to $\boldsymbol{q_m}$ to minimize
the average FOM. We apply the same restrictions as in the example of the charge density: 
$q_j\!\leq\!3.5\,{\rm fm}^{-1}$ for $^{48}$Ca and $q_j\leq2\,{\rm fm}^{-1}$ for $^{208}$Pb. Note 
that since the goal is to minimize the average mean-square error of all five generators, the 
resulting optimal values $\boldsymbol{q_m}$ only depend on 
the choice of the model. 
The starting locations for the momentum transfer in the case of $^{48}$Ca are once again fixed at 
$\boldsymbol{q_0}=[0.90,1.35,1.8,2.24,2.69]\,{\rm fm}^{-1}$, whereas for $^{208}$Pb they are now 
chosen at $\boldsymbol{q_0}=[0.63,0.94,1.26,1.57,1.88]\,{\rm fm}^{-1}$. Note that these values 
correspond to the special choice of $q_\nu\!\equiv\!\nu\pi /R_\text{cut}$ for $\nu\!\in\![2,6]$, with the
cutoff radius $R_\text{cut}\!=\!7\,{\rm fm}$ for $^{48}$Ca\,\cite{lin2015full} and 
$R_\text{cut}\!=\!10\,{\rm fm}$ for $^{208}$Pb.

\begin{figure*}[!htbp]
	\centering
		\includegraphics[width=0.9\textwidth]{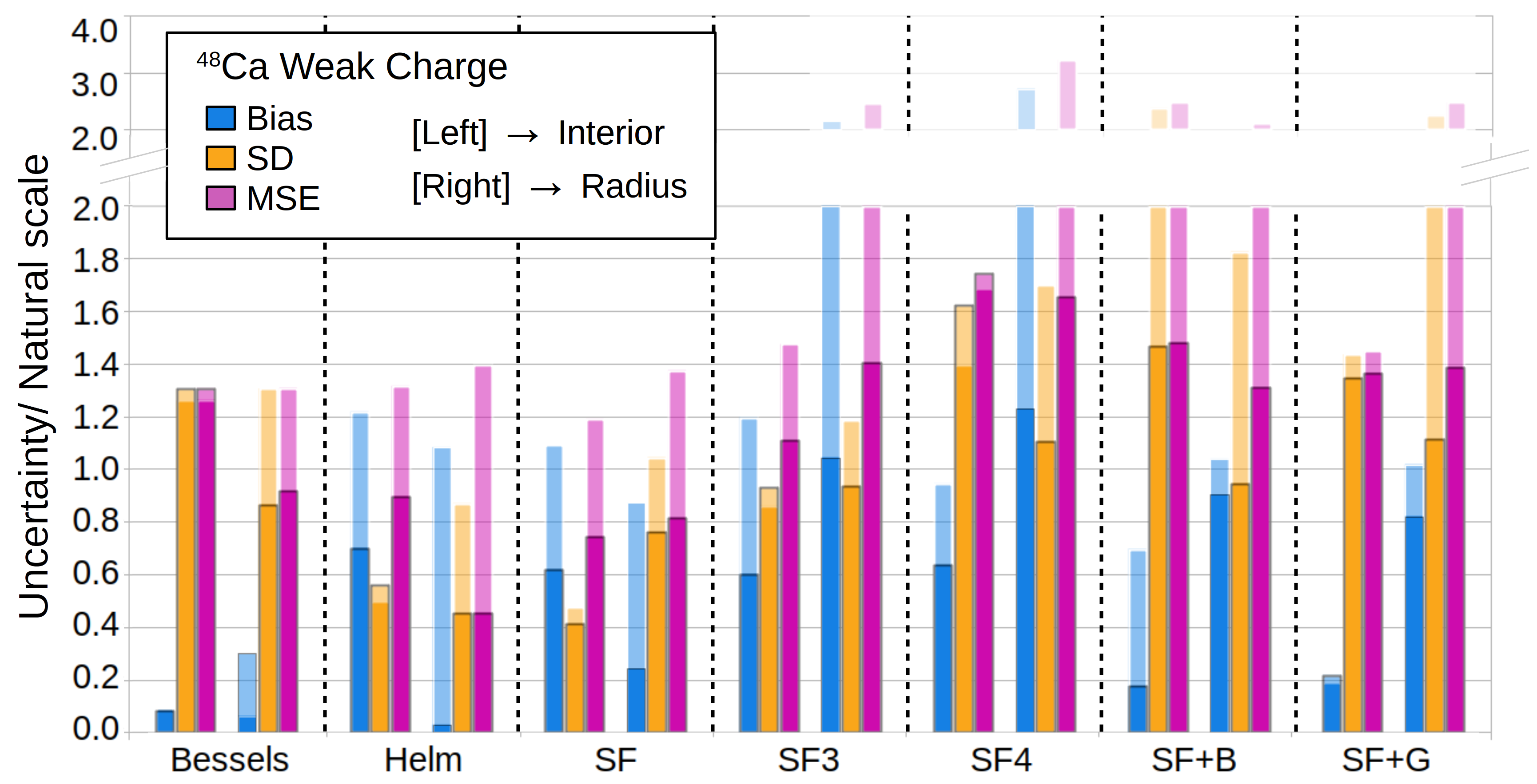}
	\caption{Comparison of the performance of the seven models used in the text for recovering the 
	interior density and weak charge radius of $^{48}$Ca as generated from a form factor ``measured"
	at five points. The three columns to the left of each model show the bias, SD, and MSE for the interior 
	density, whereas the three columns to the right display the bias, SD, and MSE for the radius. 
	All quantities have been divided by their natural scales: $\Delta \rho({\text{Ca}}) = 0.00015\,{\rm fm}^{-3}$ 
	and $\Delta R({\text{Ca}})\!=\!0.04$\,fm. The solid columns represent the optimal locations $\boldsymbol{q_m}$, 
	whereas the light borderless columns were obtained from the starting $\boldsymbol{q_0}$ points.}
	\label{Fig: Models Wk Ca48}
\end{figure*}

\begin{figure*}[!htbp]
	\centering
		\includegraphics[width=0.9\textwidth]{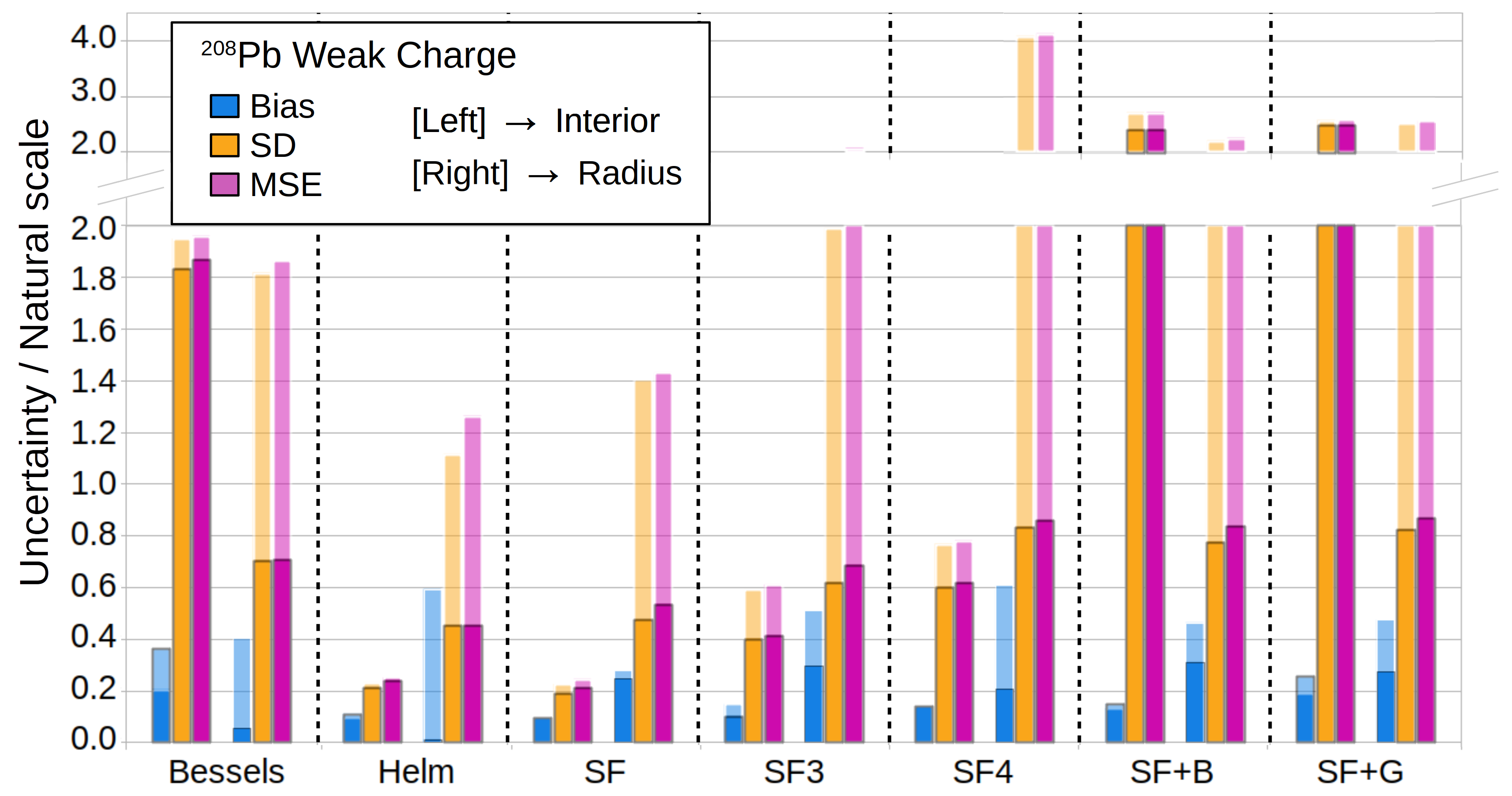}
	\caption{Comparison of the performance of the seven models used in the text for recovering the 
	interior density and weak charge radius of $^{208}$Pb as generated from a form factor ``measured"
	at five points. The three columns to the left of each model show the bias, SD, and MSE for the interior 
	density, whereas the three columns to the right display the bias, SD, and MSE for the radius. 
	All quantities have been divided by their natural scales: $\Delta \rho({\text{Pb}})\!=\!0.00008\,{\rm fm}^{-3}$ 
	and $\Delta R({\text{Pb}})\!=\!0.06$\,fm. The solid columns represent the optimal locations $\boldsymbol{q_m}$, 
	whereas the light borderless columns were obtained from the starting $\boldsymbol{q_0}$ points.}
	\label{Fig: Models Wk Pb208}
\end{figure*}

In Figures \ref{Fig: Models Wk Ca48} and\,\ref{Fig: Models Wk Pb208} we display the performance of the seven models employed in 
the text to describe the weak charge of $^{48}$Ca and $^{208}$Pb, respectively. Shown in each figure are the resulting bias, standard deviation (SD), and MSE for the interior density (three bars on the left of each panel) and the weak charge radius (three bars on the 
right of each panel).  The corresponding figures for the electric charge density are shown in Figures~\ref{Fig: Models Ch Ca48} 
and\,\ref{Fig: Models Ch Pb208} in Appendix~\,\ref{App: Tables Ca and Pb example}. We note that for a fixed model we obtained
very similar results regardless of the particular RMF generator; as an example see Figures~\ref{Fig: App Ca Wk Bess results} 
and\,\ref{Fig: App Pb Wk SF results} in Appendix~\ref{Sec: App Details about weak charge}. This suggests that the conclusions 
that we draw within each model are robust, at least within the (RMF) family of generators considered in this study. 

We want to highlight two main points from these results. First, changing the data locations (i.e., the selection of the various
momentum transfers) has a significant impact on the performance of the model. For example, in the description of the weak
charge density of $^{208}$Pb the performance of the Helm model improves by nearly a factor of two. Second, we observe
large variations in the model performance when the original fixed locations $\boldsymbol{q_0}$ are adopted. Indeed, for 
$^{48}$Ca the performance of the Bessel-Fourier expansion outperforms that of the SF+B model by about a factor of two. 
Such large discrepancy is often mitigated by selecting the optimal locations $\boldsymbol{q_m}$ for each model; see the
dramatic improvement in the description of the weak charge radius of $^{208}$Pb when the optimal locations are adopted.
Based on these two points, we can conclude that the optimal model will strongly depend on the data structure, regarding 
both locations and errors. We should expect that variance driven models (like the Bessels) will outperform bias driven models 
(like the SF) in cases where the data errors are small. For the number of experimental measurements and errors assumed in 
this example, the Helm and SF models are best suited for the simultaneous extraction of the radius and interior density of 
both nuclei. This could be expected for the case of $^{208}$Pb given that both the Helm and SF models are characterized 
by a flat interior density that provides our closest connection to the saturation density of infinite nuclear 
matter\,\cite{Horowitz:2020evx}. It might come as a surprise that these flat models outperform more flexible models like 
the Bessel-Fourier expansion which better describes the interior shell oscillations of $^{48}$Ca. The reason behind this 
finding is that we aim to minimize the MSE, which involves a combination of the bias and the variance. On \textit{average} 
(mean value), the Bessel model will provide a more genuine representation of the interior oscillations of the weak charge 
density of $^{48}$Ca. However, the noise level as quantified by the variance is so high that the expected deviation is large 
enough to make more desirable a flat description with smaller error bands. 

As an example, consider the RMF012 generator displayed as the black solid curve in 
Fig\,\ref{Fig: Ca48 Reconstruction Bias 1}(a). This RMF generator predicts a weak charge density
for $^{48}$Ca that is slightly enhanced at $r\!=\!0$ relative to the average interior density and then 
drops from the average around $r\!\approx\!1.2\,{\rm fm}$. We could then ask: how probable is it 
to conclude the opposite (i.e., $\rho(0)\!\leq\!\rho(1.2)$) after adopting the optimal locations 
$\boldsymbol{q_m}$ for the Bessel model. To answer that question, let us define the new quantity of interest 
$m\equiv\rho(1.2)-\rho(0)$ and investigate the probability that $m \geq 0$. Once the optimal 
$\boldsymbol{q_m}$ values are adopted, an average value of 
$m\!=\!-2.3\times\!10^{-4}\,{\rm fm}^{-3}$ is obtained, suggesting that $\rho(0)\!>\!\rho(1.2)$, in
agreement with the predictions from the RMF012 generator. But what about the variance in this
result? The variance of this quantity can be calculated using the TF formalism from Eq.~\eqref{Eq: TF Variance} 
as:
\begin{equation}
  \Delta m^2\!=\!\!
  \sum_{j=1}^5\!\Big(\big[\nabla\!\rho(1.2)\!-\!\nabla\!\rho(0) \big]\mathcal{H}^{-1}
  \nabla\!F(q_j,\boldsymbol{\omega}) \sigma_j^{-2}\!\Big)^2\!\sigma_j^2,
\end{equation}
where we have used Eq.\eqref{Eq: TF m Definition} and Eq.\eqref{Eq: TF omega Definition} to write the explicit 
form of $\mathcal{T\!F}_j^{m}$. Following this procedure, we obtain a standard deviation of 
$\Delta m\!=\!2.2\times10^{-4}\,{\rm fm}^{-3}$. We note that the third measurement at $q_3\!=\!1.73\,{\rm fm}^{-1}$ 
has the largest impact on the the variance, followed by appreciable contributions from 
$q_4\!=\!2.12\,{\rm fm}^{-1}$ and $q_5\!=\!2.55\,{\rm fm}^{-1}$; see Table~\ref{Tab: Opt Loc Ca Ch} in 
Appendix\,\ref{App: Tables Ca and Pb example} for the values of $\boldsymbol{q_m}$. Hence, under the
assumption that the errors are Gaussian distributed random variables, we can infers that $\rho(0)\!\leq\!\rho(1.2)$ 
in $\sim15\%$ of the experimental realizations. That is, if the experimental noise (primarily in $q_3$, $q_4$ and $q_5$) 
cannot be significantly reduced, we will conclude the incorrect oscillation structure in the interior density of $^{48}$Ca in 
one out of six experiments.

For the interior density of $^{208}$Pb, the best overall score was achieved by the SF model with a total MSE of 
$0.21\!\times\!\Delta\rho({\text{Pb}})\!=\!1.68\!\times\!10^{-5}\,{\rm  fm}^{-3}$, where $\Delta\rho({\text{Pb}})$ is 
defined in Table\,\ref{Tab: Natural Scales}. Using again RMF012 as an example of a generator, we observe that 
the total variance in $\rho(0)$ is mainly driven by the third observation, having a value of 
$|\mathcal{T\!F}_3^{\rho(0)}\sigma_3|\!=\!0.14\!\times\!\Delta\rho({\text{Pb}})$. Taking $\rho(0)$ as a representative 
value of the interior density, this implies that the measurement at $q_3\!=\!0.77\,{\rm fm}^{-1}$ should be primarily
targeted for error reduction in order to improve the uncertainty in the saturation density $\rho_0$. We underscore 
that the interior density of $^{208}$Pb is a genuine experimental observable that provides the closest connection 
to the saturation density of infinite nuclear matter. For a recent analysis on how a measurement of the interior 
density of $^{208}$Pb could constraint $\rho_0$ see Ref.\,\cite{Horowitz:2020evx}.

In the case of the weak charge radii of both nuclei, we found that they can be accurately determined using the Helm 
model: an MSE of $0.45\!\times\!\Delta R({\text{Ca}})\!=\!0.018$ fm for $^{48}$Ca and of 
$0.45\!\times\!\Delta R({\text{Pb}})\!=\!0.027$ fm for $^{208}$Pb, with both values of $\Delta R$ listed in 
Table\,\ref{Tab: Natural Scales}. In the case of $^{48}$Ca, and relying again on RMF012, the total variance in $R$ is 
uniformly distributed among the first three observations at $q_1\!=\!0.51$, $q_2\!=\!0.63$, and $q_3\!=\!0.77\,{\rm fm}^{-1}$, 
with values of $|\mathcal{T\!F}_j^{\rho(0)}\sigma_j|\!\approx\!0.2\!\times\!\Delta R({\text{Ca}})$. 
Instead, for $^{208}$Pb we found that the total variance in $R$ is driven by the two points closest to the origin,
namely, $q_1\!=\!0.37$ and $q_2\!=\!0.40\,{\rm fm}^{-1}$,  with values of 
$|\mathcal{T\!F}_j^{\rho(0)}\sigma_j|\!=\!0.29\!\times\!\Delta R({\text{Pb}})$. To improve the uncertainty in the weak charge
radii, the observations at these ``low-q" points should be targeted for error reduction. These results are hardly surprising 
given that the weak charge radius is defined in terms of the slope of the associated form factor at the origin. It is worth
noting that the weak charge radius of $^{208}$Pb, when combined with the corresponding (electric) charge radius into 
a neutron skin, provides a stringent constraint on the slope of the symmetry energy $L$---and ultimately on the radius of 
neutron stars\,\cite{Horowitz:2001ya}. In particular, a 1\% determination of the weak charge radius of $^{208}$Pb translates 
into an uncertainty of about $40$\,MeV in the slope of the symmetry energy\,\cite{Horowitz:2014bja}.

\subsection{The role of priors}\label{Sec: The role of priors}

The incorporation of priors lies at the heart of Bayesian statistics. Priors allow us to include physical biases and
intuition as well as information from previous experiments. Moreover, priors play the important role of serving as 
leverage to reduce the variance of a model at the expense of increasing its 
bias\,\cite{bishop2006pattern,sullivan2015introduction}. This can be particularly beneficial for models such as the 
Bessel-Fourier expansion or SF+G, whose MSE is largely driven by the variance given the level of noise in the
generated data.

\begin{figure*}[!htbp]
	\centering
		\includegraphics[width=0.9\textwidth]{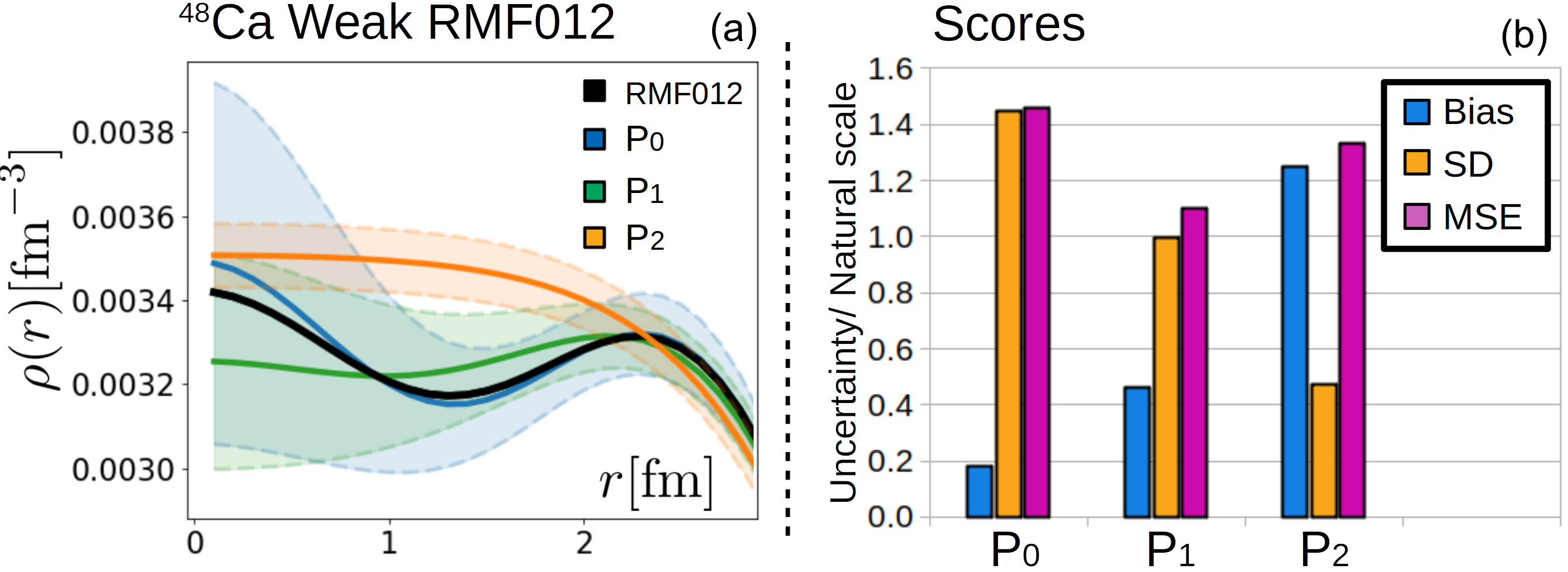}
	\caption{(a) Reconstructed weak density of $^{48}$Ca for the generator RMF012, the model SF+G and three sets of priors. $P_.0$ (in blue), $P_1$ (in green), and $P_2$. (b) Scores on the interior density in terms of bias, SD, and MSE for the three choices of prior. The numerical values have been divided by the natural scale of $\Delta \rho_{\text{Ca}}=0.00015$ fm $^{-3}$. }\label{Fig: RMF012 Example Prior}
\end{figure*}

In this section, we briefly explore the impact of an informed prior on the performance of the SF+G model as it pertains to
the weak charge density of $^{48}$Ca. As we have done earlier, we use the RMF012 generator to produce
synthetic data. The proposed locations of the measurements are the original five values of the momentum transfer: 
$\boldsymbol{q_0}\!=\![0.9,1.35,1.8,2.24,2.69]\,{\rm fm}^{-1}$. Note that the implementation of priors was discussed in 
Sec.~\ref{Sec: Bayesian Approach} and extended to the TF formalism in Sec.~\ref{Sec: Priors under TF}.
The SF+G model consists of the two-parameter symmetrized Fermi function plus three Gaussians ``bumps" to account for 
shell oscillations in the interior. The Gaussians are centered in the interior at three different locations: 
$[R_1,R_2,R_3]\!=\![0,1.3,2.6]$\,fm. We analyze three prior options for the amplitude of the Gaussians ($A_1,A_2,A_3$), 
while we leave the two intrinsic parameters of the SF model unconstrained. First, we consider a null prior that we refer to
as $P_0$. Such ``prior" effectively reproduces the original unconstrained SF+G model. Second, we consider a fairly 
uninformed prior, defined in such a way that the deviation from a flat density at the peak of each Gaussian is of the order 
of $\sim\!0.0003$ fm$^{-3}$, or roughly $10\%$ of the average interior density. We call this prior $P_1$ and is given by the following parameter centers 
and standard deviations:
\begin{equation}
   \boldsymbol{\omega^0} = [0,0,0] \hspace{3pt}{\rm and}\hspace{3pt}
  \boldsymbol{\sigma} = [0.005,0.027,0.08].
  \label{Eq: Priors SF+G} \\
\end{equation}
Finally, we consider an extremely restrictive prior ($P_2$) that forces the value of all three Gaussian amplitudes to zero
($A_1\!=\!A_2\!=\!A_3\!=\!0$), effectively reproducing the original flat SF model without oscillations. For an example on
the incorporation of priors see Appendix\,\ref{App: Details about priors SF+G}.

The reconstructed weak charge density of $^{48}$Ca for the three choices of priors is displayed in 
Fig.~\ref{Fig: RMF012 Example Prior}(a), with $P_0$ (in blue) displaying the largest variance, 
$P_1$ (in green), and $P_2$ (in orange) displaying the smallest variance but the largest bias. The
overall performance for each choice is quantified in Fig.~\ref{Fig: RMF012 Example Prior}(b). These 
results are an interesting example of the bias vs variance trade-off: the model without prior ($P_0$) 
reproduces the true (RMF012) curve almost perfectly, but displays a huge error band, whereas the 
model with the most restrictive prior ($P_2$), or effectively with the fewer number of parameters, has 
the largest bias but the smallest error bands (variance). Note that $P_0$ and $P_2$ have almost the 
same overall MSE score. Also note that since what can be measured is the form factor as a function of the momentum
transfer, the reconstructed spatial density in the interior does not have to be well constrained if there are
not enough data. This is the main reason that the orange curve ($P_2$) that fails to reproduce the 
interior oscillations, also fails to reproduce the average interior density. The model with the $P_1$ 
prior provides the best overall MSE score. Indeed, its MSE score is even better than any of the average 
MSE scores of the models studied in Sec.~\ref{Sec: Weak Charge Results}, for fixed $\boldsymbol{q_0}$ 
locations. 

We can analyze the behavior of the MSE directly from the transfer function formalism as the prior 
is modified. To do so, let us focus on the interior density $\rho(0)$. Stronger priors constrain more 
effectively $\widetilde{\mathcal{H}}^{-1}$, thereby reducing the impact of the transfer functions 
$\mathcal{T\!F}_j^{\rho(0)}$ of each data point. This effectively reduces the propagation of experimental 
uncertainty $\sigma_j$ towards the calculated variance in $\rho(0)$. The trade-off is due to the fact 
that the inclusion of a strong prior will push away the central value of $\rho(0)$ from what the central 
values of the data ($y_j$) suggest, resulting in an increase of the total bias in $\rho(0)$. Such a change 
can be written to first approximation as:
\begin{equation}
    \delta \rho(0) = \sum _k^3\mathcal{T\!F}_k^{\rho(0)} \tilde \eta_k,
\end{equation}
where the $\tilde \eta_k$ are now defined as the difference between the parameter's value without 
priors and the new prior centers $\omega_k^0$.

Appendix~\ref{App: Details about priors SF+G} includes tables with the numerical values of the transfer 
functions for $\rho(0)$ and clarifies their meaning in more detail. The important fact is that, as the prior 
strength increases from $P_0$ to $P_2$, the numerical value of the $\mathcal{T\!F}_j^{\rho(0)}$ for each 
observation $q_j$ tends to decrease---sometimes by an order of magnitude. This leads to a dramatic 
decrease in the total variance in the interior density. On the other hand, as the prior strength increases, 
the prior transfer functions $\mathcal{T\!F}_k^{\rho(0)}$ become stronger. This allows each prior center 
$\omega^{0}_k$ to push away the value of $\rho(0)$ from what the data suggest, effectively increasing 
the bias. 

The example highlights how a well chosen prior could be crucial to reduce uncertainties. However, if the prior strength is excessively
high, there is the risk of overlooking new discoveries or making erroneous conclusions. A more in depth analysis is required to optimize the prior strength and structure for each particular 
problem in order to effectively reduce the MSE for a set of given truths.

\section{Conclusions and future directions}\label{Sec: Conclusions}

In this paper, we proposed a novel statistical framework--the transfer function (TF) formalism--and applied to the extraction of 
nuclear densities from the associated form factors obtained from electron-scattering data. From this new perspective, we explored: 
model selection and model building, the impact of data locations and errors, the role of priors, and the bias vs variance trade-off. Given
the importance of the PREX and CREX campaigns at JLab in constraining the density dependence of the symmetry energy and in bridging
ab-initio descriptions to density functional theory, we focused our analysis on $^{48}$Ca and $^{208}$Pb. In particular, the two observables
of interest explored in this work were the mean square radii and interior densities of both neutron-rich nuclei. We evaluated the performance 
of seven models in faithfully reproducing these two observables, from noisy experimental data on the electric form factor and noisy 
pseudo-data generated from a variety of relativistic mean field models for the case of the weak-charge form factor. The performance of
the various models was quantified in terms of the Mean Squared Error (MSE) defined as a combined score obtained from incorporating 
both the bias and the variance.

For both the charge and weak charge densities we showed that, for the adopted noise level assumed in the data, the best performance
was obtained with the simpler SF and Helm models that are characterized by a flat interior density. More complex models such as the 
SF+G or a Fourier-Bessel expansion did not perform as well. Whereas both of these more complex models are able to reproduce the interior shell 
oscillations of both nuclei, they are hindered by a very high variance, which ultimately results in a high MSE score. In this regard, we
suggest that it will be difficult for any of the models used in this paper---at least in their present form---to faithfully reproduce the shell 
oscillation of both nuclei, particularly in the case of $^{48}$Ca where the oscillation structure is expected to be more pronounced.
Indeed, when using the Fourier-Bessel expansion as in \cite{lin2015full}, we estimated that there is a $15\%$ chance of predicting the wrong oscillating structure in the interior of $^{48}$Ca,
namely, peaks become valleys and valleys become peaks. 

In the context of experimental design, we illustrated how to use the TF formalism to identify those critical observations that are driving 
most of the uncertainty in our estimations. The identification of those critical points could help in the design of future experiments to 
allocate more resources (e.g., beam time) to those critical locations to maximize the information gained from such  experiments. 
Finally, we explored the impact of priors on the extracted weak charge density of $^{48}$Ca under the SF+G model. As the influence
of the prior increased, so did the bias while the variance was reduced---as expected from the bias vs variance trade-off. 

Going forward, there are several directions that are worth exploring. First, it would be interesting to integrate the TF formalism directly 
into model building. We believe questions such as \textit{what makes a model better than others?}, could be tackled from the TF perspective. 
Answering \textit{which} model is better at extracting data has become a central question in nuclear physics, for example in the context of
the proton puzzle. Yan et al., investigated this question and provided fundamental insights to the analysis by the PRaD collaboration.
This seminal work---which inspired a great portion of the development of the TF formalism---identified the models optimally suited to extract 
the proton radius, but did not elaborate on \textit{what} made those model successful. We believe the TF formalism could be used to make
significant advances in that direction. As shown in this paper, the TF formalism seems to be ideal to identify the delicate interplay between signal and 
noise.
Understanding the TF distribution of successful models could help not only in identifying but also in creating, some sort of ``optimal" model. 
This technique could be applied beyond density reconstruction from scattering data as implemented in this paper, to more general problems
that involve the calibration of model parameters from experimental data. 

Another fruitful direction of investigation is the role of priors and hyperparameters. Hyperparameters, such as $\gamma$ and the Gaussian locations for the SF+G, or the Bessel cut off radius $R_\text{cut}$ and number of coefficients, can drastically impact the performance of a
model. In the context of the TF formalism, we could ask questions like: \textit{Given six observations, is it better in terms of an overall 
MSE score, to have 5 or 6 adjustable Fourier-Bessel coefficients? How does the answer scale with the number of data points?} We believe 
it is possible to create a framework using the TF formalism that can tackle this type of questions in a robust and direct manner. This would 
allow to conduct a more informed search in the hyperparameter space of each model instead of just by trial and error. Our work showed that 
the incorporation of priors can have a dramatic effect on a model's performance. After all, priors are essential ingredients of the Bayesian 
formalism as they encode prior beliefs before additional experimental evidence becomes available. A more in depth study should be carried
out to identify how to optimize the hyperparameters that define the priors. To reach robust conclusions, such a research project should include more generator functions 
from other nuclear model families. We are confident that the TF formalism can guide this optimization procedure 
as well. 

Finally, a third possible application of the TF formalism is related to the recent use in nuclear physics of Bayesian frameworks for 
combining different competing models to improve over the predictions of single models \cite{neufcourt2020quantified}. Within the
context of nuclear densities, using the MSE score should allow us to test the circumstances under which the mixing of several 
models outperforms the predicting power of a single model. It would be interesting to explore in the future the generalization of 
the TF formalism to Bayesian model mixing.

\section*{Acknowledgments}

We are grateful to Edgard Bonilla for his help and critical observations during this project. We thank Prof. Antonio Linero for his guidance and key support. We thank Diogenes Figueroa for many useful conversations. We thank Prof. Douglas Higinbotham for introducing us to the bias vs variance analysis and for his encouragement at the beginning of the project. Finally, we thank Ana Posada for a careful read of the manuscript. 

This  material  is  based  upon work  supported  by  the U.S.  Department  of  Energy  Office of  Science,  Office of  Nuclear  Physics  under  Award Number  DE-FG02-92ER40750

\newpage

\appendix

\section{Mathematical proofs on the TF formalism}\label{Sec: App Math Proofs}

\subsection{Transfer Functions Structure}

This subsection presents a formal proof on the structure of the transfer functions (Eq.~\eqref{Eq: TF omega Definition}), namely that the first order change coefficients on the parameters $\boldsymbol{\omega}$ due to a perturbation on observation $y_j$ are:

\begin{equation}\label{Eq: App TF Params}
    \mathcal{T\!F}_j^{\boldsymbol{\omega}}= \mathcal{H}^{-1}\nabla F_j\sigma_j^{-2},
\end{equation}

where $\mathcal{H}^{-1}$ is the inverse of the Hessian matrix of $\chi^2/2$ defined in~\eqref{Eq: chi2 Hessian} and $\nabla F_j$ is the gradient with respect to the parameters $\boldsymbol{\omega}$ of the function $F$ being fit  evaluated at observation $y_j$.

Let us assume that we are at the minimum $\boldsymbol{\omega_0}$ of the unperturbed $\chi^2/2$. At this point, the condition of a minimum implies that the first derivative of $\chi^2/2$ with respect to all $\omega_k$ ($K$ in total) should be zero:

\begin{equation}\label{Eq: App Chi Min}
\frac{1}{2}\frac{\partial \chi^2}{\partial \omega_k}\Big|_{(\boldsymbol{\omega_0},\boldsymbol{y_0})} \equiv G_{k}( \boldsymbol{\omega},\boldsymbol{y})\big|_{(\boldsymbol{\omega_0},\boldsymbol{y_0})} = 0,
\end{equation}

where we use the notation $\boldsymbol{y}\equiv (y_1,\ ... \ y_n)$ to refer to the group of all $J$ observations, and the subscript ``$\boldsymbol{0}$" to refer to the unperturbed variables. We call $G_{k}( \boldsymbol{\omega},\boldsymbol{y})$ the first derivative of $\chi^2/2$ with respect to parameter $\omega_k$. The $G_k$ are the following functions of both the parameters and the observations:

\begin{equation}\label{Eq: App Grad Chi}
G_{k}( \boldsymbol{\omega},\boldsymbol{y})=\sum_j^J \frac{(F_j-y_j)}{\sigma_j^2}\frac{\partial F_j}{\partial \omega_k}.   
\end{equation}

Now, if we perturb observation $y_j$ by a small amount $\delta y_j$ the minimum of $\chi^2/2$ will move accordingly. If we want to preserve all $K$ equations~\eqref{Eq: App Chi Min}, (there is one equation for every parameter),
then the values of all $\omega_k$ should change a small amount as well $\delta \omega_k$ to compensate. Quantitatively, this means (to first order):

\begin{equation}
    \delta y_j \frac{\partial G_k}{\partial y_j}=-\sum_i^K \delta \omega_i \frac{\partial G_k}{\partial \omega_i}.
\end{equation}

We can arrange all $K$ equations into a matrix form: 

\begin{align*}
  \frac{\partial G_1}{\partial y_j}& \delta y_j =  - \Big( \frac{\partial G_1}{\partial \omega_1}\delta \omega_1+ \frac{\partial G_1}{\partial \omega_2} \delta \omega_2 \  ... \ + \frac{\partial G_1}{\partial \omega_K} \delta \omega_K  \Big) \\
   \frac{\partial G_2}{\partial y_j}& \delta y_j = - \Big( \frac{\partial G_2}{\partial \omega_1}\delta \omega_1+ \frac{\partial G_2}{\partial \omega_2} \delta \omega_2 \  ... \ + \frac{\partial G_2}{\partial \omega_K} \delta \omega_K  \Big)  \\
   &\;\;\vdots \notag \ \ \ \ \ \ \ \ \ \ \ \  \ \ \ \ \ \ \ \ \ \ \ \  \;\;\vdots \ \ \ \ \ \ \ \ \ \ \ \ \ \ \ \ \ \ \ \;\;\vdots \notag \\
    \frac{\partial G_K}{\partial y_j}& \delta y_j = - \Big( \frac{\partial G_K}{\partial \omega_1}\delta \omega_1+ \frac{\partial G_K}{\partial \omega_2} \delta \omega_2 \  ... \ + \frac{\partial G_K}{\partial \omega_K} \delta \omega_K  \Big) ,
\end{align*}

where, since the $G_k$ were already first derivatives of $\chi^2/2$, we can recognize the Hessian matrix $\frac{\partial G_i}{\partial \omega_k} = \mathcal{H}_{i,k}$. We also recognize $\frac{\partial G_k}{\partial y_j}=-\frac{\partial F_j}{\partial \omega_k}\sigma_j^{-2}$. We therefore have:

\begin{align*}
   -\nabla F_j\sigma_j^{-2}\delta y_j = - \mathcal{H} \boldsymbol{\delta \omega}, \\
   \big[\mathcal{H}^{-1}\nabla F_j \sigma_j^{-2}\big] \delta y_j= \boldsymbol{\delta \omega},
\end{align*}

where, since the perturbation $\delta y_j$ can be made arbitrary small, we must conclude that the quantity in brackets is what were looking for. These are the linear coefficients connecting a small change in $y_j$ with the small change in every parameter $\omega_k$, proving Eq.~\eqref{Eq: App TF Params}.

This was a constructive proof of the transfer function structure. Another approach would be to use Newton's minimization method to find the location of the new minimum of $\chi^2/2$ once we make the perturbation $y_j \rightarrow y_j + \delta y_j$. Newton's method involves the same ingredients shown in Eq.~\eqref{Eq: App TF Params}, namely the gradient and Hessian of the scalar objective function ($\chi^2/2$). The reason why the gradient does not involve all observation but just the $j$th component, is because in Eq.~\eqref{Eq: App Grad Chi} only the term multiplying $\delta y_j$ survives. Everything else gets cancelled by the definition of the minimum.

Finally, a third approach results by invoking the implicit function theorem on the minimum conditions of $\chi^2/2$. Those are $K$ equations and each one is a function of the $J$ observations and $K$ parameters, where $K\leq J$. Under these conditions, there exists a map in a vicinity around $(\boldsymbol{\omega_0},\boldsymbol{y_0})$ from the bigger set of variables (the observations) to the smaller set (the parameters). The coefficients of this linear map (first derivatives) are precisely given by  ~\eqref{Eq: App TF Params} (see \cite{spivak2018calculus} pages 41-42).

\subsection{Comparing the transfer function variance with the standard approach's variance}

This subsection presents a formal proof of the statement discussed at the end of Sec~\ref{Sec: Variance}. When the Hessian matrix~\eqref{Eq: chi2 Hessian} only involves the linear part of the model:

\begin{align*}
   & \mathcal{H}_{i,k} = \sum_{j=1}^J \frac{1}{\sigma_j^2} \Big[\Big(\frac{\partial F(q_j,\boldsymbol{\omega})}{\partial \omega_i}\Big)\Big(\frac{\partial F(q_j,\boldsymbol{\omega})}{\partial \omega_k}\Big)\Big],\numberthis \label{Eq: App chi2 Hessian}
\end{align*}

then the variance calculated using the standard approach~\eqref{Eq: Standard Fit}, and the variance calculated using the transfer function formalism~\eqref{Eq: TF Variance}, are identical, namely:

\begin{align}
   &\Delta m^2 = \nabla m \mathcal{H}^{-1} \nabla m, \label{Eq: App Var 1}\\
   &\Delta m^2 =\sum_j^J \Big[\nabla m \ \mathcal{H}^{-1} \nabla F(q_j,\boldsymbol{\omega})\sigma_j^{-2}\Big]^2 \sigma_j^2 \label{Eq: App Var 2}.
\end{align}

Let us recall that we have $J$ observations and $K$ parameters with $K\leq J$. To prove this statement, we first observe that the matrix $\mathcal{H}$ is built by the sum of $J$ tensor products between the gradients $\nabla F_j \sigma_j^{-1}$ with themselves:

\begin{equation}
    \mathcal{H}=\sum_j^J \big[\nabla F_j \sigma_j^{-1} \big] \otimes \big[\nabla F_j \sigma_j^{-1} \big].
\end{equation}

Therefore, $\mathcal{H}$ can be written as the product of a matrix $\mathcal{F}$ and its transpose as:

\begin{equation}
    \mathcal{H} = \mathcal{F} \mathcal{F}^{T},
\end{equation}

where $\mathcal{F}$ is a $K \times J$ matrix which columns are the gradients of $F$:

\begin{equation}
    \mathcal{F}= \Big(\nabla F_1 \sigma_1^{-1} \ \  \nabla F_2 \sigma_2^{-1} \ \ \cdots \ \  \nabla F_J \sigma_J^{-1}    \Big).
\end{equation}

We decompose $\mathcal{F}$ and $\mathcal{F}^{T}$ into their $QR$ decomposition (see Theorem 2 in \cite{goodall199313}):

\begin{align*}
    \mathcal{F} = R^T Q^T, \numberthis \\
    \mathcal{F}^T = QR,
\end{align*}

where $Q$ and $R$ are a $J\times K$, and a $K\times K$ matrix, respectively. The $Q$ matrix is an orthonormal matrix meaning: $Q^TQ = I_{K,K}$, where $I_{K,K}$ is the $K\times K$ identity matrix. Under these conditions, we have that:

\begin{align*}
    &\nabla m \mathcal{H}^{-1} \nabla m = \nabla m \Big[ R^T Q^T Q R \Big]^{-1} \nabla m = \numberthis \\
    &\nabla m R^{-1}(R^T)^{-1} \nabla m = 
    || (R^T)^{-1} \nabla m ||^2.
\end{align*}

Therefore, under the standard approach Eq.~\eqref{Eq: App Var 1} is calculating the norm squared of the vector $(R^{T})^{-1}\nabla m$ in $\mathbb{R}^{K}$. Let us now work with Eq.~\eqref{Eq: App Var 2} and obtain a similar structure but in $\mathbb{R}^{J}$.

Given how $\mathcal{F}$ is defined, we realize that:

\begin{equation}
    \mathcal{F}\cdot e_j = \nabla F_j \sigma_j^{-1},
\end{equation}

where $e_j = (0,0,\cdots ,1, \cdots \ ,0)$ is the $j$th vector in the canonical base of $\mathbb{R}^{J}$ with all entries as $0$ except for an entry of $1$ in position $j$.

Therefore, in Eq.~\eqref{Eq: App Var 2}, we can replace $\nabla F_j \sigma_j^{-1}$ by $\mathcal{F}\cdot e_j$ and obtain:

\begin{align*}
    &\sum_j^J \Big[\nabla m\mathcal{H}^{-1}\mathcal{F}\cdot e_j\Big]^2 = \numberthis\\
    &\sum_j^J \Big[\nabla m R^{-1} (R^T)^{-1} R^T Q^T\cdot e_j\Big]^2 =\\
    &  \sum_j^J\Big[ \Big( \nabla m R^{-1} Q^T\Big)  \cdot e_j\Big]^2 = || Q (R^T)^{-1} \nabla m||^2, 
\end{align*}

where the last term is the norm of the vector $ Q (R^T)^{-1} \nabla m$ calculated in $\mathbb{R}^{J}$. But, since the matrix $Q$ is orthonormal, it preserves norms when taking vectors from $\mathbb{R}^{K}$ to $\mathbb{R}^{J}$. We must conclude that this expression is also the norm of the vector $(R^T)^{-1} \nabla m$ in $\mathbb{R}^{K}$, which proves that Eq.~\eqref{Eq: App Var 1} and Eq.~\eqref{Eq: App Var 2} are identical.

\section{Model descriptions}\label{App: Models Details}

This section presents detailed information about the seven models we studied in this work. For each model, we provide: its analytic form (if available) as a function of its parameters in both coordinate space $\rho(r)$ and momentum space $F(q)$; a normalization condition (if any), that restricts the parameters; the mean squared radius $R$ as a function of the parameters when available, and the values of the model's hyperparameters, if any, used in this work. 

\ul{\textit{Fourier Bessel}} 

Under this formalism \cite{dreher1974determination}, the density is written as:

\begin{equation}\label{Eq: Bess}
  \rho_{FB}(r) =   H(R_\text{cut}-r)\sum _{\nu =1} ^N a_\nu j_0\big (q_\nu r\big ),
\end{equation}

where $j_0$ denotes the zeroth order spherical Bessel function of the first kind, $a_\nu$ are the free parameters, $q_\nu=\nu \pi/R_{\text{cut}}$ and $R_\text{cut}$ is such that $\rho(r)=0$ for $r>R_\text{cut}$. This last condition is enforced by the  Hevisde theta function $H$.

The form factor can be expressed analytically as:
\begin{align}
    F_{FB}(q) & = \sum_{\nu=1}^N a_\nu G_\nu(q), \ \text{where}\label{Eq: Bessel Form F} \\
    G_\nu(q) & \equiv 4 \pi (-1)^{\nu} j_0(qR_\text{cut})\frac{R_\text{cut}}{q^2 - q_\nu^2}.\label{Eq: Gnu Definition}
\end{align}

The normalization condition translates to:

\begin{equation}
    F_{FB}(0)=\sum_{\nu=1}^N (-1)^{\nu+1}\frac{4\pi R_\text{cut}}{q_\nu^2}a_\nu = 1.
\end{equation}

The mean square radius is obtained as:

\begin{equation}
R^2 = 4\pi \sum_\nu^N a_\nu  \frac{(-1)^{\nu}R_{\rm cut}^5(6-\nu^2\pi^2)}{\nu^4\pi^4}.
\end{equation}

In this work, we use $R_\text{cut}=7$ fm when analyzing $^{48}$Ca, and $R_\text{cut}=10$ fm when analyzing $^{208}$Pb. We use a total of 5 adjustable $a_\nu$ for both nuclei which translates to six Bessels ($N=6$) due to the normalization condition.

\ul{\textit{Helm Density}}

The Helm density \cite{helm1956inelastic} is:

\begin{align*}
    &\rho_H(r)=\frac{1}{2}\rho_{0H}\Big[\text{erf}\Big(\frac{r+R_0}{\sqrt{2}\sigma} \Big) -\text{erf}\Big(\frac{r-R_0}{\sqrt{2}\sigma} \Big) \Big]+\numberthis \\
    & \frac{1}{\sqrt{2\pi}}\Big( \frac{\sigma}{r}\Big)\rho_{0H}\times \\ &\Big\{\text{exp}\Big[- \frac{(r+R_0)^2}{2\sigma^2} \Big]  -\text{exp}\Big[- \frac{(r-R_0)^2}{2\sigma^2} \Big] \Big\},
\end{align*}
where:
\begin{equation}
    \rho_{0H}\equiv\frac{3}{4\pi R_0^3},
\end{equation}

where $R_0$ and $\sigma$ are the adjustable parameters, and $\text{erf}(x)$ is the error function:

\begin{equation}
    \text{erf}(x)\equiv \frac{2}{\sqrt{\pi}}\int_0^xe^{-z^2}dz.
\end{equation}

The Helm form factor is built by the product of two form factors: a uniform ``box" density inspired by the fact that nuclear matter in nuclei saturates, and a Gaussian falloff which takes into account the finite size of the nucleons:

\begin{equation}
    F_H(q)= 3 \frac{j_1(qR_0)}{qR_0} e^{-q^2\sigma^2/2},
\end{equation}

where $j_1$ is the spherical Bessel function of first order $j_1(x)=\frac{\text{sin}(x)}{x^2}-\frac{\text{cos}(x)}{x}$.

The radius is given by:

\begin{equation}
    R^2= \frac{3}{5}R_0^2+3\sigma^2.
\end{equation}

\ul{\textit{Symmetrized Fermi Function}}

The Symmetrized Fermi function \cite{sprung1997symmetrized} is constructed as: $f_{SF}(r)\equiv f_{SF}(r)+f_{SF}(-r)-1$, where $f_{SF}$ is the traditional Fermi Function \cite{woods1954diffuse}. Its density and form factor are expressed as:

\begin{equation}
    \rho_{SF}(r)=\rho_{0SF}\frac{\text{sinh}(c/a)}{\text{cosh}(r/a)+\text{cosh}(c/a)},
\end{equation}

\begin{align*}
   F_{SF}(q)=& \frac{3}{qc[(qc)^2+(\pi qa)^2]}\Big[\frac{\pi qa}{\text{sinh}(\pi qa)}\Big]\times\\\numberthis 
   &\Big[\frac{\pi qa}{\text{tanh}(\pi qa)}\text{sin}(qc)-qc\ \text{cos}(qc)  \Big],
\end{align*}

where the normalization constant is:

\begin{equation}
    \rho_{0SF}=\frac{3}{4\pi c \big(c^2+ \pi ^2 a^2 \big)},
\end{equation}

where the parameters $a$ and $c$ represent the surface diffuseness and half-density radius, respectively. The radius $R$ is given by:

\begin{equation}
    R^2= \frac{3}{5}c^2+\frac{7}{5}(\pi a)^2.
\end{equation}

\ul{\textit{Symmetrized Fermi Function of three and four parameters}}

Based on the two parameter symmetrized Fermi density, we can build a three-parameter and a four-parameter densities as:

\begin{align*}
        &\rho_{SF3}(r)=  \numberthis\label{Eq: SF3}\\
        &\rho_{0SF3}\big(1+wr^2 \big)\frac{\text{sinh}(c/a)}{\text{cosh}(r/a)+\text{cosh}(c/a)},
\end{align*}

\begin{align*}
        &\rho_{SF4}(r)= \numberthis\label{Eq: SF4}\\
        &\rho_{0SF4} \big(1+wr^2 +ur^4 \big)
        \frac{\text{sinh}(c/a)}{\text{cosh}(r/a)+\text{cosh}(c/a)}.
\end{align*}

The parameters $w$ and $u$ are introduced to add flexibility near $r=0$ to the densities and to reproduce oscillations. Due to the size of their expressions, the normalization constants $\rho_{0SF3}$ and $\rho_{0SF4}$, the form factors and the radius equations are not included here but can be calculated analytically.

\ul{\textit{Symmetrized Fermi function plus Bessels (SF+B) and plus Gaussians (SF+G)}}

By construction the two parameter symmetrized Fermi function $\rho_{SF}(r)$ exhibits a flat behaviour in the interior, so it cannot describe shell oscillations (small bumps and valleys around the saturation density background). As a different approach to adding parameters as in Eqs.~\eqref{Eq: SF3} and~\eqref{Eq: SF4}, we propose to use the following hybrid models.

SF+B:
\begin{align*}
    \rho_{SFB}(r) & = \Big(1-\sum_{\nu=1}^{N_B} (-1)^{\nu+1}\frac{4\pi R_\text{cut}}{q_\nu^2}a_\nu \Big) \rho_{SF}(r)\\
    &+ H(R_\text{cut}-r)\sum _{\nu=1}^{N_B} a_\nu \ j_0(q_{\nu}r),
\end{align*}

where $j_0$ is the spherical Bessel function of the first kind and $H$ the Heaviside theta function. We set $R_\text{cut}=3.3$ fm for $^{48}$Ca and $R_\text{cut}=5$ fm for $^{208}$Pb. $R_\text{cut}$ in this context specifies the region where we believe the interior oscillations are important. We use a total of three adjustable coefficients $a_\nu$ which in this case correspond to three Bessels ($N_B=3$) since the normalization is enforced automatically. We have therefore a total of five parameters including $a$ and $c$ from the SF model.

SF+G:

\begin{align*}
    \rho_{SFG}(r) & = \Big(1-\sum_{i=1}^{N_G} A_i \Big) \rho_{SF}(r)+ \numberthis\\
    & \frac{1}{2\pi^{3/2}\gamma^3} \sum _{i=1}^{N_G} A_i \ g(r,R_i),
\end{align*}

where $ g(r,R_i)$ is defined as:
\begin{align*}\label{Eq: g Definition}
   &g(r,R_i)  = \numberthis \\ &\frac{1}{1+2R_i^2/\gamma^2}(e^{-(r-R_i)^2/\gamma^2}+e^{-(r+R_i)^2/\gamma^2}).
\end{align*}

The amplitudes of the Gaussians $A_i$ act as adjustable parameters. In our case they are not restricted to be positive in contrast with \cite{sick1974model}. We use a total of three Gaussians ($N_G=3$), giving us five adjustable parameters including $a$ and $c$ from the SF model. The hyperparameter $\gamma$, which represents the common width of the Gaussians, is set to $\gamma=1.4$ fm for both nuclei (see Sec~\ref{Sec: Rec Bias And Opt Func} for a discussion of this value). The center of each Gaussian is denoted by $R_i$ and we chose the following values:

\begin{align*}
    &\boldsymbol{R}=[0,1.3,2.6] \ \text{fm} \ \text{for $^{48}$Ca}, \numberthis\\
    &\boldsymbol{R}=[0,2,4] \ \text{fm} \ \text{for $^{208}$Pb.}
\end{align*}

The main idea behind these expansions is that the principal behaviour of the nuclear density is modeled by the symmetrized Fermi density, while the fine details are modeled by either a sum of Bessels or Gaussians. Both $R_\text{cut}$ and the $R_i$ are chosen in such a way that they cover the region where we expect the oscillations around a flat density to be important.

The Bessel approach has the disadvantage that the total density $\rho_{SFB}(r)$ will present a discontinuity at $r=R_\text{cut}$. This ``kink",
while nonphysical, might not preclude the entire model from describing nuclear densities.

The form factors for both densities can be expressed in analytic form:

\begin{align*}
    &F_{SFB}(q) = \numberthis \\
    &\Big(1-\sum_{\nu=1}^{N_B} (-1)^{\nu+1}\frac{4\pi R_\text{cut}}{q_\nu^2}a_\nu \Big)F_{SF}(q) +\sum _{\nu=1}^{N_B} a_\nu \ G_\nu(q), 
\end{align*}

\begin{align}
    F_{SFG}(q) &= \Big(1-\sum_{i=1}^{N_G} A_i \Big)F_{SF}(q) + \sum _{i=1}^{N_G} A_i \ \tilde{g}(q,R_i),
\end{align}

where $G_\nu(q)$ is defined in Eq.~\eqref{Eq: Gnu Definition}, and $\tilde{g}(q,R_i))$ is defined as:

\begin{align*}
     & \tilde{g}(q,R_i) = \\ 
     & e^{-q^2\gamma^2/4}\frac{1}{1+2R_i^2/\gamma^2} \Big( \text{cos}(qR_i) + \frac{2R_i^2}{\gamma^2}j_0(q R_i) \Big).\numberthis \label{Eq: gtilde Definition}
\end{align*}

The SF+B radius is given by:

\begin{align*}
    R^2=&\Big(1-\sum_{\nu=1}^{N_B} (-1)^{\nu+1}\frac{4\pi R_\text{cut}}{q_\nu^2}a_\nu \Big)\Big( \frac{3}{5}c^2+\frac{7}{5}(\pi a)^2 \Big) + \\
    &4\pi \sum_{\nu=1}^{N_B} a_\nu  \frac{(-1)^{\nu}R_{\rm cut}^5(6-\nu^2\pi^2)}{\nu^4\pi^4}.\numberthis
\end{align*}

The SF+G radius is given by:

\begin{align*}
    R^2=&\Big(1-\sum_i^{N_G} A_i \Big)\Big( \frac{3}{5}c^2+\frac{7}{5}(\pi a)^2 \Big) + \\
    &\sum_{i=1}^{N_G} A_i\frac{3 \gamma ^4+4 R_i^4+12 \gamma ^2 R_i^2}{2 \left(\gamma ^2+2 R_i^2\right)}.
\end{align*}

\section{Details about the SF+G reconstruction bias analysis}\label{App: Details about the SF+G plots}

This section provides a more detailed view at the reconstruction bias for the SF+G model discussed in Sec~\ref{Sec: Rec Bias And Opt Func}. Let us recall that we are analyzing how much the placement of the fourth measurement $q_4$ impacts the change in $\rho(0)$ through the transfer function formalism:

\begin{equation}\label{Eq: App Delta Rho Eta4}
    \delta \rho(0) =  \Big[\mathcal{T\!F}_4^{(\rho(0))}\Big]\eta_{y_4}.
\end{equation}

\begin{figure*}
	\centering
		\includegraphics[width=0.9\textwidth]{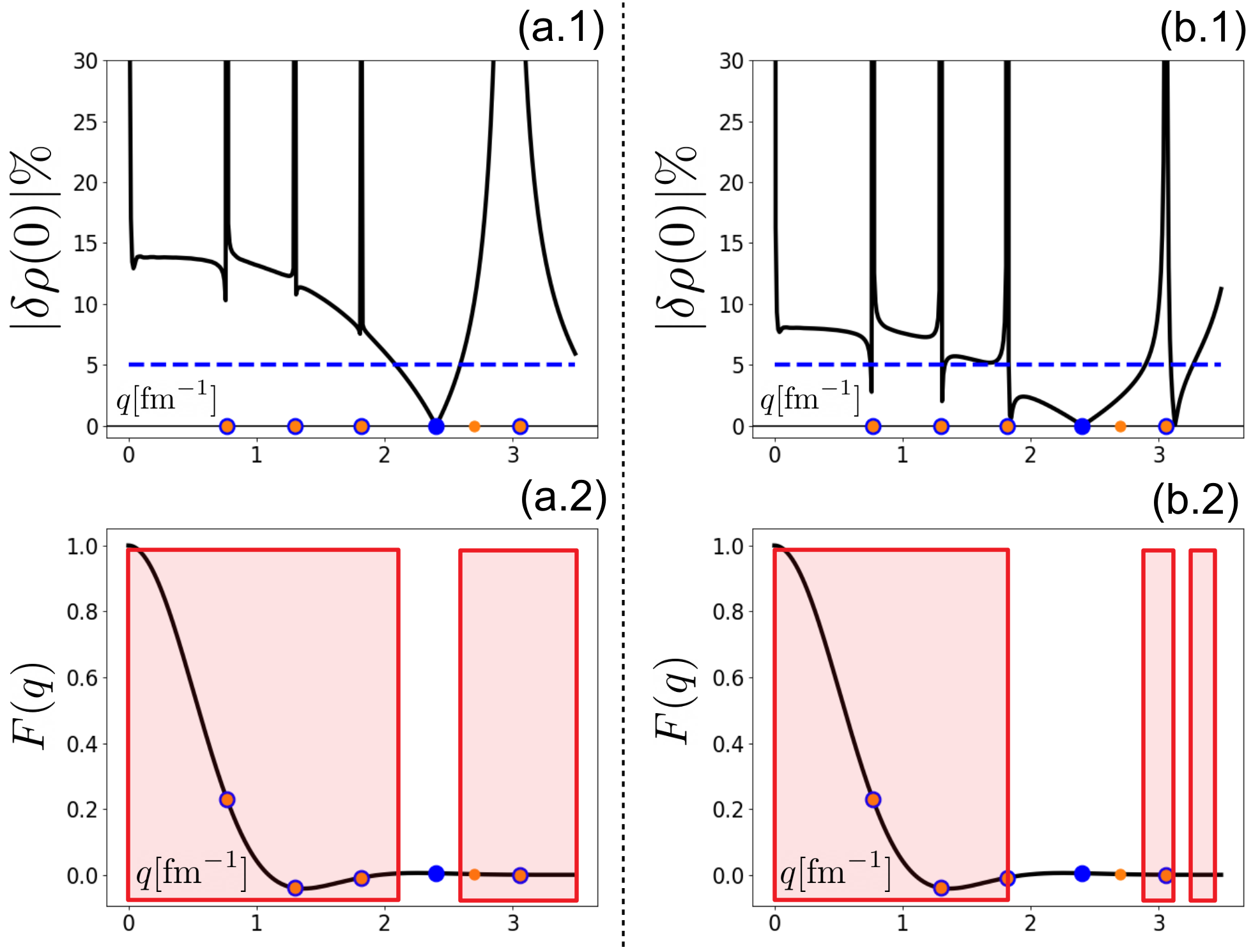}
	\caption{Bias increase in $\rho(0)$, when using the SF+G model, as a function of the location of the fourth measurement $q_4$ for $\gamma=0.7$ fm (a), and $\gamma=1.4$ fm (b). (a.1) and (b.1) show in black the calculated $\delta \rho(0)$ (as an absolute percentage) using Eq\eqref{Eq: App Delta Rho Eta4}. The blue and orange points correspond to the two original data sets while the blue dashed line represents the threshold $|\delta \rho(0)|=5\%$. (a.2) and (b.2) show the true form factor in black as well as the measurement locations for both sets. The red rectangles encompass the regions where, if $q_4$ is located, $|\delta \rho(0)|>5\%$.}\label{Fig: Details on SF+G etas}
\end{figure*}

Fig.~\ref{Fig: Details on SF+G etas} ((a.1) and (b.1)) shows this calculated $\delta \rho(0)$ in black (as an absolute percentage) as $q_4$ moves along the possible momentum transfer range $[0-3.5]$ fm$^{-1}$ while the other $q_j$ remain in position. The SF+G model with the hyperparameter $\gamma=0.7$ fm is displayed in (a.1) while $\gamma=1.4$ fm is used in (b.1). The x-axis of both plots shows the two original data sets of $\boldsymbol{q_0}=[0.77,1.30,1.82,2.41,3.06]$ fm$^{-1}$ (blue), and $\boldsymbol{q_1}=[0.77,1.30,1.82,2.7,3.06]$ fm$^{-1}$ (orange), which only differ in their $q_4$ value.

In Fig.~\ref{Fig: Details on SF+G etas} (a.1) it can be observed that for most locations the calculated $|\delta \rho(0)|\%$ exceeds the $5\%$ threshold (the blue dashed line). On the other hand, for (b.1) the black curve is overall lower, showing a wider region where $|\delta \rho(0)|\leq 5\%$. This overall reduction is the reason why we chose $\gamma=1.4$ fm to perform the analysis in Sec~\ref{Sec: Results}.

The ``spikes" where $|\delta \rho(0)|$ grows abruptly appear when we aim to move $q_4$ to the other $q_j$ locations, effectively measuring twice in the same spot. Since we have five parameters and five observations, this results in a non-invertible Hessian for $\chi^2$, which blows up the transfer function $\mathcal{T\!F}_4^{\rho(0)}$. It is interesting to note that, even though there is not an observation located at $q=0$ fm$^{-1}$, a ``spike" can still be observed. This is because by definition all model form factors must respect $F(0)=1$. Therefore a measure at $q=0$ fm$^{-1}$ provides no new information, resulting in a non-invertible Hessian as well. 

Fig.~\ref{Fig: Details on SF+G etas} ((a.2) and (b.2)) shows the true form factor (RMF012) in black and the observation locations as the blue and orange points, being (a.2) for $\gamma=0.7$ fm, and (b.2) for $\gamma=1.4$ fm. The red squares represent ``forbidden zones", regions in the $q$ space where if we place the fourth location $q_4$ we obtain a bias in $\rho(0)$ bigger than $5\%$. For (a.2) these zones occupy around $85\%$ of the total $q$ range, while for (b.2) they occupy around $70\%$.

By the construction in Sec~\ref{Sec: Rec Bias And Opt Func}, locating $q_4$ at the blue point $q=2.41$ fm$^{-1}$ will result in almost zero bias since the optimal form factor and the true form factor have the same value ($\eta_4=0$). As $q_4$ moves away from this location, we expect the bias to increase. A ``bad" model will present a bias that grows too quickly, while a ``good" model will be more tolerable. As can be observed in Fig.~\ref{Fig: Ca48 Reconstruction Bias 1}, for the same $q_4=2.7$ fm$^{-1}$, the SF+G with $\gamma=0.7$ fm (orange dashed curve) fails to be within the $\lambda_\rho$ band, while the SF+G with $\gamma=1.4$ fm (red dashed line) stays within. In Fig.~\ref{Fig: Details on SF+G etas} (a.2) the orange point is covered by the red rectangle. This is not the case for (b.2). 

It should be noted that changing the hyperparameter $\gamma$ from $0.7$ fm to $1.4$ fm can impact the locations where $\eta_j=0$ since now the optimal function will be different. Nevertheless, in the case we studied here the change in these locations was negligible, a fact that simplified our discussion.

\section{Details about the $^{48}$Ca and $^{208}$Pb charge examples and model comparison}\label{App: Tables Ca and Pb example}

This Appendix presents tables and details relevant to the $^{48}$Ca and $^{208}$Pb charge density example developed in Sec~\ref{Sec: Ca and Pb Charge Example}. It includes the results of applying the same analysis on model comparison developed in Sec~\ref{Sec: Weak Charge Results} to the electric charge densities of both nuclei instead of their weak densities counterparts.

\begin{table}[h]
\centering
\begin{tabular}{|ll|l|l|l|l|l|}
\hline
\multirow{2}{*}{$q_j$ [fm$^{-1}$]}   &  [$q_0$] & 0.90 & 1.35 & 1.80 & 2.24 & 2.69 \\ \cline{2-7} 
                      & [$q_m$] & 0.65 & 1.26 & 1.73 & 2.12 & 2.55 \\ \hline
\multirow{2}{*}{$|\mathcal{T\!F}_j^{\rho(0)}\sigma_j|$}  &  [$q_0$] & 0.84 & 1.02 & 1.44 & 1.07 & 0.65 \\ \cline{2-7} 
                      & [$q_m$] & 0.65 & 1.18 & 1.40 & 0.90 & 0.72 \\ \hline
\multirow{2}{*}{$|\mathcal{T\!F}_j^{R}\sigma_j|$} &  [$q_0$] & 0.88 & 0.80 & 0.58 & 0.32 & 0.12 \\ \cline{2-7} 
                      & [$q_m$] & 0.82 & 0.34 & 0.26 & 0.15 & 0.03 \\ \hline
\end{tabular}\caption{$^{48}$Ca electric form factor momentum transfer locations $q_j$ and transfer functions $\mathcal{T\!F}_j$ absolute values of the density at $r=0$ fm and radius for two data sets: original $\boldsymbol{q_0}$ and optimized $\boldsymbol{q_m}$. Both transfer functions have been normalized by their respective natural scales defined in Table~\ref{Tab: Natural Scales}.}\label{Tab: Ca TFs Example}
\end{table}
\begin{table}[h]
\centering
\begin{tabular}{|ll|l|l|}
\hline
\multirow{2}{*}{$q_j$ [fm$^{-1}$]}   & [$q_0$] & 0.50 & 0.80 \\ \cline{2-4} 
                      & [$q_m$] & 0.41 & 0.84 \\ \hline
\multirow{2}{*}{$|\mathcal{T\!F}_j^{\rho(0)}\sigma_j|$}  & [$q_0$] & 0.14 & 0.18 \\ \cline{2-4} 
                      & [$q_m$] & 0.17 & 0.25 \\ \hline
\multirow{2}{*}{$|\mathcal{T\!F}_j^{R}\sigma_j|$} & [$q_0$] & 0.59 & 0.48 \\ \cline{2-4} 
                      & [$q_m$] & 0.54 & 0.31 \\ \hline
\end{tabular}\caption{$^{208}$Pb electric form factor momentum transfer locations $q_j$ and transfer functions $\mathcal{T\!F}_j$ absolute values of the density at $r=0$ fm and radius for two data sets: original $\boldsymbol{q_0}$ and optimized $\boldsymbol{q_m}$. Both transfer functions have been normalized by their respective natural scales defined in Table~\ref{Tab: Natural Scales}.}\label{Tab: Pb TFs Example}
\end{table}

Tables~\ref{Tab: Ca TFs Example} and~\ref{Tab: Pb TFs Example} show for the $^{48}$Ca and $^{208}$Pb examples, respectively, the original locations $\boldsymbol{q_0}$ as well as the locations $\boldsymbol{q_m}$ that minimize the FOM defined in Eq.~\eqref{Eq: FOM}. Tables~\ref{Tab: Ca TFs Example} and~\ref{Tab: Pb TFs Example} also show the numerical values of the $\mathcal{T\!F}$ times the respective error $\sigma_j$ for the density at $r=0$ fm and the radius for both data sets $\boldsymbol{q_0}$ and $\boldsymbol{q_m}$. Let us recall that it is the total interior density (the 30 grid points in Eq.~\eqref{Eq: Grid Points}), what goes in the FOM Eq.~\eqref{Eq: FOM}. We are using $\rho(0)$ as a representative of the total interior density. All of the transfer function values have been divided by the natural scales defined in Table~\ref{Tab: Natural Scales}. 

By adding in quadrature each element in the rows of Tables~\ref{Tab: Ca TFs Example} and~\ref{Tab: Pb TFs Example} (the transfer functions values times the respective errors), the total variance in $\rho(0)$ or $R$ can be calculated from Eq.~\eqref{Eq: TF Variance}, in units of the natural scale. In this sense, each number in the table represents the contribution of that measurement to the total variance in that quantity. This allows us to identify, for example, that the variance in $R$ in the case of $\boldsymbol{q_m}$ for $^{48}$Ca is completely driven by the first observation $q_1=0.9$ fm$^{-1}$, while the contribution of the last point $q_5=2.69$ fm$^{-1}$ is negligible. From an experimental design point of view, this means that if our main goal is to reduce the uncertainty in the radius, we must allocate the resources accordingly and reduce the error bar on $q_1$ rather than reducing the error bars in the other locations.

\begin{figure*}[!htbp]
	\centering
		\includegraphics[width=0.9\textwidth]{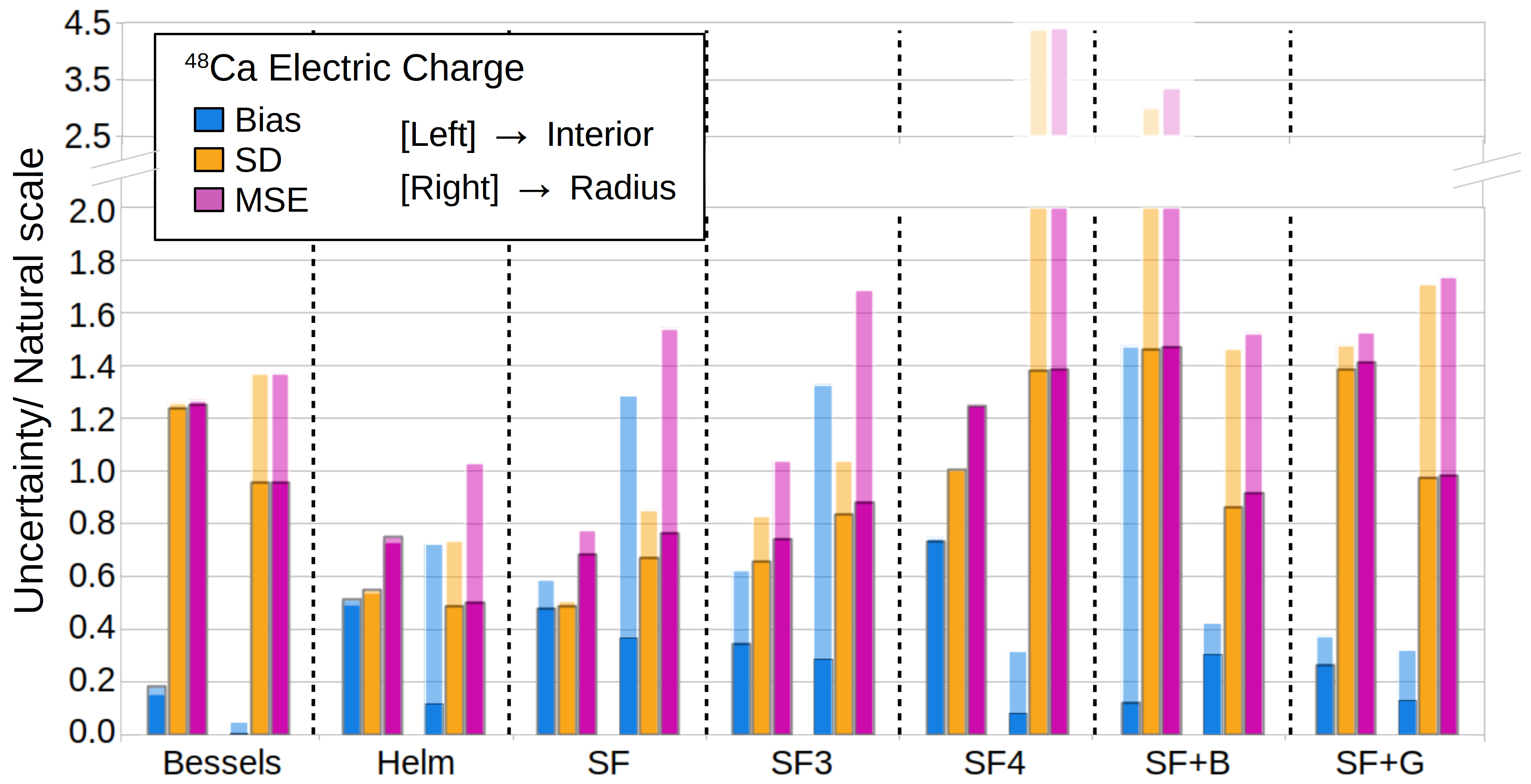}
	\caption{Seven models comparison for recovering the interior density and mean charge radius from the charge form factor data on $^{48}$Ca. The bias, SD, and MSE for each model are shown in their respective three left columns for the interior density, and the three right columns for the radius, respectively. All quantities have been divided by their natural scales: $\Delta \rho_{\text{Ca}} = 0.00015$ fm$^{-3}$ and $\Delta R_{\text{Ca}}= 0.04$ fm. The solid columns represent the optimal locations $\boldsymbol{q_m}$, while the borderless ones the starting $\boldsymbol{q_0}$.}\label{Fig: Models Ch Ca48}
\end{figure*}

\begin{figure*}[!htbp]
	\centering
		\includegraphics[width=0.9\textwidth]{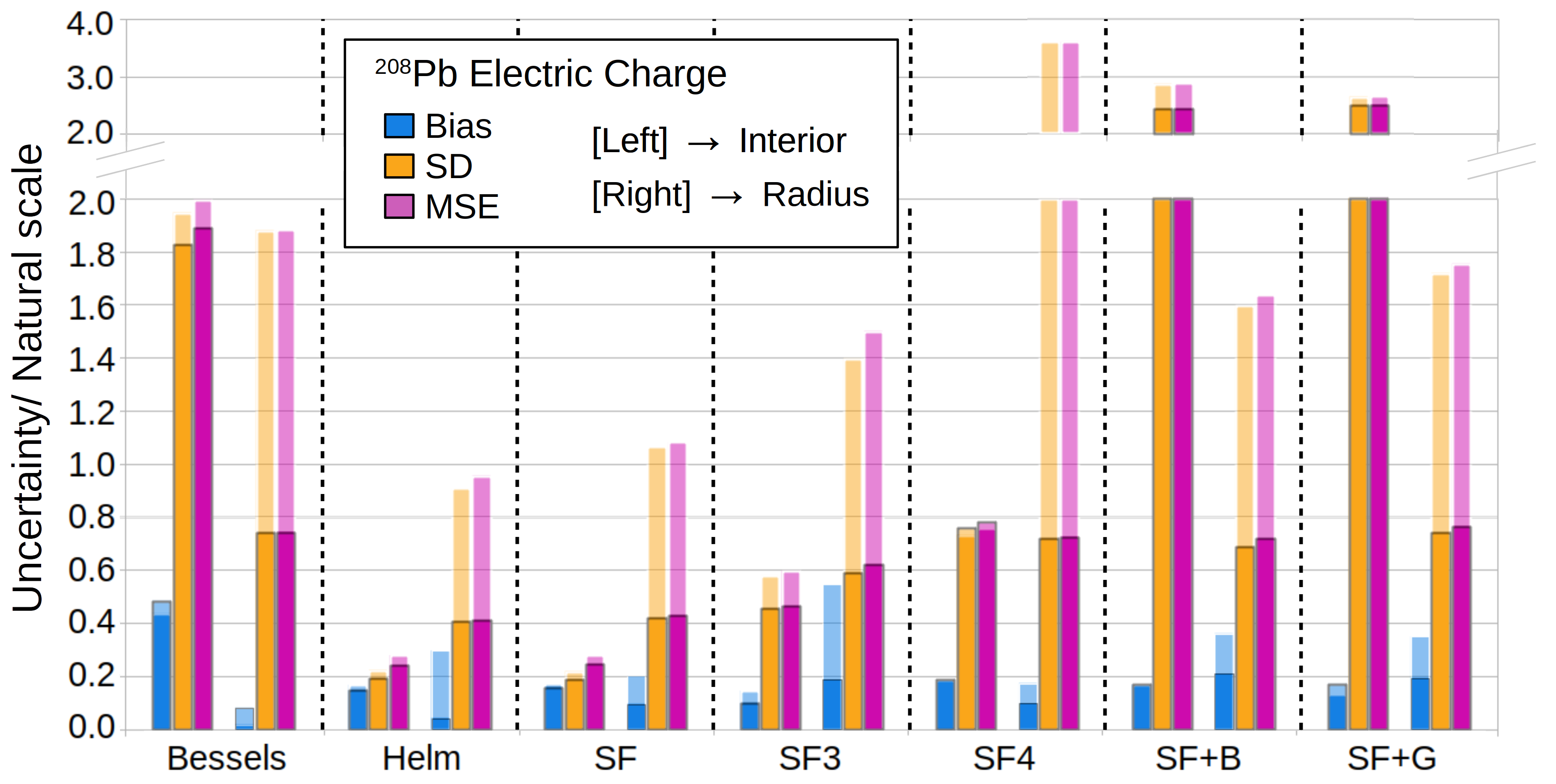}
	\caption{Seven models comparison for recovering the interior density and mean charge radius from the charge form factor data on $^{208}$Pb. The bias, SD, and MSE for each model is shown in their respective three left columns for the interior density, and the three right columns for the radius, respectively. All quantities have been divided by their natural scales: $\Delta \rho_{\text{Pb}} = 0.00008$ fm$^{-3}$ and $\Delta R_{\text{Pb}}= 0.06$ fm. The solid columns represent the optimal locations $\boldsymbol{q_m}$, while the borderless ones the starting $\boldsymbol{q_0}$.}\label{Fig: Models Ch Pb208}
\end{figure*}

Figures~\ref{Fig: Models Ch Ca48} and~\ref{Fig: Models Ch Pb208} show, for $^{48}$Ca and $^{208}$Pb, respectively, the results for comparing the seven models defined in Sec~\ref{Sec: Models, parameters and errors} to recover the interior charge density and charge radius. The numerical values shown in these Figures are written in Tables~\ref{Tab: Ca Ch Results Numbers} and \ref{Tab: Pb Ch Results Numbers}. Tables~\ref{Tab: Opt Loc Ca Ch} and~\ref{Tab: Opt Loc Pb Ch} show the associated optimal locations $\boldsymbol{q_m}$ for every model for $^{48}$Ca and $^{208}$Pb, respectively.

\begin{table}[]
\centering
\begin{tabular}{|l|l|l|l|l|l|}
\hline
$q_j$ [fm$^{-1}$]       & $q_1$   & $q_2$   & $q_3$   & $q_4$   & $q_5$   \\ \hline
Bessels & 0.65 & 1.26 & 1.73 & 2.12 & 2.55 \\ \hline
Helm    & 0.68 & 0.71 & 1.39 & 2.09 & 2.10 \\ \hline
SF      & 0.44 & 0.62 & 1.24 & 2.09 & 2.79 \\ \hline
SF3     & 0.56 & 0.90 & 1.20 & 1.84 & 2.50 \\ \hline
SF4     & 0.50 & 1.10 & 1.92 & 2.12 & 2.59 \\ \hline
SF+B    & 0.67 & 1.31 & 1.80 & 2.43 & 3.20 \\ \hline
SF+G    & 0.63 & 1.25 & 1.78 & 2.17 & 2.80 \\ \hline
\end{tabular}\caption{Optimal locations $\boldsymbol{q_m}$ for each model when optimizing the Figure of Merit in Eq.~\eqref{Eq: FOM}. These $q_j$ values correspond to the solid bars results in Fig~\ref{Fig: Models Ch Ca48} for the charge density of ${48}$Ca. All values are in units of fm$^{-1}$.}\label{Tab: Opt Loc Ca Ch}
\end{table}

\begin{table}[]
\centering
\begin{tabular}{|l|l|l|l|l|l|}
\hline
$q_j$ [fm$^{-1}$]       & $q_1$   & $q_2$   & $q_3$   & $q_4$   & $q_5$  \\ \hline
Bessels & 0.37 & 0.86 & 1.20 & 1.60 & 1.80 \\ \hline
Helm    & 0.41 & 0.43 & 0.83 & 0.88 & 1.31 \\ \hline
SF      & 0.42 & 0.43 & 0.83 & 0.87 & 0.88 \\ \hline
SF3     & 0.23 & 0.41 & 0.88 & 1.07 & 1.16 \\ \hline
SF4     & 0.25 & 0.36 & 0.70 & 1.22 & 1.51 \\ \hline
SF+B    & 0.38 & 0.87 & 1.27 & 1.75 & 1.97 \\ \hline
SF+G    & 0.36 & 0.90 & 1.29 & 1.68 & 1.91 \\ \hline
\end{tabular}\caption{Optimal locations $\boldsymbol{q_m}$ for each model when optimizing the Figure of Merit in Eq.~\eqref{Eq: FOM}. These $q_j$ values correspond to the solid bars results in Fig~\ref{Fig: Models Ch Pb208} for the charge density of $^{208}$Pb. All values are in units of fm$^{-1}$.}\label{Tab: Opt Loc Pb Ch}
\end{table}

We judge each model by their individual bias, standard deviation, and MSE both in the original locations $\boldsymbol{q_0}$ as well as the optimized locations $\boldsymbol{q_m}$ where the FOM is minimized. As we did in Sec~\ref{Sec: Weak Charge Results}, the starting locations are $\boldsymbol{q_0}=[0.9,1.35,1.8,2.24,2.69]$  fm$^{-1}$ for $^{48}$Ca and $\boldsymbol{q_0}=[0.63, 0.94, 1.26, 1.57, 1.88]$ fm$^{-1}$ for $^{208}$Pb.

\begin{table*}[]
\centering
\begin{tabular}{l|llllll|llllll|}
\cline{2-13}
                              & $\boldsymbol{q_0}$                        &                           &                           &                           &                           &      & $\boldsymbol{q_m}$                        &                           &                           &                           &                           &      \\
                              & Interior                       &                           & \multicolumn{1}{l|}{}     & Radius                       &                           &      & Interior                       &                           & \multicolumn{1}{l|}{}     & Radius                       &                           &      \\
                              & \multicolumn{1}{l|}{Bias} & \multicolumn{1}{l|}{SD}   & \multicolumn{1}{l|}{MSE}  & \multicolumn{1}{l|}{Bias} & \multicolumn{1}{l|}{SD}   & MSE  & \multicolumn{1}{l|}{Bias} & \multicolumn{1}{l|}{SD}   & \multicolumn{1}{l|}{MSE}  & \multicolumn{1}{l|}{Bias} & \multicolumn{1}{l|}{SD}   & MSE  \\ \hline
\multicolumn{1}{|l|}{Bessels} & \multicolumn{1}{l|}{0.15} & \multicolumn{1}{l|}{1.26} & \multicolumn{1}{l|}{1.27} & \multicolumn{1}{l|}{0.05} & \multicolumn{1}{l|}{1.37} & 1.37 & \multicolumn{1}{l|}{0.19} & \multicolumn{1}{l|}{1.24} & \multicolumn{1}{l|}{1.25} & \multicolumn{1}{l|}{0.00} & \multicolumn{1}{l|}{0.96} & 0.96 \\ \hline
\multicolumn{1}{|l|}{Helm}    & \multicolumn{1}{l|}{0.50} & \multicolumn{1}{l|}{0.54} & \multicolumn{1}{l|}{0.73} & \multicolumn{1}{l|}{0.72} & \multicolumn{1}{l|}{0.73} & 1.03 & \multicolumn{1}{l|}{0.51} & \multicolumn{1}{l|}{0.55} & \multicolumn{1}{l|}{0.75} & \multicolumn{1}{l|}{0.12} & \multicolumn{1}{l|}{0.49} & 0.50 \\ \hline
\multicolumn{1}{|l|}{SF}      & \multicolumn{1}{l|}{0.59} & \multicolumn{1}{l|}{0.51} & \multicolumn{1}{l|}{0.78} & \multicolumn{1}{l|}{1.28} & \multicolumn{1}{l|}{0.85} & 1.54 & \multicolumn{1}{l|}{0.48} & \multicolumn{1}{l|}{0.49} & \multicolumn{1}{l|}{0.68} & \multicolumn{1}{l|}{0.37} & \multicolumn{1}{l|}{0.67} & 0.77 \\ \hline
\multicolumn{1}{|l|}{SF3}     & \multicolumn{1}{l|}{0.62} & \multicolumn{1}{l|}{0.83} & \multicolumn{1}{l|}{1.04} & \multicolumn{1}{l|}{1.33} & \multicolumn{1}{l|}{1.04} & 1.68 & \multicolumn{1}{l|}{0.34} & \multicolumn{1}{l|}{0.66} & \multicolumn{1}{l|}{0.74} & \multicolumn{1}{l|}{0.29} & \multicolumn{1}{l|}{0.84} & 0.88 \\ \hline
\multicolumn{1}{|l|}{SF4}     & \multicolumn{1}{l|}{0.75} & \multicolumn{1}{l|}{1.00} & \multicolumn{1}{l|}{1.25} & \multicolumn{1}{l|}{0.31} & \multicolumn{1}{l|}{4.38} & 4.40 & \multicolumn{1}{l|}{0.73} & \multicolumn{1}{l|}{1.01} & \multicolumn{1}{l|}{1.25} & \multicolumn{1}{l|}{0.08} & \multicolumn{1}{l|}{1.38} & 1.38 \\ \hline
\multicolumn{1}{|l|}{SF+B}    & \multicolumn{1}{l|}{1.47} & \multicolumn{1}{l|}{3.00} & \multicolumn{1}{l|}{3.35} & \multicolumn{1}{l|}{0.42} & \multicolumn{1}{l|}{1.46} & 1.52 & \multicolumn{1}{l|}{0.12} & \multicolumn{1}{l|}{1.46} & \multicolumn{1}{l|}{1.47} & \multicolumn{1}{l|}{0.30} & \multicolumn{1}{l|}{0.86} & 0.92 \\ \hline
\multicolumn{1}{|l|}{SF+G}    & \multicolumn{1}{l|}{0.37} & \multicolumn{1}{l|}{1.48} & \multicolumn{1}{l|}{1.52} & \multicolumn{1}{l|}{0.32} & \multicolumn{1}{l|}{1.71} & 1.74 & \multicolumn{1}{l|}{0.26} & \multicolumn{1}{l|}{1.39} & \multicolumn{1}{l|}{1.41} & \multicolumn{1}{l|}{0.13} & \multicolumn{1}{l|}{0.97} & 0.98 \\ \hline
\end{tabular}

\caption{Numerical values for the data displayed in Fig~\ref{Fig: Models Ch Ca48}: model comparison for the electric charge density of $^{48}$Ca.}\label{Tab: Ca Ch Results Numbers}
\end{table*}

\begin{table*}[]
\centering
\begin{tabular}{l|llllll|llllll|}
\cline{2-13}
                              & $\boldsymbol{q_0}$                        &                           &                           &                           &                           &      & $\boldsymbol{q_m}$                        &                           &                           &                           &                           &      \\
                              & Interior                       &                           & \multicolumn{1}{l|}{}     & Radius                       &                           &      & Interior                       &                           & \multicolumn{1}{l|}{}     & Radius                       &                           &      \\
                              & \multicolumn{1}{l|}{Bias} & \multicolumn{1}{l|}{SD}   & \multicolumn{1}{l|}{MSE}  & \multicolumn{1}{l|}{Bias} & \multicolumn{1}{l|}{SD}   & MSE  & \multicolumn{1}{l|}{Bias} & \multicolumn{1}{l|}{SD}   & \multicolumn{1}{l|}{MSE}  & \multicolumn{1}{l|}{Bias} & \multicolumn{1}{l|}{SD}   & MSE  \\ \hline
\multicolumn{1}{|l|}{Bessels} & \multicolumn{1}{l|}{0.44} & \multicolumn{1}{l|}{1.95} & \multicolumn{1}{l|}{2.00} & \multicolumn{1}{l|}{0.02} & \multicolumn{1}{l|}{1.88} & 1.88 & \multicolumn{1}{l|}{0.48} & \multicolumn{1}{l|}{1.82} & \multicolumn{1}{l|}{1.89} & \multicolumn{1}{l|}{0.08} & \multicolumn{1}{l|}{0.74} & 0.74 \\ \hline
\multicolumn{1}{|l|}{Helm}    & \multicolumn{1}{l|}{0.17} & \multicolumn{1}{l|}{0.22} & \multicolumn{1}{l|}{0.28} & \multicolumn{1}{l|}{0.30} & \multicolumn{1}{l|}{0.91} & 0.96 & \multicolumn{1}{l|}{0.14} & \multicolumn{1}{l|}{0.19} & \multicolumn{1}{l|}{0.24} & \multicolumn{1}{l|}{0.04} & \multicolumn{1}{l|}{0.41} & 0.41 \\ \hline
\multicolumn{1}{|l|}{SF}      & \multicolumn{1}{l|}{0.17} & \multicolumn{1}{l|}{0.22} & \multicolumn{1}{l|}{0.28} & \multicolumn{1}{l|}{0.20} & \multicolumn{1}{l|}{1.06} & 1.08 & \multicolumn{1}{l|}{0.16} & \multicolumn{1}{l|}{0.19} & \multicolumn{1}{l|}{0.24} & \multicolumn{1}{l|}{0.09} & \multicolumn{1}{l|}{0.42} & 0.43 \\ \hline
\multicolumn{1}{|l|}{SF3}     & \multicolumn{1}{l|}{0.15} & \multicolumn{1}{l|}{0.58} & \multicolumn{1}{l|}{0.60} & \multicolumn{1}{l|}{0.55} & \multicolumn{1}{l|}{1.40} & 1.50 & \multicolumn{1}{l|}{0.10} & \multicolumn{1}{l|}{0.45} & \multicolumn{1}{l|}{0.46} & \multicolumn{1}{l|}{0.19} & \multicolumn{1}{l|}{0.59} & 0.62 \\ \hline
\multicolumn{1}{|l|}{SF4}     & \multicolumn{1}{l|}{0.19} & \multicolumn{1}{l|}{0.73} & \multicolumn{1}{l|}{0.76} & \multicolumn{1}{l|}{0.17} & \multicolumn{1}{l|}{3.59} & 3.59 & \multicolumn{1}{l|}{0.18} & \multicolumn{1}{l|}{0.76} & \multicolumn{1}{l|}{0.78} & \multicolumn{1}{l|}{0.09} & \multicolumn{1}{l|}{0.72} & 0.72 \\ \hline
\multicolumn{1}{|l|}{SF+B}    & \multicolumn{1}{l|}{0.17} & \multicolumn{1}{l|}{2.86} & \multicolumn{1}{l|}{2.86} & \multicolumn{1}{l|}{0.36} & \multicolumn{1}{l|}{1.60} & 1.64 & \multicolumn{1}{l|}{0.17} & \multicolumn{1}{l|}{2.43} & \multicolumn{1}{l|}{2.44} & \multicolumn{1}{l|}{0.21} & \multicolumn{1}{l|}{0.69} & 0.72 \\ \hline
\multicolumn{1}{|l|}{SF+G}    & \multicolumn{1}{l|}{0.13} & \multicolumn{1}{l|}{2.64} & \multicolumn{1}{l|}{2.64} & \multicolumn{1}{l|}{0.35} & \multicolumn{1}{l|}{1.72} & 1.75 & \multicolumn{1}{l|}{0.17} & \multicolumn{1}{l|}{2.50} & \multicolumn{1}{l|}{2.51} & \multicolumn{1}{l|}{0.19} & \multicolumn{1}{l|}{0.74} & 0.76 \\ \hline
\end{tabular}

\caption{Numerical values for the data displayed in Fig~\ref{Fig: Models Ch Pb208}: model comparison for the electric charge density of $^{208}$Pb.}\label{Tab: Pb Ch Results Numbers}
\end{table*}

We observe similar trends as the one displayed by Figures~\ref{Fig: Models Wk Ca48} and~\ref{Fig: Models Wk Pb208} when comparing models for the weak charge densities. Changing the data locations can result in an important reduction of the MSE, for example by a factor of two by the SF in the $^{48}$Ca radius. There are also significant variations in performance among the models, for example a factor of three between the SF+B and SF+G when compared to the Helm model when extracting the interior density of $^{208}$Pb for $\boldsymbol{q_0}$. Once again, for the data range and errors we have assumed, the Helm and SF model seem to outperform all of the other options in both nuclei. 

Finally, it is interesting to analyze the distribution of optimal locations $\boldsymbol{q_m}$ for the seven models in Tables~\ref{Tab: Opt Loc Ca Ch} and~\ref{Tab: Opt Loc Pb Ch}. Let us recall that we limited the maximum value of any $q_j$ to be less than $3.5$ fm$^{-1}$ for $^{48}$Ca and less than $2$ fm$^{-1}$ for $^{208}$Pb. Although there is no clear pattern, both the SF and Helm models seem to have overall smaller values of $q_j$, while the models involving Bessels seem to be more in the high end. Depending on the experiment details and constraints, some regions of the $q$ space would be easier to access than others. In that case, a more detailed analysis could be done to optimize a modified version of the FOM in which we take into account the experimental budget. It is very likely that the FOM would not be extremely sensitive to the exact locations of $\boldsymbol{q_m}$. Therefore, an adjustment of each $q_j$ could result in a notorious reduction of the experimental budget while the FOM deteriorates just a small amount.

\section{Details about the $^{48}$Ca and $^{208}$Pb weak charge analysis}\label{Sec: App Details about weak charge}

This appendix presents tables and details relevant to the $^{48}$Ca and $^{208}$Pb weak charge density model comparison developed in Sec~\ref{Sec: Weak Charge Results}. The numerical values associated with Figures~\ref{Fig: Models Wk Ca48} and~\ref{Fig: Models Wk Pb208} are shown in Tables~\ref{Tab: Numerical Wk Ca48 Model Comparison} and \ref{Tab: Numerical Wk Pb208 Model Comparison}, respectively. Tables~\ref{Tab: Opt values Ca Wk} and~\ref{Tab: Opt values Pb Wk} show the associated optimal locations $\boldsymbol{q_m}$ for every model for $^{48}$Ca and $^{208}$Pb, respectively.

\begin{table}[]
\centering
\begin{tabular}{|l|l|l|l|l|l|}
\hline
$q_j$ [fm$^{-1}$]       & $q_1$   & $q_2$   & $q_3$   & $q_4$   & $q_5$  \\ \hline
Bessels & 0.71 & 1.32 & 1.81 & 2.28 & 2.65 \\ \hline
Helm    & 0.51 & 0.63 & 0.77 & 1.44 & 2.93 \\ \hline
SF      & 0.74 & 0.92 & 1.36 & 1.67 & 2.50 \\ \hline
SF3     & 0.54 & 1.32 & 1.69 & 1.87 & 2.42 \\ \hline
SF4     & 0.44 & 1.41 & 2.10 & 2.82 & 2.88 \\ \hline
SF+B    & 0.61 & 1.26 & 1.78 & 2.32 & 3.03 \\ \hline
SF+G    & 0.57 & 1.18 & 1.69 & 2.06 & 2.68 \\ \hline
\end{tabular}\caption{Optimal locations $\boldsymbol{q_m}$ for each model when optimizing the Figure of Merit in Eq.~\eqref{Eq: FOM}. These $q_j$ values correspond to the solid columns results in Fig~\ref{Fig: Models Wk Ca48} for the five generated weak charge densities of $^{48}$Ca. All values are in units of fm$^{-1}$.}\label{Tab: Opt values Ca Wk}
\end{table}

\begin{table}[]
\centering
\begin{tabular}{|l|l|l|l|l|l|}
\hline
$q_j$ [fm$^{-1}$]       & $q_1$   & $q_2$   & $q_3$   & $q_4$   & $q_5$  \\ \hline
Bessels & 0.37 & 0.84 & 1.21 & 1.61 & 1.80 \\ \hline
Helm    & 0.37 & 0.40 & 0.84 & 1.23 & 1.24 \\ \hline
SF      & 0.37 & 0.40 & 0.77 & 0.84 & 1.36 \\ \hline
SF3     & 0.38 & 0.43 & 0.87 & 1.03 & 1.17 \\ \hline
SF4     & 0.32 & 0.52 & 0.68 & 0.81 & 1.09 \\ \hline
SF+B    & 0.37 & 0.84 & 1.25 & 1.70 & 1.94 \\ \hline
SF+G    & 0.32 & 0.82 & 1.23 & 1.65 & 1.85 \\ \hline
\end{tabular}\caption{Optimal locations $\boldsymbol{q_m}$ for each model when optimizing the Figure of Merit in Eq.~\eqref{Eq: FOM}. These $q_j$ values correspond to the solid columns results in Fig~\ref{Fig: Models Wk Pb208} for the five generated weak charge densities of $^{208}$Pb. All values are in units of fm$^{-1}$.}\label{Tab: Opt values Pb Wk}
\end{table}

\begin{table*}[]
\centering

\begin{tabular}{l|llllll|llllll|}
\cline{2-13}
                              & $\boldsymbol{q_0}$                        &                           &                           &                           &                           &      & $\boldsymbol{q_m}$                        &                           &                           &                           &                           &      \\
                              & Interior                       &                           & \multicolumn{1}{l|}{}     & Radius                       &                           &      & Interior                       &                           & \multicolumn{1}{l|}{}     & Radius                       &                           &      \\
                              & \multicolumn{1}{l|}{Bias} & \multicolumn{1}{l|}{SD}   & \multicolumn{1}{l|}{MSE}  & \multicolumn{1}{l|}{Bias} & \multicolumn{1}{l|}{SD}   & MSE  & \multicolumn{1}{l|}{Bias} & \multicolumn{1}{l|}{SD}   & \multicolumn{1}{l|}{MSE}  & \multicolumn{1}{l|}{Bias} & \multicolumn{1}{l|}{SD}   & MSE  \\ \hline
\multicolumn{1}{|l|}{Bessels} & \multicolumn{1}{l|}{0.09} & \multicolumn{1}{l|}{1.26} & \multicolumn{1}{l|}{1.26} & \multicolumn{1}{l|}{0.06} & \multicolumn{1}{l|}{1.31} & 1.31 & \multicolumn{1}{l|}{0.08} & \multicolumn{1}{l|}{1.30} & \multicolumn{1}{l|}{1.31} & \multicolumn{1}{l|}{0.30} & \multicolumn{1}{l|}{0.87} & 0.92 \\ \hline
\multicolumn{1}{|l|}{Helm}    & \multicolumn{1}{l|}{1.22} & \multicolumn{1}{l|}{0.50} & \multicolumn{1}{l|}{1.32} & \multicolumn{1}{l|}{1.08} & \multicolumn{1}{l|}{0.87} & 1.39 & \multicolumn{1}{l|}{0.70} & \multicolumn{1}{l|}{0.56} & \multicolumn{1}{l|}{0.90} & \multicolumn{1}{l|}{0.03} & \multicolumn{1}{l|}{0.45} & 0.45 \\ \hline
\multicolumn{1}{|l|}{SF}      & \multicolumn{1}{l|}{1.09} & \multicolumn{1}{l|}{0.47} & \multicolumn{1}{l|}{1.19} & \multicolumn{1}{l|}{0.87} & \multicolumn{1}{l|}{1.04} & 1.37 & \multicolumn{1}{l|}{0.62} & \multicolumn{1}{l|}{0.41} & \multicolumn{1}{l|}{0.74} & \multicolumn{1}{l|}{0.24} & \multicolumn{1}{l|}{0.76} & 0.81 \\ \hline
\multicolumn{1}{|l|}{SF3}     & \multicolumn{1}{l|}{1.20} & \multicolumn{1}{l|}{0.86} & \multicolumn{1}{l|}{1.47} & \multicolumn{1}{l|}{2.14} & \multicolumn{1}{l|}{1.18} & 2.44 & \multicolumn{1}{l|}{0.60} & \multicolumn{1}{l|}{0.93} & \multicolumn{1}{l|}{1.11} & \multicolumn{1}{l|}{1.04} & \multicolumn{1}{l|}{0.94} & 1.40 \\ \hline
\multicolumn{1}{|l|}{SF4}     & \multicolumn{1}{l|}{0.94} & \multicolumn{1}{l|}{1.39} & \multicolumn{1}{l|}{1.68} & \multicolumn{1}{l|}{2.71} & \multicolumn{1}{l|}{1.70} & 3.22 & \multicolumn{1}{l|}{0.63} & \multicolumn{1}{l|}{1.62} & \multicolumn{1}{l|}{1.74} & \multicolumn{1}{l|}{1.23} & \multicolumn{1}{l|}{1.10} & 1.65 \\ \hline
\multicolumn{1}{|l|}{SF+B}    & \multicolumn{1}{l|}{0.70} & \multicolumn{1}{l|}{2.37} & \multicolumn{1}{l|}{2.47} & \multicolumn{1}{l|}{1.04} & \multicolumn{1}{l|}{1.83} & 2.10 & \multicolumn{1}{l|}{0.17} & \multicolumn{1}{l|}{1.47} & \multicolumn{1}{l|}{1.48} & \multicolumn{1}{l|}{0.90} & \multicolumn{1}{l|}{0.95} & 1.31 \\ \hline
\multicolumn{1}{|l|}{SF+G}    & \multicolumn{1}{l|}{0.19} & \multicolumn{1}{l|}{1.44} & \multicolumn{1}{l|}{1.45} & \multicolumn{1}{l|}{1.02} & \multicolumn{1}{l|}{2.25} & 2.47 & \multicolumn{1}{l|}{0.22} & \multicolumn{1}{l|}{1.35} & \multicolumn{1}{l|}{1.36} & \multicolumn{1}{l|}{0.82} & \multicolumn{1}{l|}{1.11} & 1.38 \\ \hline
\end{tabular}

\caption{Numerical values for the data displayed in Fig~\ref{Fig: Models Wk Ca48}: model comparison for the five generated weak charge densities of $^{48}$Ca.}\label{Tab: Numerical Wk Ca48 Model Comparison}
\end{table*}

\begin{table*}[]
\centering
\begin{tabular}{l|llllll|llllll|}
\cline{2-13}
                              & $\boldsymbol{q_0}$                        &                           &                           &                           &                           &      & $\boldsymbol{q_m}$                        &                           &                           &                           &                           &      \\
                              & Interior                       &                           & \multicolumn{1}{l|}{}     & Radius                       &                           &      & Interior                       &                           & \multicolumn{1}{l|}{}     & Radius                       &                           &      \\
                              & \multicolumn{1}{l|}{Bias} & \multicolumn{1}{l|}{SD}   & \multicolumn{1}{l|}{MSE}  & \multicolumn{1}{l|}{Bias} & \multicolumn{1}{l|}{SD}   & MSE  & \multicolumn{1}{l|}{Bias} & \multicolumn{1}{l|}{SD}   & \multicolumn{1}{l|}{MSE}  & \multicolumn{1}{l|}{Bias} & \multicolumn{1}{l|}{SD}   & MSE  \\ \hline
\multicolumn{1}{|l|}{Bessels} & \multicolumn{1}{l|}{0.20} & \multicolumn{1}{l|}{1.95} & \multicolumn{1}{l|}{1.96} & \multicolumn{1}{l|}{0.40} & \multicolumn{1}{l|}{1.82} & 1.86 & \multicolumn{1}{l|}{0.36} & \multicolumn{1}{l|}{1.83} & \multicolumn{1}{l|}{1.87} & \multicolumn{1}{l|}{0.05} & \multicolumn{1}{l|}{0.70} & 0.71 \\ \hline
\multicolumn{1}{|l|}{Helm}    & \multicolumn{1}{l|}{0.10} & \multicolumn{1}{l|}{0.23} & \multicolumn{1}{l|}{0.25} & \multicolumn{1}{l|}{0.59} & \multicolumn{1}{l|}{1.11} & 1.26 & \multicolumn{1}{l|}{0.11} & \multicolumn{1}{l|}{0.21} & \multicolumn{1}{l|}{0.24} & \multicolumn{1}{l|}{0.01} & \multicolumn{1}{l|}{0.45} & 0.45 \\ \hline
\multicolumn{1}{|l|}{SF}      & \multicolumn{1}{l|}{0.09} & \multicolumn{1}{l|}{0.22} & \multicolumn{1}{l|}{0.24} & \multicolumn{1}{l|}{0.28} & \multicolumn{1}{l|}{1.40} & 1.43 & \multicolumn{1}{l|}{0.09} & \multicolumn{1}{l|}{0.19} & \multicolumn{1}{l|}{0.21} & \multicolumn{1}{l|}{0.25} & \multicolumn{1}{l|}{0.47} & 0.53 \\ \hline
\multicolumn{1}{|l|}{SF3}     & \multicolumn{1}{l|}{0.15} & \multicolumn{1}{l|}{0.59} & \multicolumn{1}{l|}{0.61} & \multicolumn{1}{l|}{0.51} & \multicolumn{1}{l|}{1.99} & 2.06 & \multicolumn{1}{l|}{0.10} & \multicolumn{1}{l|}{0.40} & \multicolumn{1}{l|}{0.41} & \multicolumn{1}{l|}{0.29} & \multicolumn{1}{l|}{0.62} & 0.69 \\ \hline
\multicolumn{1}{|l|}{SF4}     & \multicolumn{1}{l|}{0.14} & \multicolumn{1}{l|}{0.77} & \multicolumn{1}{l|}{0.78} & \multicolumn{1}{l|}{0.61} & \multicolumn{1}{l|}{4.08} & 4.13 & \multicolumn{1}{l|}{0.14} & \multicolumn{1}{l|}{0.60} & \multicolumn{1}{l|}{0.62} & \multicolumn{1}{l|}{0.21} & \multicolumn{1}{l|}{0.83} & 0.86 \\ \hline
\multicolumn{1}{|l|}{SF+B}    & \multicolumn{1}{l|}{0.13} & \multicolumn{1}{l|}{2.70} & \multicolumn{1}{l|}{2.70} & \multicolumn{1}{l|}{0.46} & \multicolumn{1}{l|}{2.19} & 2.24 & \multicolumn{1}{l|}{0.15} & \multicolumn{1}{l|}{2.40} & \multicolumn{1}{l|}{2.41} & \multicolumn{1}{l|}{0.31} & \multicolumn{1}{l|}{0.78} & 0.84 \\ \hline
\multicolumn{1}{|l|}{SF+G}    & \multicolumn{1}{l|}{0.19} & \multicolumn{1}{l|}{2.56} & \multicolumn{1}{l|}{2.57} & \multicolumn{1}{l|}{0.47} & \multicolumn{1}{l|}{2.50} & 2.54 & \multicolumn{1}{l|}{0.26} & \multicolumn{1}{l|}{2.48} & \multicolumn{1}{l|}{2.49} & \multicolumn{1}{l|}{0.27} & \multicolumn{1}{l|}{0.82} & 0.87 \\ \hline
\end{tabular}

\caption{Numerical values for the data displayed in Fig~\ref{Fig: Models Wk Pb208}: model comparison for the five generated weak charge densities of $^{208}$Pb.}\label{Tab: Numerical Wk Pb208 Model Comparison}
\end{table*}

\begin{figure*}[!htbp]
	\centering
		\includegraphics[width=0.9\textwidth]{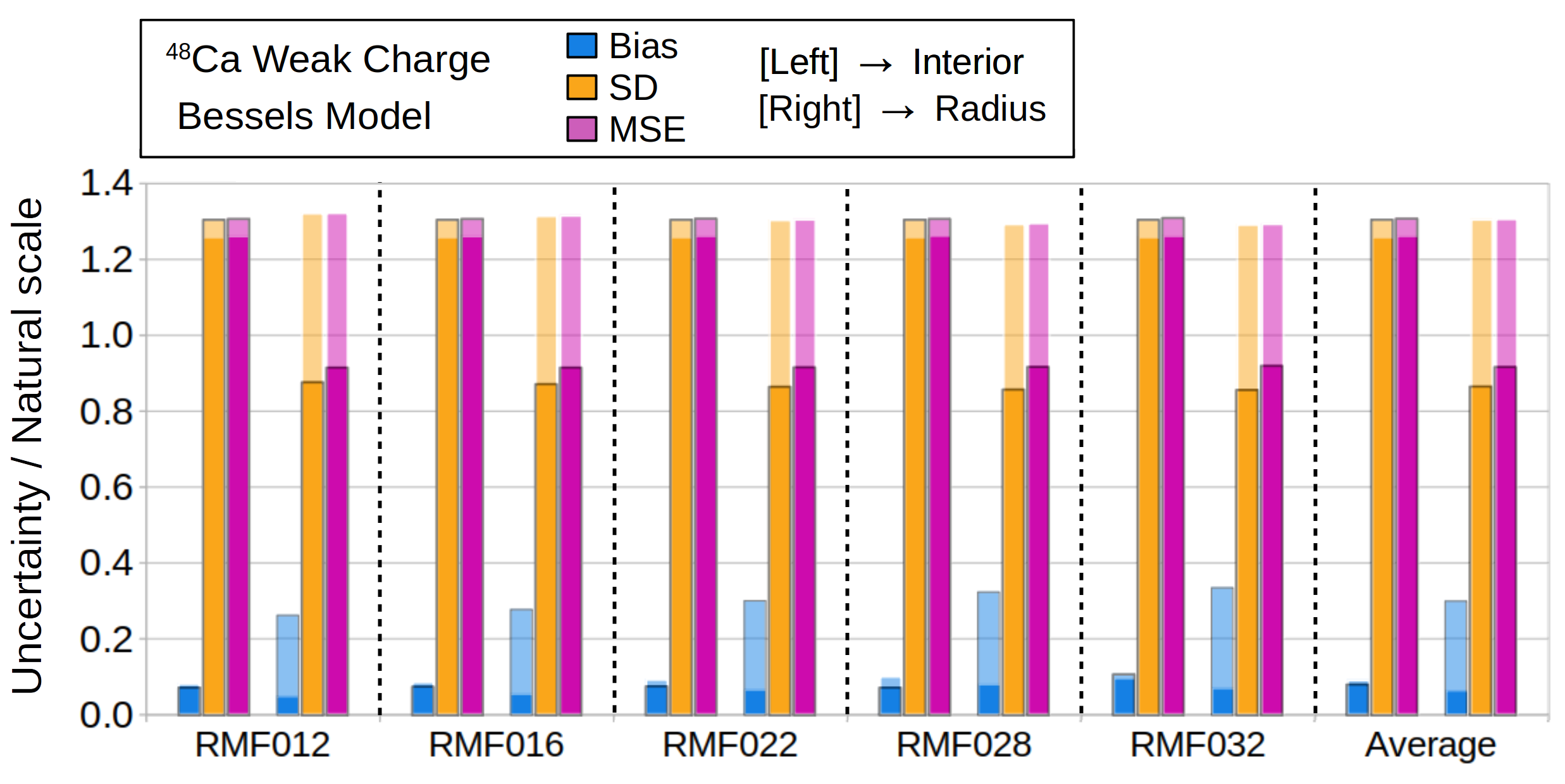}
	\caption{Bessel model bias, standard deviation (SD), and MSE across the five different generators for the weak charge of $^{48}$Ca. The last group of columns shows the square average of the five generators. Within each generator, the first three columns refer to the interior density while the second three to the radius. All quantities have been divided by their natural scales defined in Table~\ref{Tab: Natural Scales}.}\label{Fig: App Ca Wk Bess results}
\end{figure*}

\begin{figure*}[!htbp]
	\centering
		\includegraphics[width=0.9\textwidth]{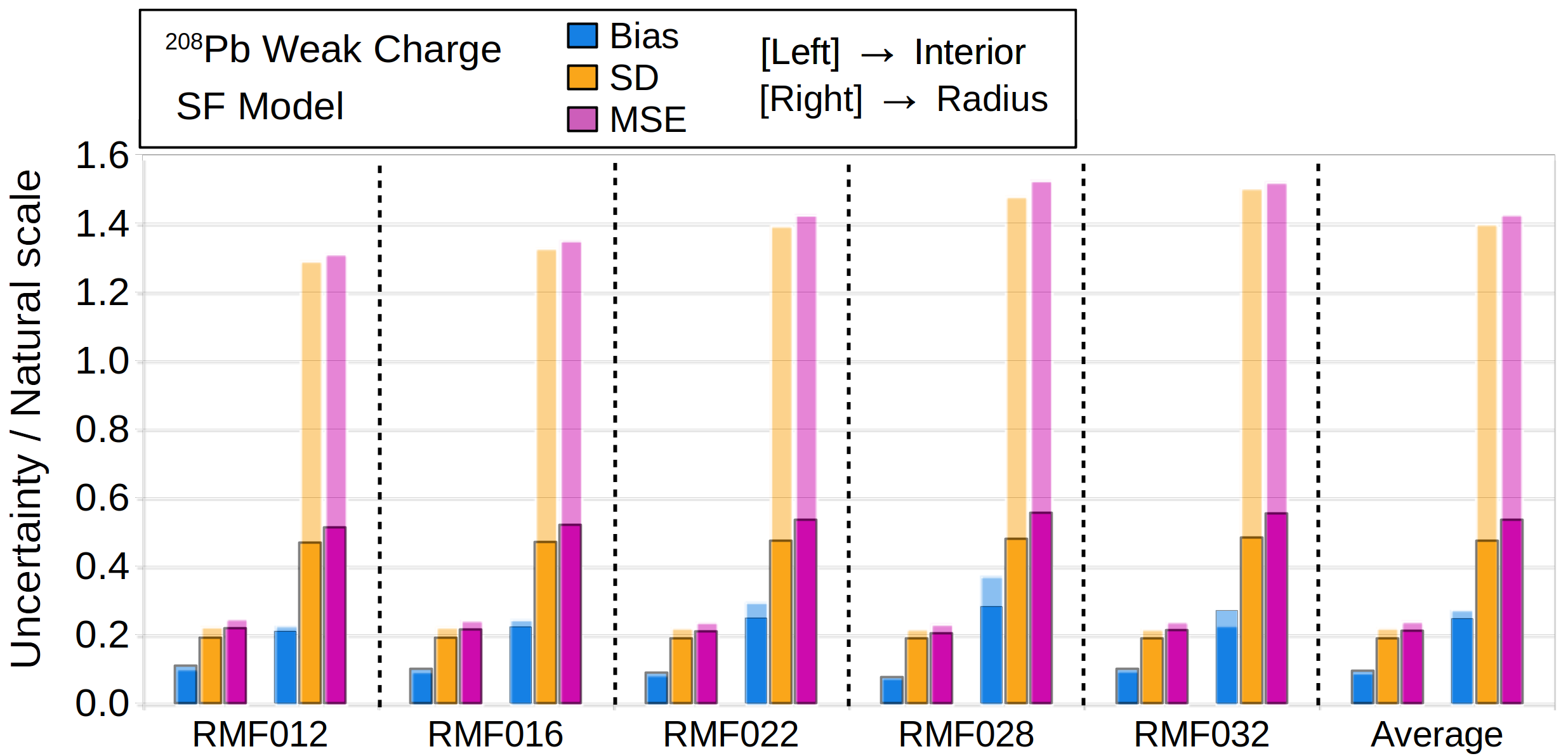}
	\caption{SF model bias, standard deviation (SD) and MSE across the five different generators for the weak charge of $^{208}$Pb. The last group of columns shows the square average of the five generators. Within each generator, the first three columns refer to the interior density while the second three to the radius. All quantities have been divided by their natural scales defined in Table~\ref{Tab: Natural Scales}.}\label{Fig: App Pb Wk SF results}
\end{figure*}

As done in Appendix~\ref{App: Tables Ca and Pb example}, we can analyze the distribution of these optimal locations $\boldsymbol{q_m}$. We observe a similar structure from their charge counterparts: no clear pattern but both SF and Helm models seem to have overall smaller values of $q_j$, while the models involving Bessels seems to be more in the high end. 

Since parity violating experiments are extremely expensive and challenging, the experimental constraints and budget should definitely be considered in a more detailed analysis to optimize a modified version of the FOM. We again anticipate that the FOM will not be extremely sensitive to the exact locations of $\boldsymbol{q_m}$, and that an adjustment of each $q_j$ could result in a huge impact on the budget while the FOM deteriorates just a small amount. This analysis could also change our conclusions regarding the optimal models. For example, for the weak results of $^{208}$Pb, the Helm model seems to have a better FOM combination than the SF, but it could be that the $\boldsymbol{q_m}$ from the SF model are more experimentally accessible than those from the Helm model.

Finally, Figures~\ref{Fig: App Ca Wk Bess results} and~\ref{Fig: App Pb Wk SF results} show, as an example, the detailed results of the Bessels model for $^{48}$Ca and the SF model for $^{208}$Pb, respectively, across the five different generators used in Sec~\ref{Sec: Weak Charge Results}. The last columns on each figure show the squared average (see Eq.~\ref{Eq: Average MSE}) of each quantity (bias, SD, and MSE), which corresponds to the single values displayed in Figures~\ref{Fig: Models Wk Ca48} and \ref{Fig: Models Wk Pb208} in the main text. As can be seen, the variations among different generators are very small. We interpret this as a sign that our conclusions are robust at least within the family of generators we considered in this study.

\section{Details about the role of priors}\label{App: Details about priors SF+G}

This section presents tables with the numerical values of the transfer functions of both the data and the priors related to what was discussed in Sec~\ref{Sec: The role of priors}. Let us recall that we are using the SF+G model with three scenarios for the prior. In terms of their strength these are: 1) $P_0$: no prior (unconstrained SF+G); 2) $P_1$: the specified prior in Eq.~\eqref{Eq: Priors SF+G}; and 3) $P_2$: a very restrictive prior which makes the SF+G practically behave as the SF model without Gaussians.

\begin{table*}[]
\centering
\begin{tabular}{|l|l|l|l|l|l|}
\hline
& $q_1$     & $q_2$     & $q_3$    & $q_4$     & $q_5$     \\ \hline
$P_0$ & 1.06 & 0.46 & 2.38 & 0.35 & 1.10 \\ \hline
$P_1$   & 0.68 & 0.78 & 0.64 & 0.47 & 0.20 \\ \hline
$P_2$       & 0.40 & 0.16 & 0.16 & 0.09 & 0.20 \\ \hline
\end{tabular}\caption{Absolute value transfer function values for the density at $r=0$ fm ( $|\mathcal{T\!F}_j^{\rho(0)}\sigma_j|/\Delta \rho_{\text{Ca}}$) for the five locations $\boldsymbol{q_0}=[0.9,1.35,1.8,2.24,2.69]$ [fm$^{-1}]$ in units of the natural scale $\Delta \rho_{\text{Ca}}$. Each $\mathcal{T\!F}$ has been multiplied by their respective $\sigma_j$ to somehow represent a fraction of the total standard deviation SD. The three different prior options are explored for the SF+G model.}\label{Tab: App TF data prior SF+G}
\end{table*}

Table~\ref{Tab: App TF data prior SF+G} shows the numerical value of $|\mathcal{T\!F}_j^{\rho(0)}\sigma_j|/\Delta \rho_{\text{Ca}}$ for the five different locations and for the three different priors. Let us recall that the FOM in Eq.~\eqref{Eq: FOM}, is calculated using the entire interior density, but in this section we are focusing in $\rho(0)$ as a representative. For each scenario, the total contribution of the data in the variance on the density at $r=0$, namely $\Delta \rho(0)^2$, is obtained by adding the numbers in Table~\ref{Tab: App TF data prior SF+G} in quadrature\footnote{The position $r=0$ fm is only one of the 30 grid points in Eq.~\eqref{Eq: Grid Points}. The total interior MSE will receive contributions not only from $\rho(0)$.}. In this sense, each number in the table reflects how much that particular data point uncertainty $\sigma_j$ contributes to the total band that surrounds $\rho(0)$ in Fig.~\ref{Fig: RMF012 Example Prior}, in units of the natural scale $\Delta \rho_{\text{Ca}}$ from Table~\ref{Tab: Natural Scales}.

It can be observed that as the prior strength increases, the influence of each data point uncertainty $\sigma_j$ tends to decrease, in some cases even by an order of magnitude. This reflects the fact that models constrained by priors, which are in some sense less complex, will present a smaller variance. From the transfer function point of view, this is driven by a more constrained $\widetilde{\mathcal{H}}$ (see Eq.~\eqref{Eq: chi2 tilde Hessian}).

To get a complete picture in terms of standard deviation, we must also look at the variance introduced by the addition of the prior itself. In Eq.~\eqref{Eq: Prior Form}, we can see that each Gaussian prior has an associated center $\omega_k^0$, and ``error" $\sigma_k$ which are analogous to the true observations $y_j$ and true error $\sigma_j$ of the data in Eq.~\eqref{Eq: chi^2 Definition}. Although $\widetilde{\mathcal{H}}$ is more constrained, we now have to take into account the transfer functions ($\mathcal{T\!F}_k^{\rho(0)}$) associated with the fact that in principle $\omega_k^0$ could fluctuate around its center by as much as $\sigma_k$ (to 1 sigma). 

The scenario without prior $P_0$ does not present these transfer functions. The scenario with the extremely constrained prior $P_2$ does not present the transfer functions either since it is as if the parameters associated with the Gaussians were not there in the first place (also, we have observed that numerically $\mathcal{T\!F}_k^{\rho(0)} \sigma_k\rightarrow 0$ when $\sigma_k \rightarrow 0$). Table~\ref{Tab: App prior tfs} shows in the case of the intermediate prior $P_1$ the numerical value of these  $\mathcal{T\!F}_k^{\rho(0)}$ times their respective uncertainty $\sigma_k$ in terms of the natural scale $\Delta \rho_{\text{Ca}}$. The total variance in $\rho(0)$ for the model with this prior is calculated by adding in quadrature these values plus the ones associated with the data in Table~\ref{Tab: App TF data prior SF+G}.

\begin{table}[!htbp]
\centering
\begin{tabular}{|l|l|l|l|}
\hline
      & $A_1$     & $A_2$      & $A_3$      \\ \hline
$P_1$ & 0.89 & 0.24 & 0.58 \\ \hline
\end{tabular}\caption{Absolute value prior transfer function values for the density at $r=0$ fm ( $|\mathcal{T\!F}_k^{\rho(0)}\sigma_k|/\Delta \rho_{\text{Ca}}$) for the three amplitudes in units of the natural scale $\Delta \rho_{\text{Ca}}$. Each $\mathcal{T\!F}$ has been multiplied by their respective $\sigma_k$ (see Eq.~\eqref{Eq: Priors SF+G}) to somehow represent a fraction of the total SD. The finite prior is the only one considered.}\label{Tab: App prior tfs}
\end{table}

We can also describe the induced bias by the inclusion of a prior, i.e., how the estimated central value of $\rho(0)$ is impacted by the new prior. This description is done in terms of the prior transfer functions $\mathcal{T\!F}_k^{\rho(0)}$ by analyzing how the parameters $\boldsymbol{\omega}_{P_0}$ (those obtained in the absence of a prior) move to either $\boldsymbol{\omega}_{P_1}$ or $\boldsymbol{\omega}_{P_2}$ (the fitted parameters when using prior $P_1$ or $P_2$). The reasoning is similar to the discussion about the $\eta_j$ and how they moved the fitted parameters $\boldsymbol{\omega}$ away from the optimal value $\boldsymbol{\omega}_\text{Opt}$. 

Suppose that we are currently at $\boldsymbol{\omega}_{P_0}$ and we add a prior term to $\chi^2$ (converting it to $\widetilde{\chi}^2$ defined in Eq.~\eqref{Eq: chi2 tilde}) in such a way that the centers $\omega_k^0$ fall exactly at the value of their respective parameters $\boldsymbol{\omega}_{P_0}\{k\}$, then the central value of the fitted parameters will not change by the addition of that prior. As a concrete example, let us assume that when the data are fitted, the value of the first unconstrained parameters (A1) is $\boldsymbol{\omega_{P_0}}\{1\}=0.005$. Then, if we add to the total $\chi^2$ a prior term of the form:

\begin{equation}
    \frac{(\omega_1-0.005)^2}{\sigma_1^2},
\end{equation}

and fit the parameters again, we will obtain the same values for all the parameters (the Hessian would be more restricted, but the center location will be intact). Now, let us imagine that we ``perturb" the value of the center $\omega_k^0$ (in the example 0.005), and move it to the original location of the prior we want to enforce (either $\boldsymbol{\omega}_{P_1}$ or $\boldsymbol{\omega}_{P_2}$). This change, which we call $\tilde\eta_k$, will now produce a change in the value of the parameters and therefore, in $\rho(0)$:

\begin{equation}
    \delta \rho(0) = \mathcal{T\!F}_k^{\rho(0)} \tilde\eta_{\omega_k},\label{Eq: App TF priors}
\end{equation}

where $\tilde\eta_k\equiv \omega_k^0 -\boldsymbol{\omega_{P_0}}\{k\} $, the difference between the parameter's value without priors and the new prior centers.  Table~\ref{Tab: App Priors TF} shows in units of $\Delta \rho_{\text{Ca}}$, the predicted change $\delta \rho(0)_k = \mathcal{T\!F}_k^{\rho(0)} \tilde\eta_{\omega_k}$ driven by the inclusions of the prior in the three amplitudes of the Gaussians for $P_1$ and $P_2$. 

To obtain the total predicted change in $\rho(0)$ (in units of $\Delta \rho_{\text{Ca}}$), we must add all numbers in each row. Due to the fact that they alternate signs for this example, the total change in the case with the stronger prior $P_2$ ends up being smaller than the intermediate strength. However, we can appreciate that as the prior strength increases, the influence of its transfer functions increases by a factor between $2$ and $4$. We interpret this as an increase in the bias of the model as compared to its priorless counterpart.

\begin{table}[!htbp]
\centering
\begin{tabular}{|l|l|l|l|}
\hline
      & $A_1$     & $A_2$      & $A_3$      \\ \hline
$P_1$ & 5.1 & -0.17 & -2.7 \\ \hline
$P_2$    & 10 & 0.64  & -11 \\ \hline
\end{tabular}\caption{Prior transfer function values for the density at $r=0$ fm ( $\mathcal{T\!F}_k^{\rho(0)} \eta_{\omega_k}/\Delta \rho_{\text{Ca}}$) for the three amplitudes in units of $\Delta \rho_{\text{Ca}}$. Each $\mathcal{T\!F}$ has been multiplied by their respective $\tilde\eta_{k}$ to represent their fraction of the prior-induced bias.}\label{Tab: App Priors TF}
\end{table}

Finally, note that Eq\eqref{Eq: App TF priors} is just a linear approximation. Nonlinear models will deviate from the predictions of this linear approximation if the parameters change considerably. In this particular case, the predictions on $\delta \rho(0)$ are within $\approx 40\%$ of the true change $\rho(0)$ obtained when re-fitting the parameters with the new priors. Even though the numerical accuracy is not perfect, these types of analysis can help in better estimating the bias vs variance trade-off when including priors.


\break

\bibliography{RefCombined.bib}

\end{document}